\documentclass[11pt,a4paper,oneside]{article}

\usepackage{amsmath}    
\usepackage{amssymb}    
\usepackage{amsfonts}   
\usepackage{mathtools}  
\usepackage{physics}    
\usepackage{siunitx}    

\usepackage{graphicx}   
\usepackage{float}      
\usepackage{subcaption} 
\usepackage{wrapfig}    
\usepackage{tikz}       
\usepackage{pgfplots}   
\pgfplotsset{compat=newest}

\usepackage{booktabs}   
\usepackage{multirow}   
\usepackage{array}      
\usepackage{caption}    
\usepackage{pdflscape}
\usepackage{rotating} 

\usepackage[utf8]{inputenc}     
\usepackage[T1]{fontenc}        
\usepackage{microtype}          
\usepackage{setspace}           
\usepackage{geometry}           
\usepackage{xcolor}

\usepackage{url}
\usepackage{hyperref}                       
\hypersetup{
    colorlinks=true,
    linkcolor=blue,
    filecolor=magenta,
    urlcolor=blue,
    citecolor=red
}

\usepackage{amsthm}     

\DeclareMathOperator{\sinc}{sinc}

\DeclareUnicodeCharacter{03B1}{$\alpha$} 
\DeclareUnicodeCharacter{2009}{$\thinspace$} 
\DeclareUnicodeCharacter{22C6}{$\star$} 
\DeclareUnicodeCharacter{2212}{\textminus}
\DeclareUnicodeCharacter{039B}{$\Lambda$}
\DeclareUnicodeCharacter{2131}{\mathcal{F}}
\DeclareUnicodeCharacter{03A9}{$\Omega$}
\DeclareUnicodeCharacter{223C}{\sim}
\DeclareUnicodeCharacter{0304}{\= }
\DeclareUnicodeCharacter{FE20}{-}  
\DeclareUnicodeCharacter{FE21}{-}  

\title{\textbf{On the Field Theoretical Description of an Alternative Model to Generalized Chaplygin Gas and its Thermodynamic Behaviour}}
\author{
    \textbf{Tamal Mukhopadhyay}~\href{https://orcid.org/0000-0001-9843-906X}{\includegraphics[height=1em]{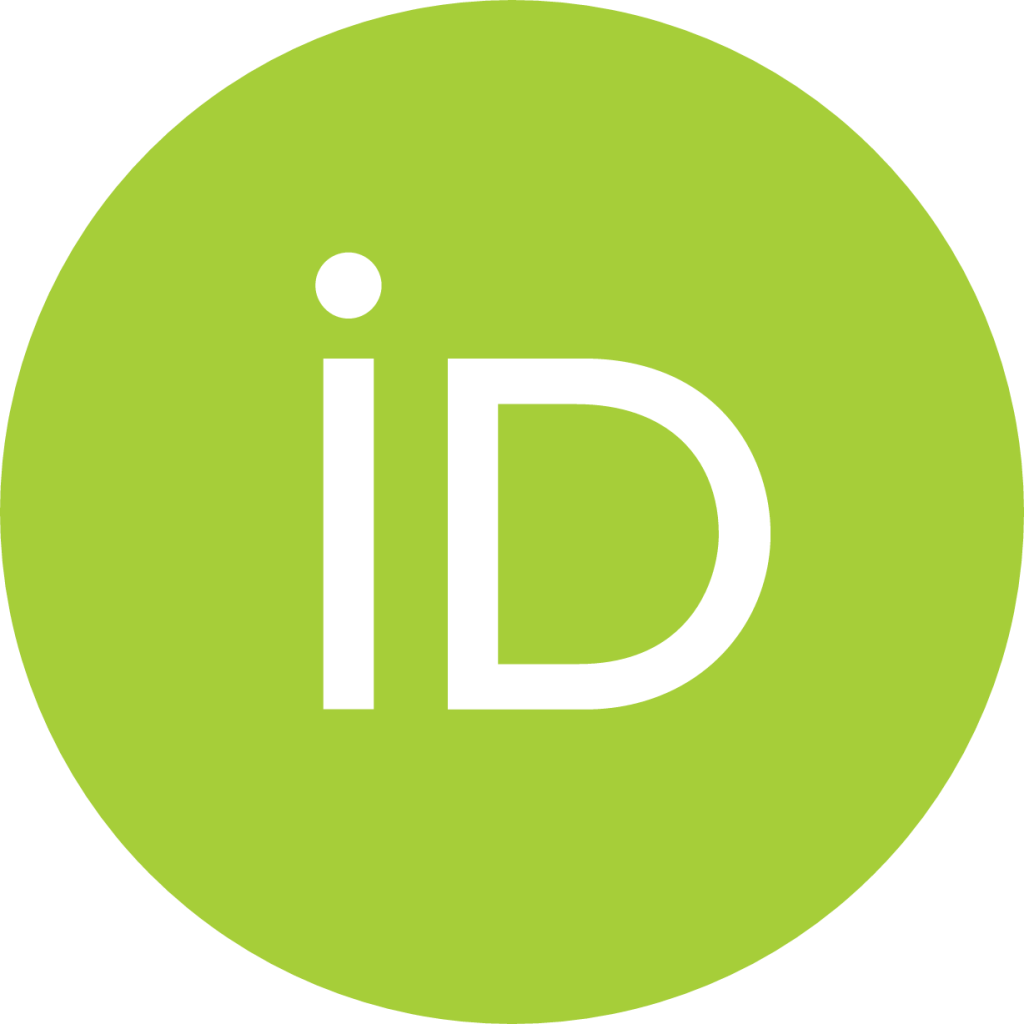}}\thanks{\href{mailto:tamalmukhopadhyay7@gmail.com}{tamalmukhopadhyay7@gmail.com}}, 
    \textbf{Banadipa Chakraborty}~\href{https://orcid.org/0009-0008-2378-0544}{\includegraphics[height=1em]{ORCID_iD_svg.png}}\thanks{\href{mailto:jhelum.chakraborty@gmail.com}{jhelum.chakraborty@gmail.com}} \\ 
    \small Department of Physics, Sister Nivedita University,\\
    \small DG Block (Newtown) 1/2, Action Area I, Kolkata-700156, India.
    \and
    \textbf{Ujjal Debnath}~\href{https://orcid.org/0000-0002-2124-8908}{\includegraphics[height=1em]{ORCID_iD_svg.png}}\thanks{\href{mailto:ujjaldebnath@gmail.com}{ujjaldebnath@gmail.com}} \\ 
    \small Department of Mathematics, Indian Institute of Engineering Science and Technology, \\ 
    \small Shibpur, Howrah-711103, India.
    \and     
     \textbf{Anirudh Pradhan}~
     \href{https://orcid.org/0000-0002-1932-8431}
     {\includegraphics[height=1em]{ORCID_iD_svg.png}}     \thanks{\href{mailto:pradhan.anirudh@gmail.com}{pradhan.anirudh@gmail.com}} \\ 
    \small Centre for Cosmology, Astrophysics and Space Science (CCASS), GLA University. \\ 
    \small Mathura-281406, Uttar Pradesh, India.
}

\date{\today}

\begin{document}

\maketitle

\begin{center}
    \hrule\vspace{0.5em}
    \begin{abstract}
        This paper investigates a newly proposed fluid description of dark energy within the framework of the late-time accelerated expansion of the universe. Our primary objective is to explore the theoretical foundation of the proposed equation of state by establishing its correspondence with well-known scalar field models, such as quintessence, k-essence, and DBI-essence. Through this correspondence, we reconstruct key field parameters, including the scalar field $\phi$ and scalar potential $V(\phi)$, and analyse their evolutionary behaviour across cosmic time. The study also evaluates the model’s physical consistency and cosmological implications by examining fundamental energy conditions—Null Energy Condition (NEC), Dominant Energy Condition (DEC), and Strong Energy Condition (SEC). Furthermore, we conduct a comprehensive stability analysis to ensure the robustness of the model and investigate its thermodynamic properties, including possible phase transitions using entropy and Gibbs free energy. To assess the observational viability of the model, we compare its predictions against recent datasets, including Cosmic Chronometers (CC), Baryon Acoustic Oscillation (BAO), and Supernova Type-Ia from the Pantheon+SH0ES compilation and Union 2.1, as well as recent DESI and DESY5 data. Our analysis demonstrates that the proposed fluid model aligns well with observational constraints, reproduces the late-time acceleration of the universe, and offers a compelling alternative to the standard $\Lambda$CDM model while maintaining consistency with current data.
    \end{abstract}
    \vspace{0.5em}\hrule
\end{center}

\section{Introduction}\label{Sect:Intro}
Observational evidence confirms that the universe is undergoing a late-time accelerated expansion~\cite{riess1998observational,Perlmutter_1998,perlmutter1999measurements}, compelling physicists to investigate possible explanations for this phenomenon. The unknown energy component responsible, termed dark energy, must violate the strong energy condition $(\rho+3p<0)$ to drive cosmic acceleration.  Though, an effort to revive the good old cosmological constant $\Lambda$ was made, it faced several unavoidable drawbacks~\cite{velten2014aspects,yoo2012theoretical,weinberg2001cosmological}. In response, physicists made an approach to introduce additional dynamical degrees of freedom through scalar field models where the kinetic or the potential term of the Lagrangian drives the universe from a dust-like matter dominated era to the dark energy dominated accelerated expansion stage. This kind of dynamic dark energy field was first introduced as quintessence in~\cite{peebles1988cosmology, Caldwell_1998,zlatev1999quintessence,carroll1998quintessence, Tsujikawa_2013}.

Beyond quintessence, more general scalar field theories such as k-essence~\cite{armendariz2001essentials} and DBI-essence~\cite{DBI_essence} have been introduced. The k-essence framework extends quintessence by allowing non-canonical kinetic terms, yielding a Lagrangian $P(X,\phi)$ with $X = \frac{1}{2} \dot{\phi}^2$. This leads to self-regulating dynamics that naturally drive acceleration without fine-tuning, while permitting a variable sound speed to ensure perturbative stability. In contrast, DBI-essence, inspired by string theory, employs a Dirac-Born-Infeld (DBI) action with a warp factor, modifying the evolution through a Lorentz-boost-like $\gamma$ parameter. This introduces non-trivial dynamical behaviour, distinguishing it from quintessence by exhibiting an effective fluid description with modified kinetic terms. These scalar field approaches provide theoretical motivation for dynamical dark energy within high-energy physics frameworks~\cite{chakraborty_dbi,debnath_dbi,Chattopadhyay_2010,Rendall_2006,bandyopadhyay2012k,Gangopadhyay_2010,PANDA2025103059}.

Another promising alternative is the fluid-based equation of state (EoS) approach, particularly the Chaplygin gas model, which unifies dark matter and dark energy under a single framework~\cite{kamenshchik2001alternative,gorini2005chaplygin}. This model interpolates between pressureless dust at early times and a negative-pressure fluid at late times. Various modifications, including the Generalized Chaplygin Gas (GCG)~\cite{bento2002generalized}, Modified Chaplygin Gas (MCG)~\cite{Benoum_mcg}, and Variable Modified Chaplygin Gas (VMCG)~\cite{Panigrahi_2016}, have been proposed to better accommodate observational constraints. A more generalized EoS framework for dark energy was introduced in~\cite{Nojiri_2005}, while its behaviour near future singularities was studied in~\cite{Nojiri_2004,stefani__2005}.

The gravitational-thermodynamic relationship has emerged as a pivotal research domain in theoretical physics. The connection between black hole physics and thermodynamics revealed profound insights into spacetime's geometric-energetic interplay \cite{jacobson1995thermodynamics}. Studying entropy-area correspondence in specialized spacetime configurations uncovered deep structural connections between geometric and thermal properties. Pioneering work demonstrated how surface gravitational characteristics connect with thermal measurements in complex geometries \cite{bekenstein1973black}. Significant theoretical advances emerged when researchers explored parallels between cosmological evolution equations and entropy descriptions in quantum field theories \cite{verlinde2000holographic}. Gibbons and Hawking's groundbreaking research illuminated de Sitter spacetime's thermodynamic properties \cite{gibbons1977cosmological}, establishing foundations for horizon dynamics studies. Gong and colleagues \cite{gong2007j} identified the apparent horizon as the most relevant boundary for thermodynamic analysis, based on relationships between Friedmann equations and thermodynamic principles. This becomes crucial in dark energy-dominated accelerating universes, where thermodynamic laws remain valid at the apparent horizon but break down at the event horizon \cite{wang2006thermodynamics}. Contemporary research examines thermodynamic behaviours across different energy dominance regimes, particularly: Standard Chaplygin gas models \cite{myung2011thermodynamics}, Generalized Chaplygin gas configurations \cite{santos2006thermodynamic}, Modified Chaplygin gas frameworks \cite{Debnath_2004, bandyopadhyay2010laws, santos2007thermodynamic}, Variable modified Chaplygin gas scenarios \cite{chakraborty2019evolution}, and Modified Chaplygin-Jacobi/Abel gas variants \cite{CHAKRABORTY_MCJG_MCAG}. Recent studies have revealed the irreversible nature of thermodynamic processes in various cosmological contexts through information-theoretic principles \cite{odintsov2024landauerprinciplecosmologylink}. Investigations into Chaplygin gas-dominated universes \cite{izquierdo2006dark} verified the generalized second law's validity during early expansion phases, while comparative research \cite{chakraborty2019thermodynamics} assessed this law's validity at both cosmological horizons in FLRW worlds with different Chaplygin gas models.

The correspondence between scalar field models and fluid-based EoS descriptions is of fundamental importance in both gravitational physics and particle physics. From a gravitational perspective, different scalar field Lagrangians correspond to different modifications of Einstein’s field equations, affecting the structure of cosmic expansion, perturbations, and energy conditions. By mapping a given fluid equation of state to an equivalent scalar field model, one can investigate whether the fluid can emerge from an underlying fundamental theory, rather than being purely phenomenological. Physically, the field-theoretical description of the Chaplygin gas has significant implications. It provides a mathematically consistent framework for modelling cosmic acceleration, naturally emerges from string theoretical constructs, and unifies the properties of dark energy and dark matter. As such, the Chaplygin gas remains a cornerstone model for studying the evolution of the universe and testing cosmological theories against observational data. In recent work, Hova et al.~\cite{Hova_model} constructed an alternative fluid model by employing a tachyonic field Lagrangian of a non-BPS D3-brane to describe the background dynamics of the universe. This formulation suggests that the Chaplygin gas-like equation of state may arise from a variety of scalar field models, beyond just the tachyonic framework. Indeed, several studies~\cite{Debnath_2004,PhysRevD.98.043515,RUDRA_2013,sharif2012solvable,Debnath_2011} have explored field-theoretic origins of Chaplygin gas models using quintessence, k-essence, and DBI-essence scalar fields, demonstrating that various field-theoretic formulations can mimic the Chaplygin gas behaviour. Given this rich theoretical background, it is natural to ask whether the fluid model proposed by Hova et al. can also be realized as an effective description arising from scalar field dynamics. To systematically explore this possibility, we construct explicit correspondences between different scalar field models and the proposed fluid EoS. This analysis allows us to examine the fundamental theoretical origins of the model and evaluate whether it can be embedded within a broader field-theoretic framework. The following subsections present a detailed derivation of the correspondence between the proposed fluid model and well-known scalar field approaches, including quintessence, k-essence, and DBI-essence. The subsequent subsections present a detailed analysis of these scalar field analogs and their physical implications. This approach allows us to probe:
\begin{itemize}
    \item The consistency of fluid models with fundamental field theories.
    \item The violation or satisfaction of key energy conditions in alternative gravity theories.
    \item The impact of different Lagrangian structures on the stability and evolution of cosmic perturbations.
\end{itemize}

From a particle physics perspective, scalar fields naturally arise in high-energy theories, including supergravity, string theory, and effective field theories. The connection between a fluid EoS and scalar fields provides insights into whether a given dark energy model has a viable embedding within fundamental physics. In particular:
\begin{itemize}
    \item Quintessence-like correspondence may suggest an origin from fields such as pseudo-Nambu-Goldstone bosons or moduli fields in extra-dimensional theories. Pseudo-Nambu–Goldstone bosons (pNGBs) arise when a global symmetry is spontaneously broken, leading to naturally light scalar fields that can serve as candidates for dark energy. Such fields can acquire small masses through explicit symmetry breaking, making them suitable for quintessence models. In extra-dimensional scenarios, moduli fields associated with the shape of compact extra dimensions can behave as quintessence fields, driving cosmic acceleration \cite{Peloso_2003}.
    \item K-essence-like correspondence could arise from higher-derivative interactions in effective field theories. K-essence extends quintessence by introducing a non-canonical kinetic term $P(X,\phi)$, where $X = \frac{1}{2} \dot{\phi}^2$. These higher-derivative interactions can emerge naturally in low-energy effective field theories derived from high-energy physics. Such terms allow for variable sound speeds and self-adjusting dynamics, potentially resolving fine-tuning issues associated with cosmic acceleration \cite{armendariz2001essentials}.
    \item DBI-like correspondence has direct implications for brane-world scenarios, where dark energy could emerge from the motion of a brane in a higher-dimensional bulk. The Dirac–Born–Infeld (DBI) action, inspired by string theory, describes the dynamics of D-branes moving through a warped extra-dimensional space. In a cosmological setting, the DBI scalar field can be interpreted as the position of a brane moving in a higher-dimensional bulk, leading to an effective dark energy component with a modified equation of state \cite{e14071203}.
    \item The field-theoretic description permits the systematic inclusion of quantum corrections and higher-order interactions, offering insights into the renormalization behaviour and vacuum stability of the theory, which are essential for establishing a consistent and predictive framework. Quantum corrections can significantly affect the mass and self-interactions of a scalar field, altering its cosmological implications. Ensuring vacuum stability through renormalization techniques is crucial for preventing instabilities in dark energy models and maintaining theoretical consistency \cite{Kohri_2017}.

\end{itemize}

This paper investigates an alternative fluid model for dark energy, recently proposed by Hova et al.~\cite{Hova_model}, as a potential replacement for the Generalized Chaplygin Gas (GCG). Unlike traditional GCG models, which smoothly interpolate between dust-like and dark energy behaviour, this new model modifies the equation of state using a $\sinc$ function, altering both the thermodynamic and dynamical properties of the universe. Our key objectives are:
\begin{itemize}
    \item To establish a field-theoretic correspondence for this fluid model using quintessence, k-essence, and DBI-essence.
    \item To investigate the thermodynamic properties, including entropy conditions and stability.
    \item To test the validity of the Generalized Second Law of Thermodynamics (GSL).
    \item To compare the model’s predictions with cosmological observations, including CC, BAO, and Supernova Type-Ia datasets.
\end{itemize}

The structure of the paper is as follows. In Section~\ref{Sect:Model_description}, we introduce the alternative fluid model and its equation of state. Section~\ref{Sect:Scalar_fields} is dedicated to establishing its scalar field correspondence. In Section~\ref{Sect:Thermodynamic_analysis}, we perform a detailed thermodynamic analysis, while Section~\ref{Sect:GSL} examines the validity of the GSL. Section~\ref{Sect:Observational_constraints} presents observational constraints based on CC, BAO, and Supernova Type-Ia data. Finally, Section~\ref{Sect:Conclusion} summarizes our findings and discusses their broader implications for cosmology and fundamental physics.

\section{Brief Description of The Model}\label{Sect:Model_description}
Hoava recently introduced a phenomenological fluid description of dark energy which can be an alternative to the Generalized Chaplygin gas with the following equation of state~\cite{Hova_model}, 
\begin{equation}\label{Hoava_EoS}
    p= -\rho+\rho \sinc\left(\frac{\mu\pi\rho^0}{\rho}\right)=-\rho+\dfrac{\rho^2}{\mu\pi\rho^0}\sin{\left(\dfrac{\mu\pi\rho^0}{\rho}\right)},
\end{equation}
where, $\sinc$ represents the sine cardinal function with $\sinc{(x)} = \frac{\sin{x}}{x}$. The functional form of the above EoS is just like an expansion which is centered around the vacuum equation of state $(p=-\rho)$ \cite{stefani__2005} and closely resembles with the general form $p=-\rho+f(\rho)$ where, setting $f(\rho)=0$ recovers the cosmological constant scenario. The above equation of state is indeed a homogeneous one as it does not contain any function of Hubble parameter~\cite{Nojiri_2005}.
Here, $\mu$ is a fine tuning parameter significantly affecting the behaviour of the fluid and $\rho^0$ is the current energy density of the fluid. Hoava, in his original paper chooses $\mu=0.876$, considering the stellar age of the universe~\cite{Hova_model}. Others also reported the value of $\mu$ to be $0.843^{+0.014}_{-0.015}$ in a flat universe using the observational Hubble data and Supernova Type Ia data~\cite{hernandez2019cosmological} and $0.897^{+0.0058}_{-0.0059}$ using Cosmic Microwave Background (CMB), Cosmic Chronometers (CC), Pantheon and R18 datasets~\cite{Yang2019}. The role of $\mu$ in this model not only confined to restrict the estimated age of the universe but also to the overall behaviour of the fluid and thus the evolution dynamics of the universe itself which we will see at the later sections. Now, there is some physical considerations that can be applied to this equation of state. Firstly for the earlier epochs where $\rho>>\rho^0$ the eq.~\eqref{Hoava_EoS} gives the pressure $p=0$. On the other hand, at the later epochs where, $\rho \approx \rho^0$, the eq.~\eqref{Hoava_EoS} yields,
\begin{equation}\label{Hova_EoS_laterepoch}
    p = -\rho + \dfrac{\rho^2}{\mu\pi\rho^0}\left[\dfrac{\mu\pi\rho^0}{\rho}-\dfrac{1}{3!}\left(\dfrac{\mu\pi\rho^0}{\rho}\right)^3\cdots\right] \approx -\dfrac{1}{3!}\dfrac{(\mu\pi\rho^0)^2}{\rho}\quad .
\end{equation}
This is exactly the equation of state of Chaplygin gas with $A=\frac{(\mu\pi\rho^0)^2}{3!}$. Thus, the model EoS can acts as pressureless dust like matter in earlier epochs and as a dark energy component with negative pressure at later epochs and hence, the above equation of state in \eqref{Hoava_EoS} can unify the both dark sectors of the universe (dark matter \& dark energy), at least phenomenologically.
\\In a spatially flat Friedmann-Lema\^{\i}tre-Robertson-Walker (FLRW) universe, the dynamics of cosmic expansion are governed by the Friedmann equations. The first Friedmann equation is expressed as 
\begin{equation}\label{friedmann_1}
H^2 = \frac{8\pi G}{3} (\rho_m + \rho_{de}),
\end{equation}
where $H = \frac{\dot{a}}{a}$ is the Hubble parameter (the rate of expansion of the universe), $\rho_m$ is the energy density of matter, and $\rho_{de}$ is the energy density of dark energy. This equation encapsulates the total energy density of the universe and its relation to the expansion rate. The second Friedmann equation, 
\begin{equation}\label{friedmann_2}
\frac{\ddot{a}}{a} = -\frac{4\pi G}{3} \left(\rho_m + \rho_{de} + 3p_{de} \right), 
\end{equation}
describes the acceleration of the universe. Here, $p_{de}$ is the pressure of dark energy, which can take negative values to drive accelerated expansion. Together, these equations highlight the influence of energy density and pressure on the cosmic expansion.
\\The conservation equations for the energy densities of matter and dark energy follow from the general continuity equation, $\dot{\rho} + 3H(\rho + p) = 0$, which ensures the local conservation of energy in the expanding universe. For the matter component, the conservation equation reduces to 
\begin{equation}\label{matter_conservation}
\dot{\rho}_m + 3H (\rho_m+p_m) = 0,
\end{equation}
leading to the solution $$\rho_m(z) = \rho_{m0} (1+z)^{3(1+\omega_m)},$$ where, $\omega_m$ is the equation of state parameter for matter and using the relation $a(t)=\frac{1}{1+z}$. This result reflects the dilution of matter density due to the expansion of the universe. For dark energy, the conservation equation is 
\begin{equation}\label{DE_conservation}
\dot{\rho}_{de} + 3H (\rho_{de} + p_{de}) = 0.
\end{equation}
\\The Hubble parameter, which measures the rate of expansion of the universe as a function of the redshift $z$, can be derived from the first Friedmann equation. It is given by 
\begin{equation}\label{Hubble_parameter}
H(z) = H_0 \sqrt{\Omega_{m0}(1+z)^{3(1+\omega_m)} + \Omega_{de0} f(z)},
\end{equation}
where $H_0$ is the present-day Hubble parameter, $\omega_m$ is the equation of state parameter for matter, $\Omega_{m0} = \frac{8\pi G \rho_{m0}}{3H_0^2}$ is the present-day matter density parameter, and $\Omega_{de0}$ is the present-day dark energy density parameter. The term $f(z)$ encapsulates the evolution of dark energy density as a function of the scale factor, reflecting the possible time variation of dark energy. The dependence on $(1+z)^{3(1+\omega_m)}$ for matter arises due to its dilution with the expansion, as matter density scales inversely with volume.
\\Now, using the equation~\eqref{Hoava_EoS}, the solution of the dark energy conservation equation~\eqref{DE_conservation} yields~\cite{Hova_model},
\begin{equation}\label{rho_Hova}
    \rho = \dfrac{\mu \pi \rho^0}{2 \arctan\left[a^3 \tan(\frac{\mu \pi}{2})\right]}\quad .
\end{equation}

\section{Correspondence with the Dark Energy Scalar Fields}\label{Sect:Scalar_fields}
The Chaplygin gas represents a compelling model in cosmology, bridging fundamental field theory and cosmic dynamics. This unique form of matter, originally introduced through an exotic equation of state, is characterized by its ability to unify dark energy and dark matter within a single framework. This section presents a detailed exploration of its theoretical underpinnings and field-theoretical formulation.
\\The Chaplygin gas is fundamentally described by a scalar field action, expressed as:
\begin{equation}\label{Standard_CG_action}
S[g,\phi] = -\int dt d^3x \sqrt{-g} \left(\frac{1}{2}\partial_\mu\phi\partial^\mu\phi - V(\phi)\right),
\end{equation}
where $g$ denotes the determinant of the metric tensor and $\phi$ represents the scalar field. This action encapsulates the spacetime dynamics of the system, with the scalar potential $V(\phi)$ playing a central role in determining the behaviour of the Chaplygin gas. A canonical form of the potential is given by:
\begin{equation}
V(\phi) = \phi^2\ln\phi^2 + V_0,
\end{equation}
where $V_0$ is a constant term. This specific form of the potential is instrumental in generating the unique properties of the Chaplygin gas.

A key theoretical motivation for the Chaplygin gas arises from its connection to Tachyonic scalar fields, which naturally emerge in string theoretical contexts. In this framework, the scalar field $\phi$ exhibits tachyonic behaviour and couples to a U(1) gauge field defined on the world volume of a non-BPS D3-brane~\cite{Joseph_A_Minahan_2000}. This interpretation establishes a strong link between the Chaplygin gas model and string theory, suggesting a fundamental origin for the equation of state. Gorini et al.~\cite{Gorini_2004} proposed an alternative formulation where the potential remains constant, given by:
\begin{equation}
V(\phi) \sim A^{1/2} = V_0,
\end{equation}
where $A$ is a parameter intrinsic to the Chaplygin equation of state:
\begin{equation}
p = -\frac{A}{\rho},
\end{equation}
with $p$ representing pressure and $\rho$ the energy density. This equation highlights the Chaplygin gas's dual role as a candidate for dark matter and dark energy.
Building on these theoretical foundations, the subsequent subsections explore how our alternative fluid model can be interpreted through correspondences with well-known scalar field models, including quintessence, k-essence, and DBI-essence. This comparative analysis aims to reveal the underlying field-theoretic origins of the fluid equation of state, offering deeper insights into its dynamical behaviour and cosmological implications.

\subsection{Quintessence Scalar Field}\label{Subsect:quintessence}
\begin{figure}[htbp]
    \centering
    \begin{subfigure}[b]{0.45\textwidth}
        \centering
        \includegraphics[width=\textwidth]{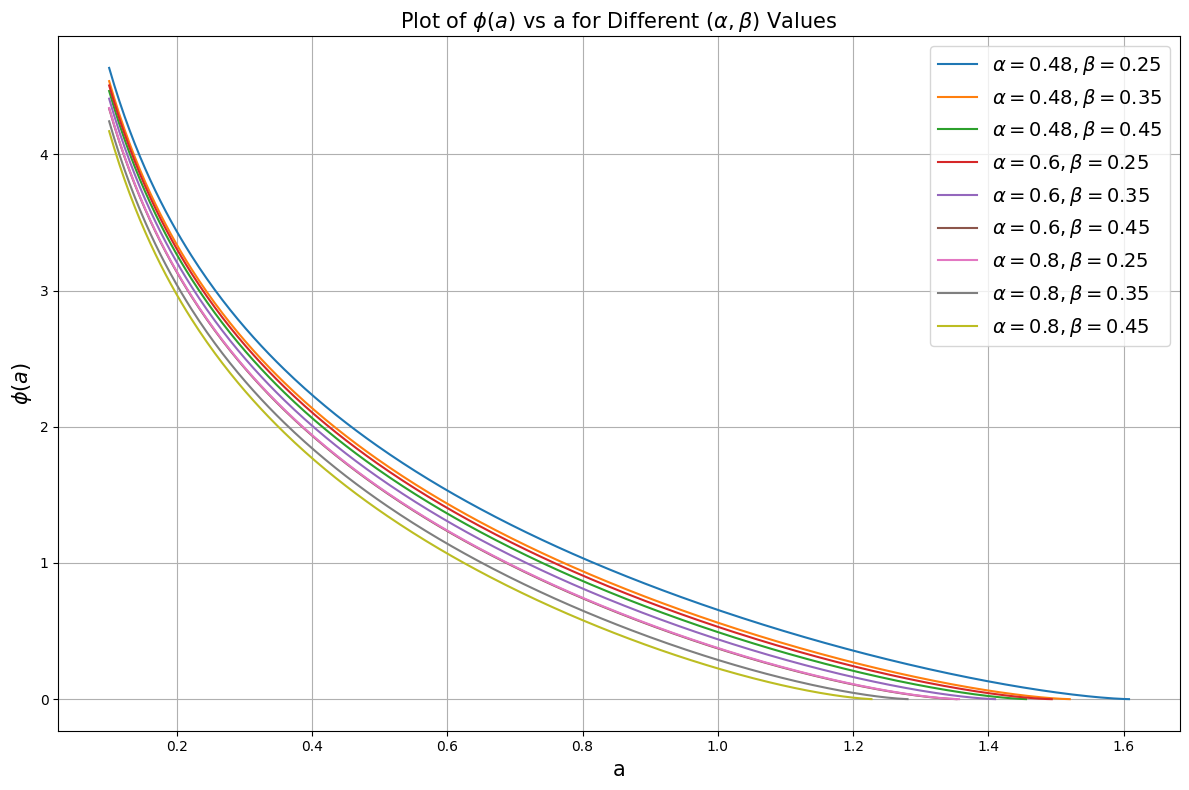}
        \caption{The variation of the scalar field with the scale factor}
        \label{fig:quintessence_phi_a}
    \end{subfigure}
    \hfill
    \begin{subfigure}[b]{0.45\textwidth}
        \centering
        \includegraphics[width=\textwidth]{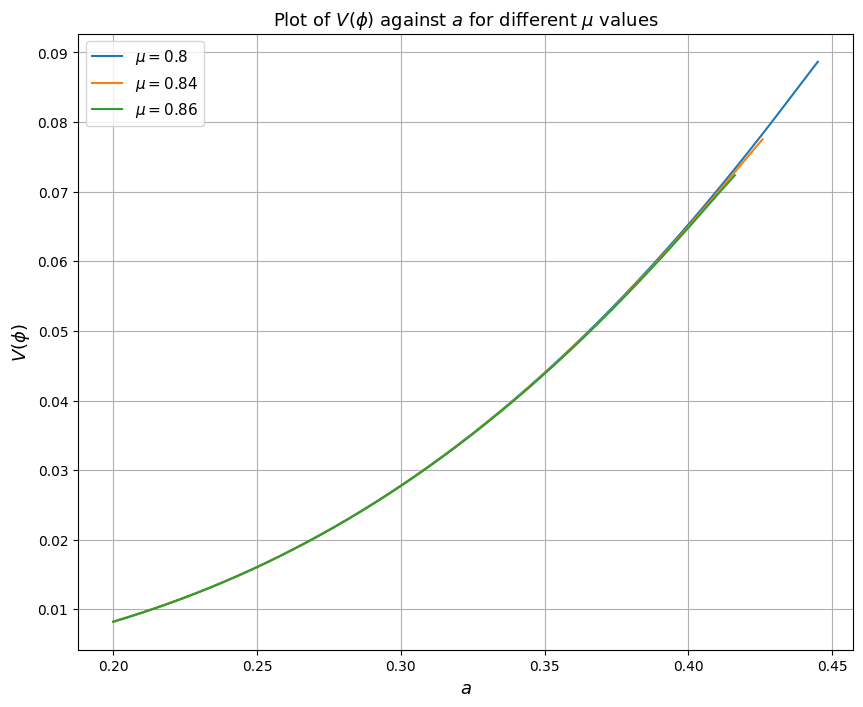}
        \caption{Variation of the scalar potential with the scale factor}
        \label{fig:quintessence_V_a}
    \end{subfigure}
    \caption{Here, in the first plot we can see that the scalar field $\phi$ is decreasing with scale factor and reaches the minimum at a certain point. In the second plot, the scalar potential is an increasing function of the scale factor. An interesting remark can be seen that the parameter $\mu$ (in terms of $\alpha$ and $\beta$) affects the slope (or the decrease rate) of the $\phi$ vs. $a(t)$ plot but it does not affect much the scalar potential variation with the scale factor in the second plot. }
    \label{fig:quintessence_phi_V_a}
\end{figure}

First, we have assumed the self interacting potential with a minimally coupled scalar field $\phi$ having the Lagrangian,
\begin{equation}\label{quintessence_lagrangian}
    \mathcal{L}=\dfrac{1}{2}\dot{\phi}^2-V(\phi).
\end{equation}
The pressure and energy density of this field description is given by,
\begin{gather}
    \rho_{\phi} = \dfrac{1}{2}\dot{\phi}^2+V(\phi), \label{quinetssence_density}\\
    p_{\phi}=\dfrac{1}{2}\dot{\phi}^2-V(\phi).\label{quintessence_pressure}
\end{gather}
Now, to obtain the specific form of scalar potential $V(\phi)$ and scalar field $\phi$ we have to equate the equations.~\eqref{quinetssence_density} and \eqref{quintessence_pressure} with equations.~\eqref{rho_Hova} and \eqref{Hoava_EoS} respectively. By doing so, we have obtained,
\begin{gather}
    V(\phi) = -\dfrac{\sqrt{2\pi}a\sqrt{\frac{-\pi a^6\mu\rho^0\tan^2(\frac{\pi\mu}{2}) + 1}{a^2}}\sqrt{\frac{\mu\rho^0}{\arctan(a^3\tan(\frac{\pi\mu}{2}))}}}{4} + \frac{\pi\mu\rho^0}{2\arctan(a^3\tan(\frac{\pi\mu}{2}))} \label{quintessence_V},\\
    \dot{\phi} = \frac{\sqrt{2\pi}a\sqrt{\frac{-\pi a^6\mu\rho^0\tan^2(\frac{\pi\mu}{2}) + 1}{a^2}}\sqrt{\frac{\mu\rho^0}{\arctan(a^3\tan(\frac{\pi\mu}{2}))}}}{2}.\label{quintessence_dotphi}
\end{gather}
Now, from the equation~\eqref{quintessence_dotphi} and the first Friedmann equation~\eqref{friedmann_1}, we have
\begin{equation}\label{quintessence_dphida}
    \dfrac{d\phi}{da}=\sqrt{\dfrac{3}{a^2}\left[1-\dfrac{4\left(\frac{\beta}{2a^3\alpha}\right)^2}{\beta}\right]},
\end{equation}
where, $\alpha$ and $\beta$ is expressed as,

\begin{gather}
    \alpha = \tan{\left(\dfrac{\mu\pi}{2}\right)},\\
    \beta = \mu\pi\rho^0.
\end{gather}
Now, we have integrated the above equation using the initial conditions as, for $a \to 0$, $\phi \to \infty$ and for $a\to \infty$, $\phi \to 0$. The scalar field $\phi$ as a function of the scale factor is obtained as,
\begin{equation}\label{quintessence_phi}
    \phi = \phi_0 + \frac{a\sqrt{\frac{1}{a^2} - a^4\alpha^2\beta}(\sqrt{1 - a^6\alpha^2\beta} - \tanh^{-1}{\sqrt{1 - a^6\alpha^2\beta}})}{\sqrt{3 - 3a^6\alpha^2\beta}}.
\end{equation}
The equation~\eqref{quintessence_V} shows that $V(\phi) = \dfrac{\dot{\phi}}{2}+\frac{\pi\mu\rho^0}{2\arctan(a^3\tan(\frac{\pi\mu}{2}))}$. Now for the smaller values of the scale factor, $V(\phi) = \frac{\pi\mu\rho^0}{2\arctan(a^3\tan(\frac{\pi\mu}{2}))}$ and for the large values of the scale factor $V(\phi) = \dfrac{\dot{\phi}}{2}$. The graphical variation of $V(\phi)$ and $\phi$ against $a(t)$ is presented here in the fig.~\ref{fig:quintessence_phi_V_a}. The expression of the scalar potential $V(\phi)$ as a function of scalar field $\phi$ is very complicated to achieve so we will solve this numerically and presented in the following figure~\ref{fig:quintessence_V_phi}.
\begin{figure}[htbp]
    \centering
    \begin{subfigure}[b]{0.65\textwidth}
        \centering
        \includegraphics[width=\textwidth]{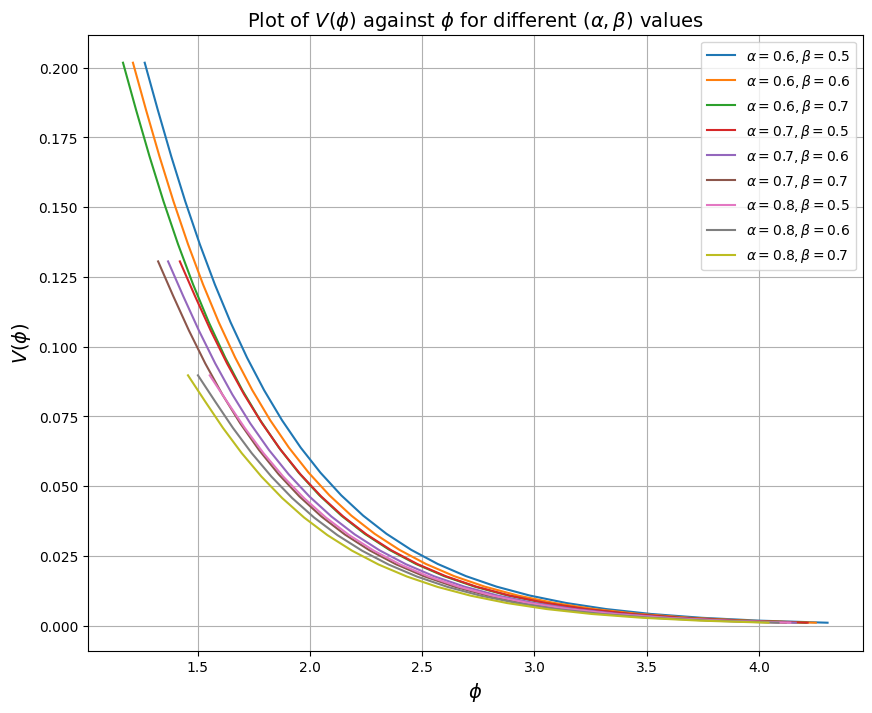}
        \caption{The variation of the scalar potential with the scalar field $\phi$}
        \label{fig:quintessence_V_phi}
    \end{subfigure}
    \hfill
    \begin{subfigure}[b]{0.8\textwidth}
        \centering
        \includegraphics[width=\textwidth]{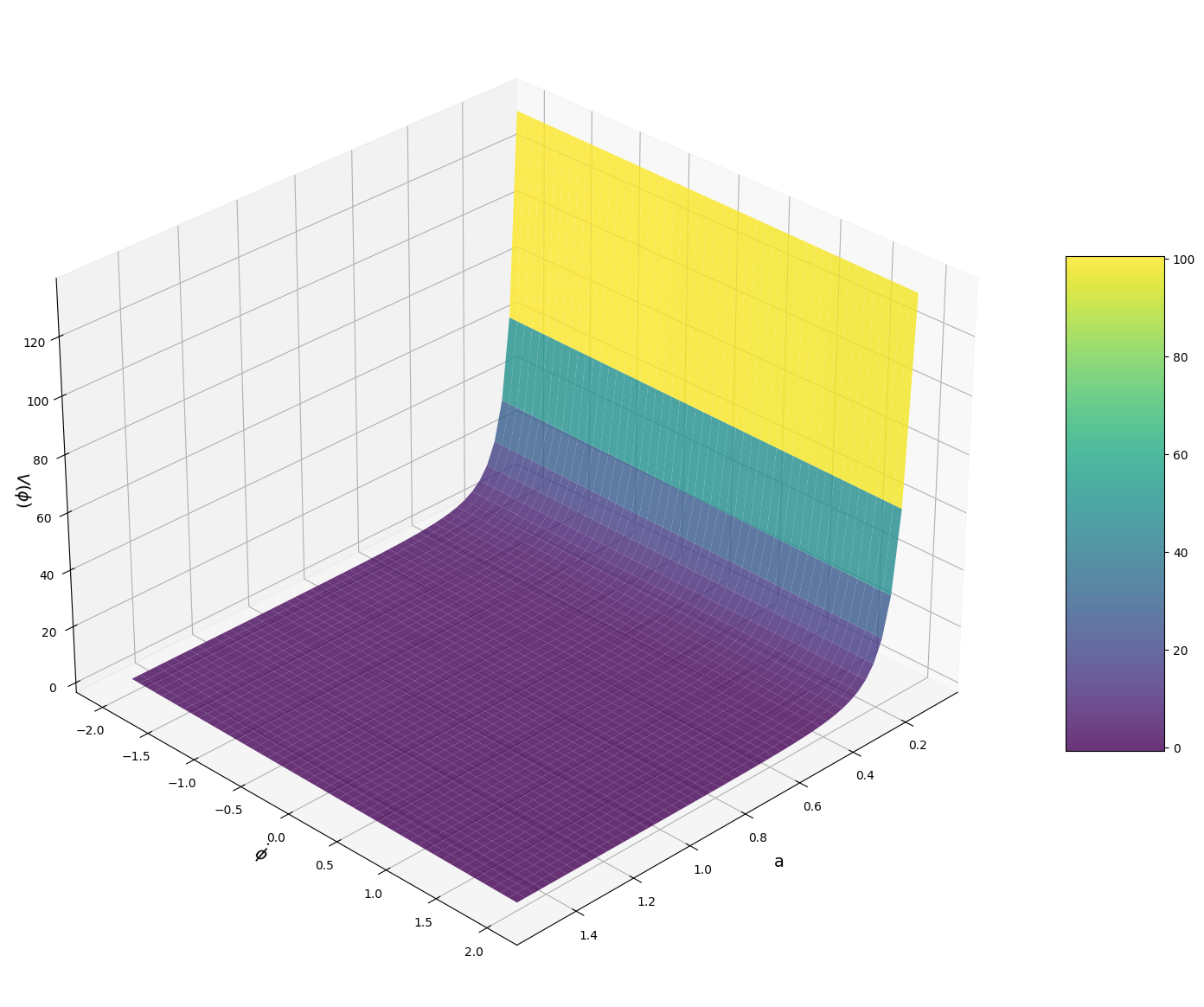}
        \caption{Variation of the scalar potential with the scale factor and the time derivative of the scalar field}
        \label{fig:quintessence_V_dotphi_a}
    \end{subfigure}
    \caption{Here, in the first plot we can see that the scalar potential $\phi$ is decreasing with scalar field and the decrease rate is faster than the slow rolling potential and reaches the minimum at a certain point. The slope of the $V(\phi)$ against $\phi$ variation is greatly affected by the choice of parameters $\alpha$ and $\beta$ (and that is the value of $\mu$). In the second plot, the scalar potential is an increasing function of the scale factor and time derivative of the scalar field. }
    \label{fig:quintessence_V_phi_dotphi_a}
\end{figure}
\\The equation~\eqref{quintessence_V} imparts that for the very small values of the scale factor the scalar potential $V(\phi)$ have a discontinuity. Again, for the equation~\eqref{quintessence_phi} and \eqref{quintessence_V}, the scalar field and the scalar potetial remains analytical except for $a^6\alpha^2\beta = 1$, $a^6 \alpha^2 \beta \notin \left(0,1\right]$ and $\mu = 2m + 1, \quad m \in \mathbb{Z}$. Thus, both the scalar potential and the scalar field mimicking the above described fluid model is analytical throughout the evolution of the scale factor, except for the above specified domains. Thus this alternative model can be originated from an minimally coupled scalar field with a self interacting potential.

\subsection{k-essence Scalar Field}\label{Subsect:k_essence}
Next, we will consider another type of scalar field Lagrangian which includes a non-canonical kinetic term and hence known as k-essence scalar field. The action of the k-essence field is given as~\cite{k_essence},
\begin{equation}\label{k_essence_action}
    S = \int d^4x \sqrt{-g}\left[R+2K(X,\phi)\right],
\end{equation}
where, $X = -0.5g^{\mu\nu}\nabla_{\mu}\phi\nabla_{\nu}\phi$. The pressure and energy density corresponding to the k-essence field is given as,
\begin{gather}
    \rho = 2XK_X-K, \label{k_essence_rho}\\
    p = K. \label{k_essence_pressure}
\end{gather}
Now, The scaling relation is obtained as,
\begin{equation}\label{k_essence_scaling_relation}
    XK_X^2 = ka^{-6}.
\end{equation}
Here, for our calculation we will consider a purely kinetic k-essence scalar field with $K=F(X)$~\cite{myrzakulov2016cosmologicalmodelsnoncanonicalscalar}. From equation~\eqref{k_essence_rho} we can obtain that,
\begin{equation}\label{k_essence_F}
    F = D\sqrt{X}+\dfrac{\sqrt{X}}{2}\int\dfrac{\rho}{X^{1.5}}dX.
\end{equation}
Here, D is an arbitrary integration constant. Now, the above equation coupled with the eq.~\eqref{k_essence_pressure} will give us the expressions of $F(X)$ and $\phi$. But, due to complexity of the equations, we solved them numerically and presented in the following figure~\ref{fig:k_essence_dotphi_phi_a} and \ref{fig:k_essence_F(X)_X_a}.
\begin{figure}
    \centering
    \includegraphics[width=1\linewidth]{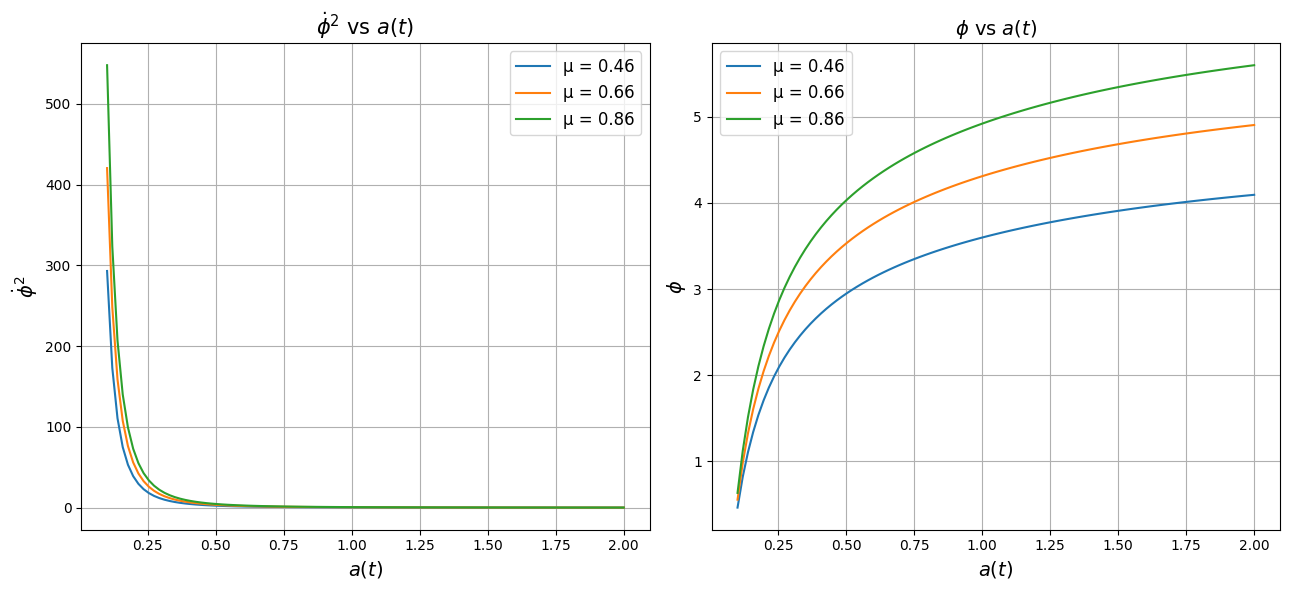}
    \caption{Plot of $\dot{\phi}^2$ and $\phi$ against the scale factor $a(t)$.}
    \label{fig:k_essence_dotphi_phi_a}
\end{figure}
In the fig.~\ref{fig:k_essence_dotphi_phi_a} it is readily evident that the time derivative of the scalar field $\dot{\phi}^2$ is decreasing with the scale factor which physically implies that the rate of change of dark energy k-essence field mimicking the proposed fluid EoS is decreasing and after a certain period it will be nearly zero making the scalar field $\phi$ increasing in a more linear way which is definitely evident from the plot of $\phi$ against scale factor $a(t)$. Also, this figure shows that the increase of the scalar field $\phi$ with the scale factor $a(t)$ which is definitely a different scenario than that of the quintessential minimally coupled self-interacting field in the sect.~\ref{Subsect:quintessence}. But similar to the to quintessence model, here the free parameter $\mu$ controls the slope of the $\phi$ vs. $a(t)$ plot as well.

\begin{figure}
    \centering
    \includegraphics[width=1\linewidth]{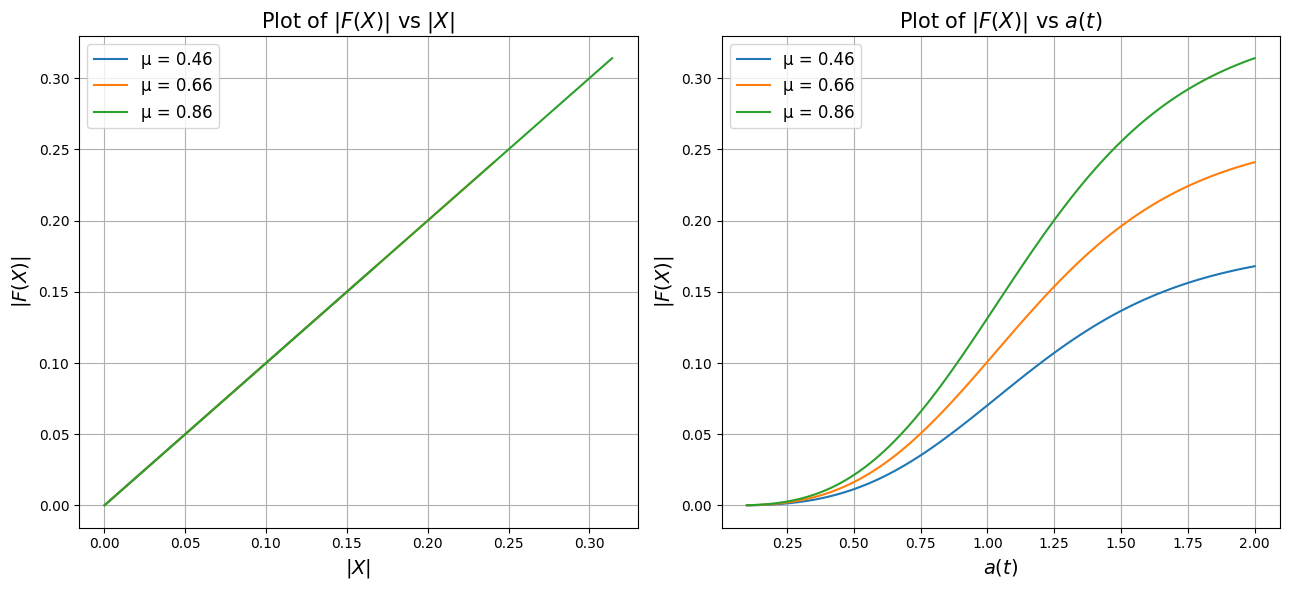}
    \caption{Plot of variation of $F(X)$ vs $X$ and $F(X)$ vs scale factor $a(t)$.}
    \label{fig:k_essence_F(X)_X_a}
\end{figure}
\begin{figure}
    \centering
    \includegraphics[width=1\linewidth]{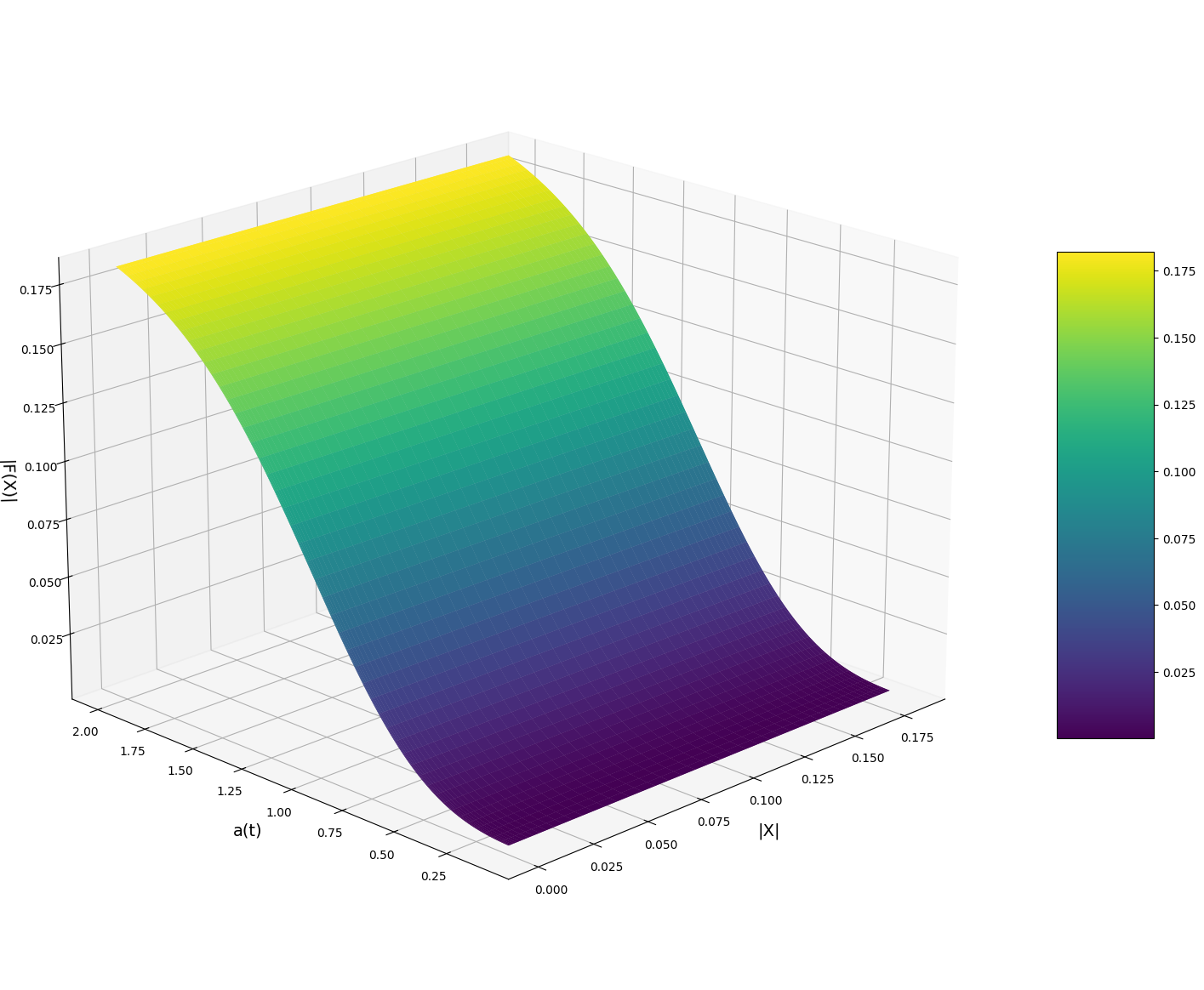}
    \caption{3D Plot of variation of $F(X)$ against $X$ and scale factor $a(t)$.}
    \label{fig:k_essence_F(X)_X_a_3D}
\end{figure}

In the fig.~\ref{fig:k_essence_F(X)_X_a} we can see that, the the kinetic term $F(X)$ is increasing linearly with X and it is independent of the value of $\mu$. Also, the second plot shows that the increase of the non-canonical kinetic term $F(X)$ with the increase in scale factor and more the value of $\mu$, more the $|F(X)|$ vs. $a(t)$ plot becomes steeper. The fig.~\ref{fig:k_essence_F(X)_X_a_3D} shows that $F(X)$ behaves as a rolling potential when it is plotted with $X$ and $a(t)$. Notably, the kinetic term $F(X)$ is responsible for the accelerated expansion of the universe and the the increase $F(X)$ in fig.~\ref{fig:k_essence_F(X)_X_a} is consistent with the physical perspective.

\subsection{DBI-essence}\label{Subsect:DBI_essence}
\begin{figure}[htbp]
    \centering
    \begin{subfigure}[b]{0.45\textwidth}
        \centering
        \includegraphics[width=\textwidth]{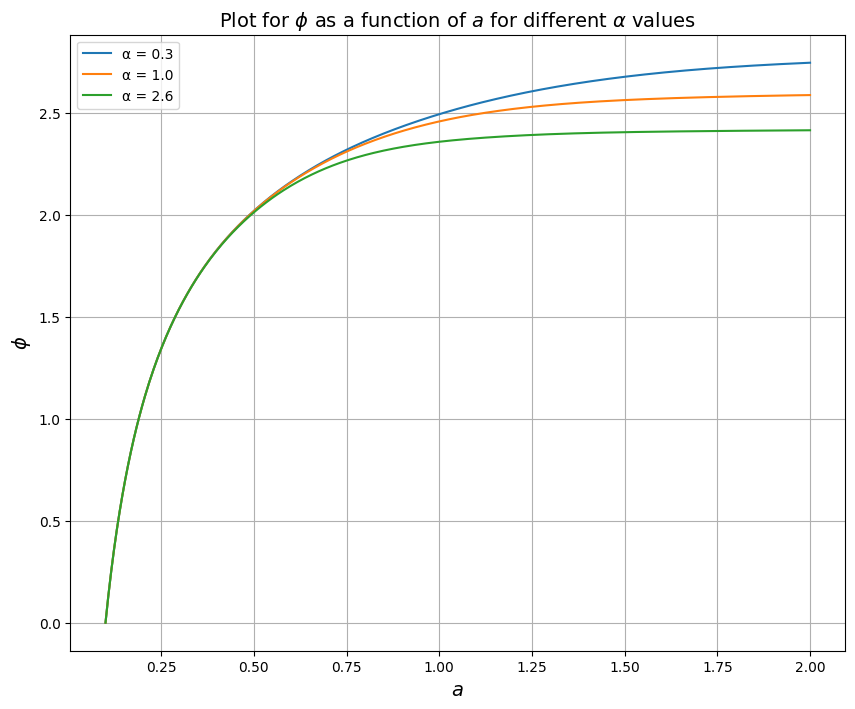}
        \caption{The plot of variation of the scalar field $\phi$ against the scale factor $a(t)$ for constant $\gamma$.}
        \label{fig:DBI_essence_phi_a_constgamma}
    \end{subfigure}
    \hfill
    \begin{subfigure}[b]{0.45\textwidth}
        \centering
        \includegraphics[width=\textwidth]{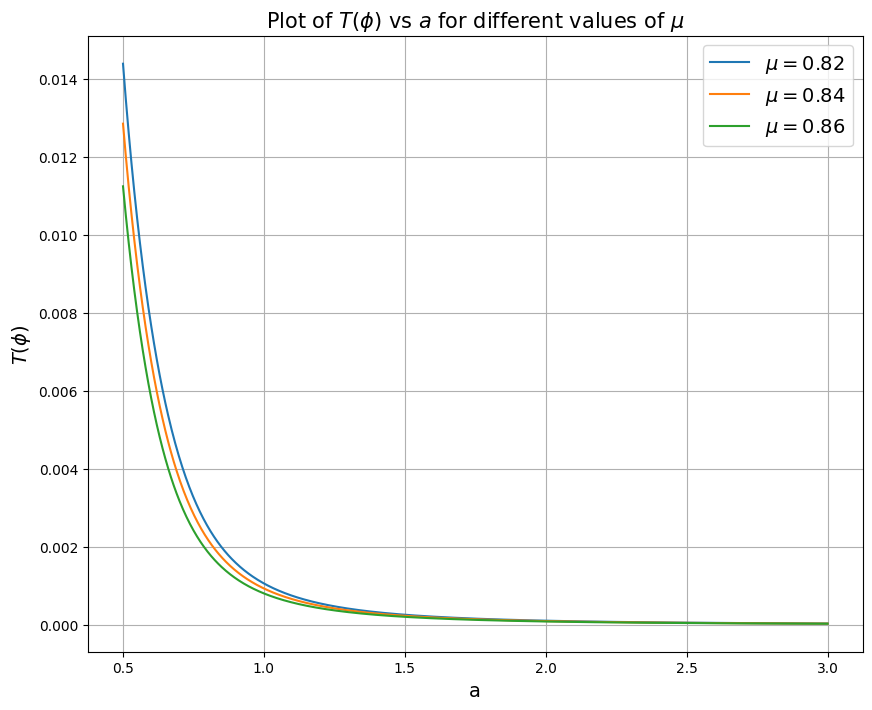}
        \caption{The variation of the warp brane tension $T(\phi)$ against scale factor $a(t)$ for constant $\gamma$.}
        \label{fig:DBI_essence_T_a_constgamma}
    \end{subfigure}
    \hfill
    \begin{subfigure}[b]{0.45\textwidth}
        \centering
        \includegraphics[width=\textwidth]{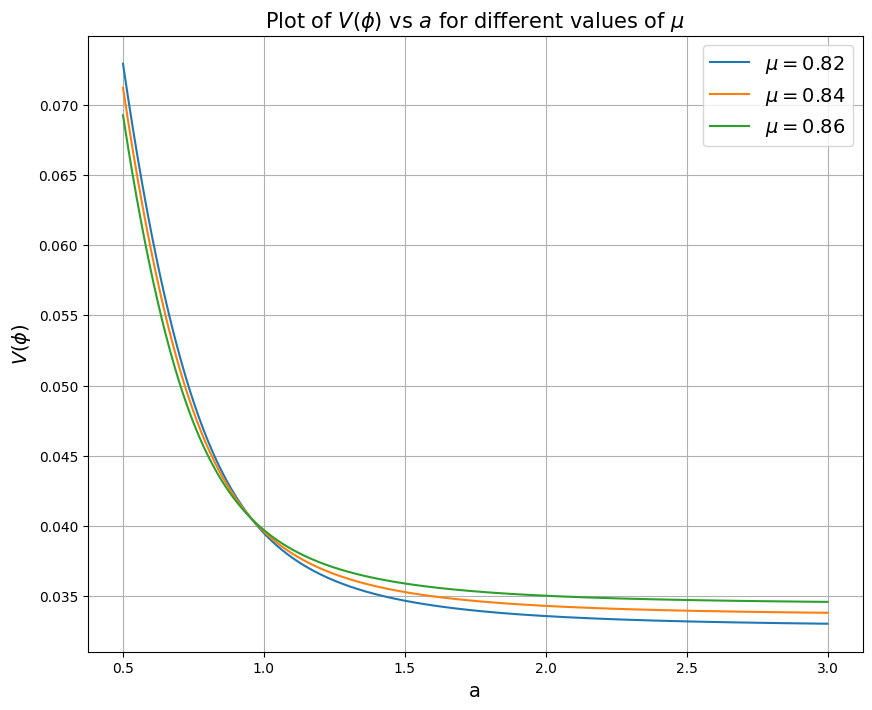}
        \caption{The plot of the scalar potential $V(\phi)$ against scale factor $a(t)$ for constant $\gamma$.}
        \label{fig:DBI_essence_V_a_constgamma}
    \end{subfigure}
    \caption{Here in this figure. (a) it is seen that the scalar field $\phi$ is increasing with the scale factor which is similar to the behaviour we saw in case of k-essence scalar field. Though the increase rate is not as much affected by the values of $\alpha$ (and hence, $\mu$) as we have seen in the k-essence scenario. (b) it is evident from the plot that the warp brane tension $T(\phi)$ is decreasing with the scale factor and becomes zero after a certain time. (c) it shows that the scalar potential $V(\phi)$ is decreasing with scale factor. The impact of $\mu$ in the variations of $T(\phi)$ and $V(\phi)$ is not only to control the slope but also to determine the asymptote of the plot behaviour.}
    \label{fig:DBI_essence_constgamma_phi_V_T}
\end{figure}

\begin{figure}[htbp]
    \centering
    \begin{subfigure}[b]{0.40\textwidth}
        \centering
        \includegraphics[width=\textwidth]{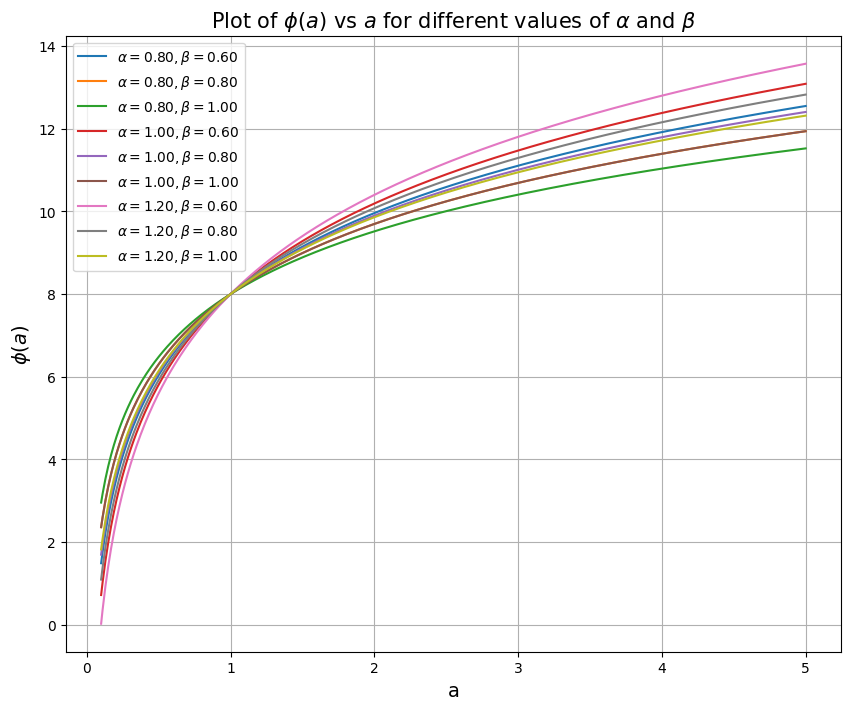}
        \caption{The plot of variation of the scalar field $\phi$ against the scale factor $a(t)$ for variable $\gamma$.}
        \label{fig:DBI_essence_phi_a_vargamma}
    \end{subfigure}
    \hfill
    \begin{subfigure}[b]{0.45\textwidth}
        \centering
        \includegraphics[width=\textwidth]{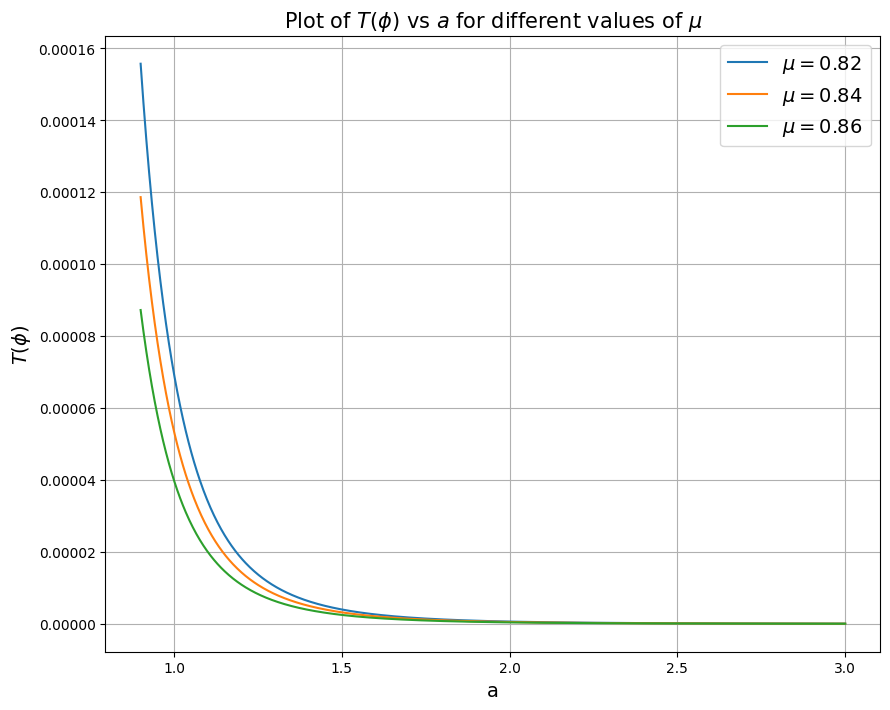}
        \caption{The variation of the warp brane tension $T(\phi)$ against scale factor $a(t)$ for variable $\gamma$.}
        \label{fig:DBI_essence_T_a_vargamma}
    \end{subfigure}
    \hfill
    \begin{subfigure}[b]{0.45\textwidth}
        \centering
        \includegraphics[width=\textwidth]{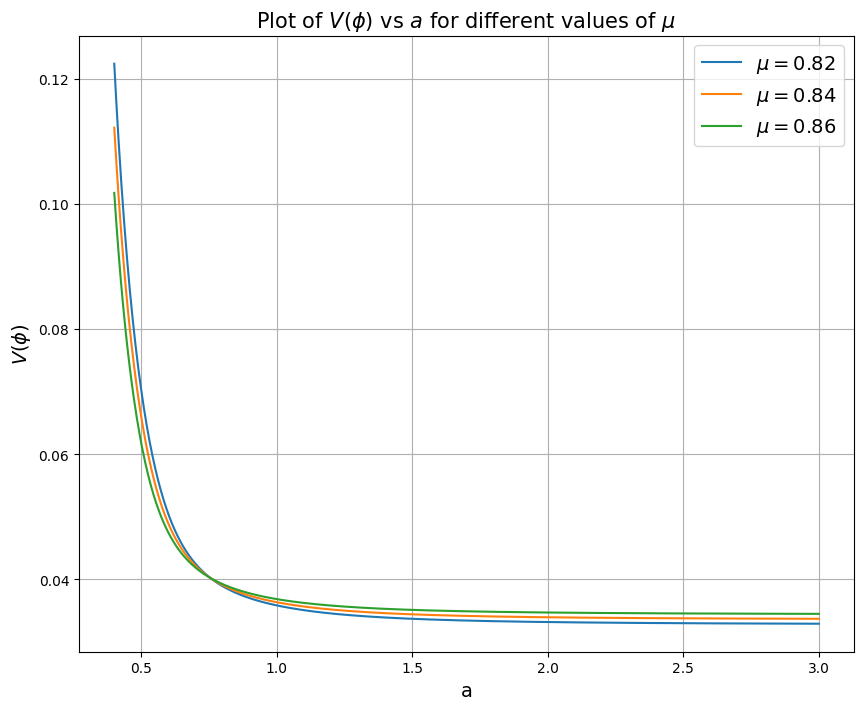}
        \caption{The plot of the scalar potential $V(\phi)$ against scale factor $a(t)$ for variable $\gamma$.}
        \label{fig:DBI_essence_V_a_vargamma}
    \end{subfigure}
    \caption{Here in this figure. (a) it is seen that the scalar field $\phi$ is increasing with the scale factor for variable $\gamma$ also. But the variation of the increase rate with $\alpha$ and $\beta$ (i.e. with $\mu$) is much prominent than that in the case of constant $\gamma$. (b) it is evident from the plot that the warp brane tension $T(\phi)$ is decreasing with the scale factor and becomes zero after a certain time. The $\mu$ dependency is more prominent here than that of constant $\gamma$ model. (c) it shows that the scalar potential $V(\phi)$ is decreasing for variable $\gamma$ with scale factor and the decrease rate is more than that of in the constant $\gamma$ scenario.}
    \label{fig:DBI_essence_phi_T_V_vargamma}
\end{figure}

\begin{figure}[htbp]
    \centering
    \begin{subfigure}[b]{0.75\textwidth}
        \centering
        \includegraphics[width=\textwidth]{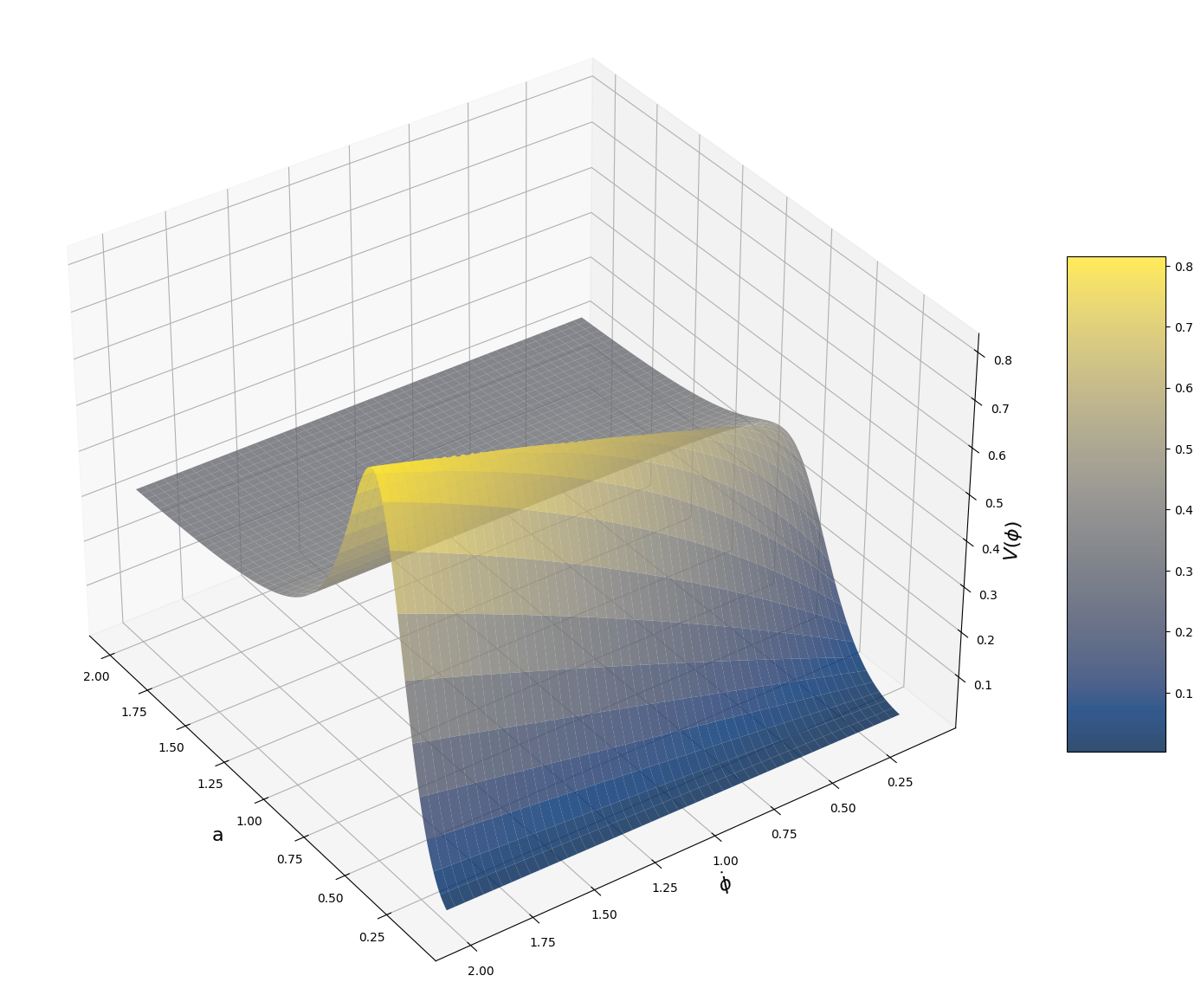}
        \caption{The Variation of the scalar potential with the scale factor and the time derivative of the scalar field for the constant $\gamma$ scenario.}
        \label{fig:DBI_essence_V_dotphi_a_3D_constgamma}
    \end{subfigure}
    \hfill
    \begin{subfigure}[b]{0.75\textwidth}
        \centering
        \includegraphics[width=\textwidth]{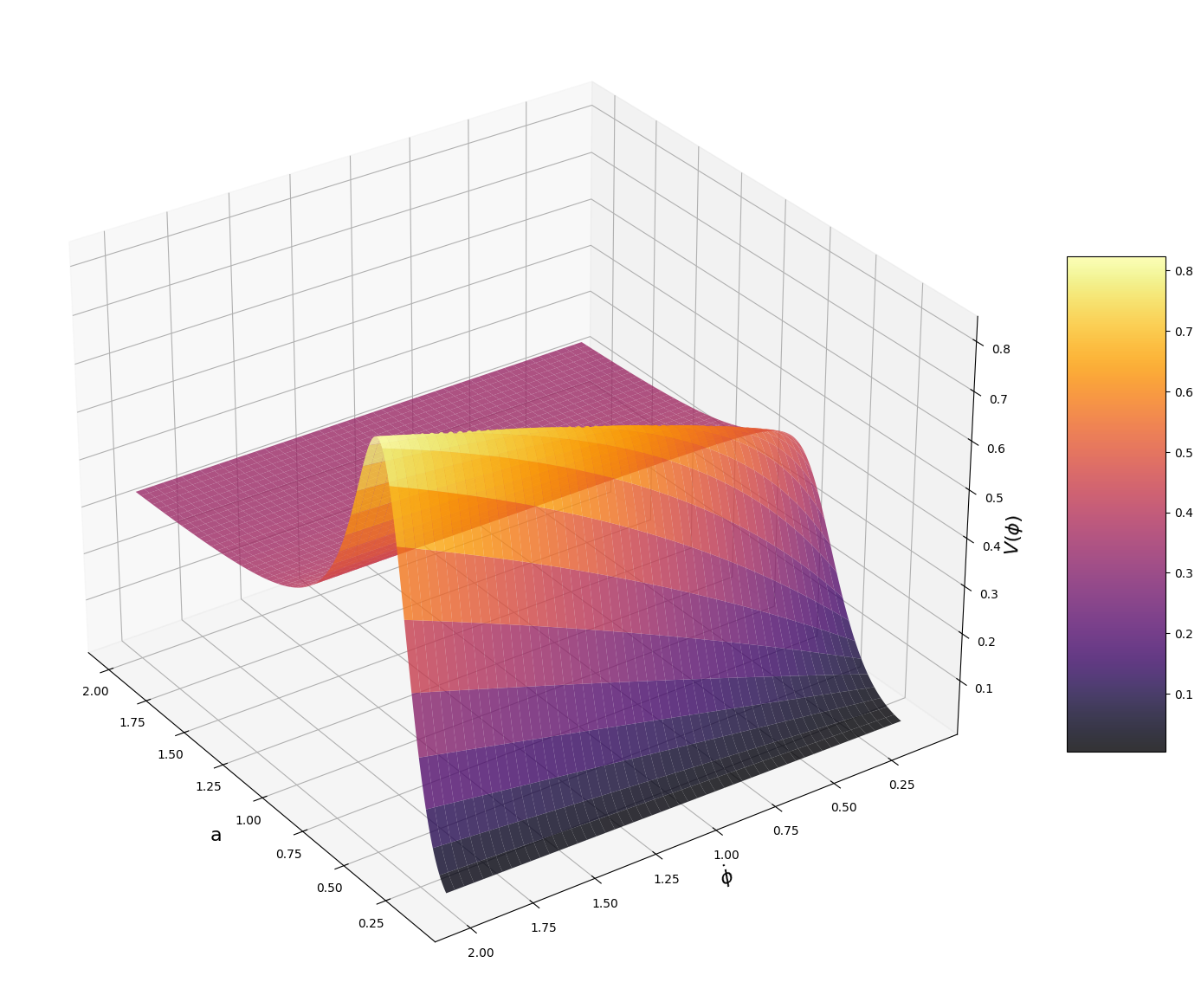}
        \caption{Variation of the scalar potential with the scale factor and the time derivative of the scalar field for the variable $\gamma$ scenario.}
        \label{fig:DBI_essence_V_dotphi_a_3D_vargamma}
    \end{subfigure}
    \caption{Here, the two figures show the similar type of behaviour with the $\dot{\phi}$ and the scale factor $a(t)$ for both the constant and variable $\gamma$. But, variation rate of scaler field $(V(\phi)$ is more flat in case constant $\gamma$ than that of variable $\gamma$ as seen from the both plots.}
    \label{fig:DBI_essence_V_dotphi_a_3D_vargamma_constgamma}
\end{figure}

Next, we will consider another promising dark energy candidate, the DBI-essence field and the corresponding Lagrangian is given by~\cite{DBI_essence},
\begin{equation}\label{DBI_essence_lagrangian}
    \mathcal{L}= T(\phi)\left(\dfrac{1}{\gamma}-1\right)+V(\phi).
\end{equation}
Here, $T(\phi)$ is the warped brane tension and $V(\phi)$ is the self interacting potential. The $\gamma$ term represents the reminiscent from the usual relativistic Lorentz factor and is given by, 
\begin{equation}\label{DBI_essence_gamma}
    \gamma=\dfrac{1}{\sqrt{1-\frac{\dot{\phi}^2}{T(\phi)}}}.
\end{equation}
The pressure and energy density of a DBI-essence scalar field is given as,
\begin{gather}
    \rho_{\phi} = (\gamma-1)T(\phi)+V(\phi), \label{DBI_essence_rho} \\
    p_{\phi} = \dfrac{\gamma-1}{\gamma}T(\phi)-V(\phi). \label{DBI_essence_pressure}
\end{gather}
From the eq.~\eqref{DBI_essence_rho}, \eqref{DBI_essence_pressure} and \eqref{DBI_essence_gamma}, we can express $T(\phi)$ and $V(\phi)$ as,
\begin{gather}
    T(\phi)=\dfrac{\dot{\phi}^2(\rho_{\phi}+p_{\phi})^2}{(\rho_{\phi}+p_{\phi})^2-\dot{\phi}^4} \label{DBI_essence_T},\\
    V(\phi)=\dfrac{\dot{\phi}^2\rho_{\phi}-p_{\phi}(\rho_{\phi}+p_{\phi})}{\dot{\phi}^2+(\rho_{\phi}+p_{\phi})}. \label{DBI_essence_V}
\end{gather}
Now, we will consider two cases, one is for constant $\gamma$ and another is the variable $\gamma$.\\\\
\textbf{Case I:}
Here, we will consider the $\gamma$ to be constant. For this case we can obtain,
\begin{equation}\label{DBI_essence_phidot_constgamma}
    \dot{\phi}^2 = \dfrac{1}{\gamma}(\rho_{\phi}+p_{\phi}).
\end{equation}
Now, to draw a correspondence with the theoretical fluid model under consideration, we will use the pressure and energy density of the fluid mentioned in eq.~\eqref{rho_Hova} and \eqref{Hoava_EoS} in place of $p_{\phi}$ and $\rho_{\phi}$ in equations~\eqref{DBI_essence_rho}, \eqref{DBI_essence_pressure}. Putting these values we can obtain,
\begin{equation}\label{DBI_essence_dphida_constgamma}
    \dfrac{d\phi}{da}=\sqrt{\dfrac{1}{3 \gamma a^3 \left( 1 + a^6 \alpha^2 \right)}}.
\end{equation}
Solving this equation we can obtain,
\begin{equation}\label{DBI_essence_phi_constgamma}
    \phi = \phi_0 +\frac{\sqrt{3} \, \Gamma \left( -\frac{1}{12} \right) \, {}_2F_1 \left( \left( -\frac{1}{12}, \frac{1}{2} \right), \left( \frac{11}{12} \right), a^6 \alpha^2 e^{i \pi} \right)}{18 \sqrt{a} \sqrt{\gamma} \Gamma \left( \frac{11}{12} \right)}.
\end{equation}
Here, ${}_2F_1$ represents the hypergeometric function, $\phi_0$ is the integration constant and $\Gamma$ is the gamma function. Interestingly, this scalar fields $\phi$ tends to $0$ for $a\to 0$ and diverges without bound at $a\to\infty$. The behaviour of this type is previously seen in the k-essence scalar field scenario. Next, we have calculated the self interacting potential $V(\phi)$ and brane tension $T(\phi)$ and obtained that,
\begin{gather}
    V(\phi) = \frac{\beta}{4} \cdot \frac{[-\gamma\beta\sin(2\theta) - 2\theta]\sin(2\theta) + 8\theta^3}{(\gamma\beta\sin(2\theta) + 4\theta^2)\theta^2}, \label{DBI_essence_V_constgamma} \\ 
    T(\phi) = \frac{\gamma \beta \sin(2\theta)}{4(\gamma^2 - 1) \theta^2}.\label{DBI_essence_T_constgamma}
\end{gather}
where, \begin{equation}\label{DBI_essence_alpha_beta}
    \alpha = \tan(\frac{\mu\pi}{2}), \quad \beta = \mu\pi\rho^0, \quad \theta = \tan^{-1}(\alpha a^3).
\end{equation}
Now, we have plotted the scalar field, the warp tension and the potential against the scale factor $a(t)$ in fig.~\ref{fig:DBI_essence_constgamma_phi_V_T}. The equation~\eqref{DBI_essence_V_constgamma} is analytic except for $\mu = 2m + 1, \quad m \in \mathbb{Z}$. Again, the warp brane tension is analytic except at $\gamma=1$. We have also showed the variation of potential $V(\phi)$ against $\dot{\phi}$ and $a(t)$ in a 3D plot in fig.~\ref{fig:DBI_essence_V_dotphi_a_3D_constgamma}. 
\\\\ \textbf{Case II:} In this scenario, we will consider a variable $\gamma$. Let us assume one simple case of $\gamma$ variation that is $\gamma = \dfrac{1}{\dot{\phi}}$, which is also described in ref~\cite{Debnath_2011}. Then,  we can easily write that,
\begin{equation}\label{DBI_essence_dotphi_vargamma}
    \dot{\phi}^2 = (\rho_{\phi}+p_{\phi})^2.
\end{equation}
Now, in order to make a correspondence with the theoretical fluid model of our consideration, we equate the corresponding pressure and energy densities and by using this we can write that,
\begin{equation}\label{DBI_essence_dphida_vargamma}
    \dfrac{d\phi}{da} = \frac{\sqrt{6 \alpha}}{a \sqrt{\beta}}.
\end{equation}
Here, the $\alpha$ and $\beta$ have the same definition as \eqref{DBI_essence_alpha_beta}. The solution of this equation reads,
\begin{equation}\label{DBI_essence_phi_vargamma}
    \phi = \frac{\sqrt{6} \sqrt{\alpha} \, \ln(a)}{\sqrt{\beta}} + \phi_0.
\end{equation}
Here, $\phi_0$ is an integration constant with arbitrary value. The asymptotic behaviour of the scalar field $\phi$ is quite similar to that of in \textbf{Case I}. The scalar field $\phi$ is analytical everywhere except at $\beta=\mu\pi\rho^0=0$. After that, we have again calculated the warp brane tension $T(\phi)$ and scalar potential $V(\phi)$.
\begin{gather}
    T(\phi) = \frac{\mu^2 \pi^2 \rho_0^2 \sin^2\left[2 \arctan\left(a^3 \tan\left(\frac{\mu \pi}{2}\right)\right)\right]}{-\mu^2 \pi^2 \rho_0^2 \sin^2\left[2 \arctan\left(a^3 \tan\left(\frac{\mu \pi}{2}\right)\right)\right] + 16 \left[\arctan\left(a^3 \tan\left(\frac{\mu \pi}{2}\right)\right)\right]^4}, \label{DBI_essence_T_vargamma}\\
    V(\phi) = \frac{\mu \pi \rho_0 \left[-\sin\left(2 \arctan\left(a^3 \tan\left(\frac{\mu \pi}{2}\right)\right)\right) + 4 \arctan\left(a^3 \tan\left(\frac{\mu \pi}{2}\right)\right)\right]}{8 \left[\arctan\left(a^3 \tan\left(\frac{\mu \pi}{2}\right)\right)\right]^2}.\label{DBI_essence_V_vargamma}
\end{gather}
The scalar potential $V(\phi)$ and the warp brane tension $T(\phi)$ is analytic except for the values of $\mu = 2m + 1, \quad m \in \mathbb{Z}$. This imparts an important constraints in the values of $\mu$ and is also present in the \textbf{Case I} and the quintessential scalar field in sect.~\ref{Subsect:quintessence}. Now, we have plotted the scalar field, the warp tension and the potential against the scale factor $a(t)$ in fig.~\ref{fig:DBI_essence_phi_T_V_vargamma} for the variable $\gamma$ scenario. We have also showed the variation of potential $V(\phi)$ against $\dot{\phi}$ and $a(t)$ in a 3D plot in fig.~\ref{fig:DBI_essence_V_dotphi_a_3D_vargamma}. 
\\Thus, we can successfully construct the DBI-essence scalar field and scalar potential considering the constant $\gamma$ and a particular variation of $\gamma$. The behaviour of the scalar field is similar to that of k-essence scenario in both the cases. The potential and the warp brane tension have some unique behaviour though, for the constant $\gamma$ and the variable $\gamma$ scenario.

The alternative fluid model proposed by Hova et al. introduces an EoS correction term involving a $\sinc$ function, leading to a modification of the standard Chaplygin gas framework. The presence of the parameter $\mu$, which appears in both the field potential reconstruction and thermodynamic relations, serves as a key dynamical tuning factor. Unlike the Generalized Chaplygin Gas (GCG), which smoothly interpolates between dust-like matter and dark energy, this model introduces an oscillatory correction term that may have implications for the equation of state evolution at intermediate redshifts. To understand the field-theoretic origins of this fluid, we establish a correspondence with well-known scalar field models such as quintessence, k-essence, and DBI-essence. Each correspondence is established by reconstructing the field potential $V(\phi)$ and kinetic terms, ensuring consistency with the fluid equation of state. We summarize our results in Table.~\ref{tab:scalar_field_comparison}, which outlines how the derived potentials for different scalar field models compare under this framework.

\begin{sidewaystable}
\centering
\caption{Comparative Analysis of Different Scalar Field Models for Dark Energy}
\renewcommand{\arraystretch}{1.5}
\begin{tabular}{|p{3cm}|p{5.5cm}|p{5.5cm}|p{5.5cm}|p{5.5cm}|}
\hline
\textbf{Properties} & \textbf{Quintessence Field} & \textbf{k-essence Field} & \textbf{DBI-essence (Const. $\gamma$)} & \textbf{DBI-essence (Var. $\gamma$)} \\
\hline
\textbf{Scalar Field $\phi$ behaviour} & 
Decreases with scale factor; reaches minimum at certain point; $\phi \to \infty$ as $a \to 0$ and $\phi \to 0$ as $a \to \infty$ &
Increases with scale factor; becomes more linear at large $a$ &
Increases with scale factor; behaviour similar to k-essence; $\phi \to 0$ as $a \to 0$ and $\phi$ diverges as $a \to \infty$ &
Increases logarithmically with scale factor; steeper slope with higher $\mu$ values \\
\hline
\textbf{Potential $V(\phi)$ behaviour} & 
Increases with scale factor; analytical except when $a^6\alpha^2\beta = 1$, $a^6 \alpha^2 \beta \notin \left(0,1\right]$ and $\mu = 2m + 1, m \in \mathbb{Z}$ &
Not applicable (purely kinetic model) &
Decreases with scale factor; analytical except when $\mu = 2m + 1, m \in \mathbb{Z}$ &
Decreases with scale factor; faster decrease rate than constant $\gamma$ case; analytical except when $\mu = 2m + 1, m \in \mathbb{Z}$ \\
\hline
\textbf{Kinetic Term behaviour} & 
$\dot{\phi}$ derived from the potential; follows equating conditions with fluid model &
$F(X)$ increases linearly with $X$; independent of $\mu$; resembles rolling potential in 3D plot &
$T(\phi)$ (warp tension) decreases with scale factor, approaches zero asymptotically; analytical except at $\gamma=1$ &
$T(\phi)$ decreases more rapidly with scale factor; stronger $\mu$ dependency than constant $\gamma$ case \\
\hline
\textbf{Parameter $\mu$ Effect} & 
Affects slope/decrease rate of $\phi$ vs. $a$ plot; minimal effect on $V(\phi)$ vs. $a$ &
Controls slope of $\phi$ vs. $a$ plot; higher $\mu$ makes $|F(X)|$ vs. $a$ plot steeper &
Minimal effect on $\phi$ rate of increase; affects both slope and asymptotic behaviour of $T(\phi)$ and $V(\phi)$ &
Prominent effect on increase rate of $\phi$; stronger influence on $T(\phi)$ and $V(\phi)$ behaviour \\
\hline
\textbf{Physical Significance} & 
Represents traditional dark energy with decreasing field strength over cosmic expansion; potential energy dominates &
Acceleration driven by non-canonical kinetic term; $\dot{\phi}^2$ decreases with expansion, showing kinetic energy dissipation &
Warped brane tension decreases with expansion; constant relativistic factor constrains field dynamics &
Variable relativistic factor ($\gamma = 1/\dot{\phi}$) allows more rapid potential evolution; more efficient energy transfer mechanism \\
\hline
\textbf{Analytical Constraints} & 
Discontinuity in $V(\phi)$ for very small $a$; analytical constraints on parameter $\mu$ &
No major analytical constraints identified &
Analytical constraints for $\mu = 2m + 1, m \in \mathbb{Z}$ and $\gamma=1$ &
Analytical constraints for $\mu = 2m + 1, m \in \mathbb{Z}$ and $\beta=0$ \\
\hline
\end{tabular}
\label{tab:scalar_field_comparison}
\end{sidewaystable}

\section{Thermodynamic Analysis}\label{Sect:Thermodynamic_analysis}
\begin{figure}[htbp]
    \centering
    \begin{subfigure}[b]{0.45\textwidth}
        \centering
        \includegraphics[width=\textwidth]{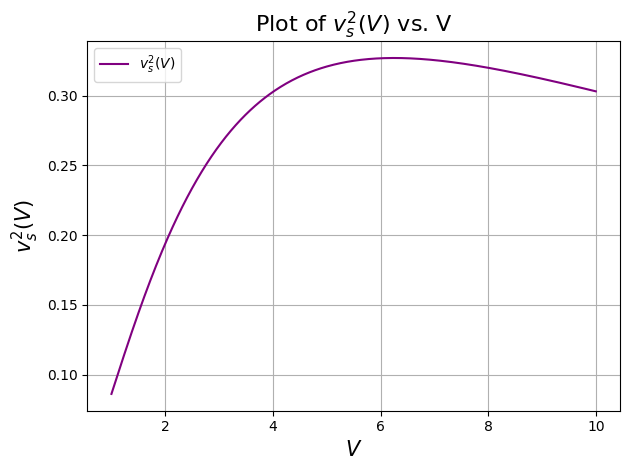}
        \caption{The variation of the square speed of sound $v_s^2$ with the volume $V$}
        \label{fig:v_square_V}
    \end{subfigure}
    \hfill
    \begin{subfigure}[b]{0.45\textwidth}
        \centering
        \includegraphics[width=\textwidth]{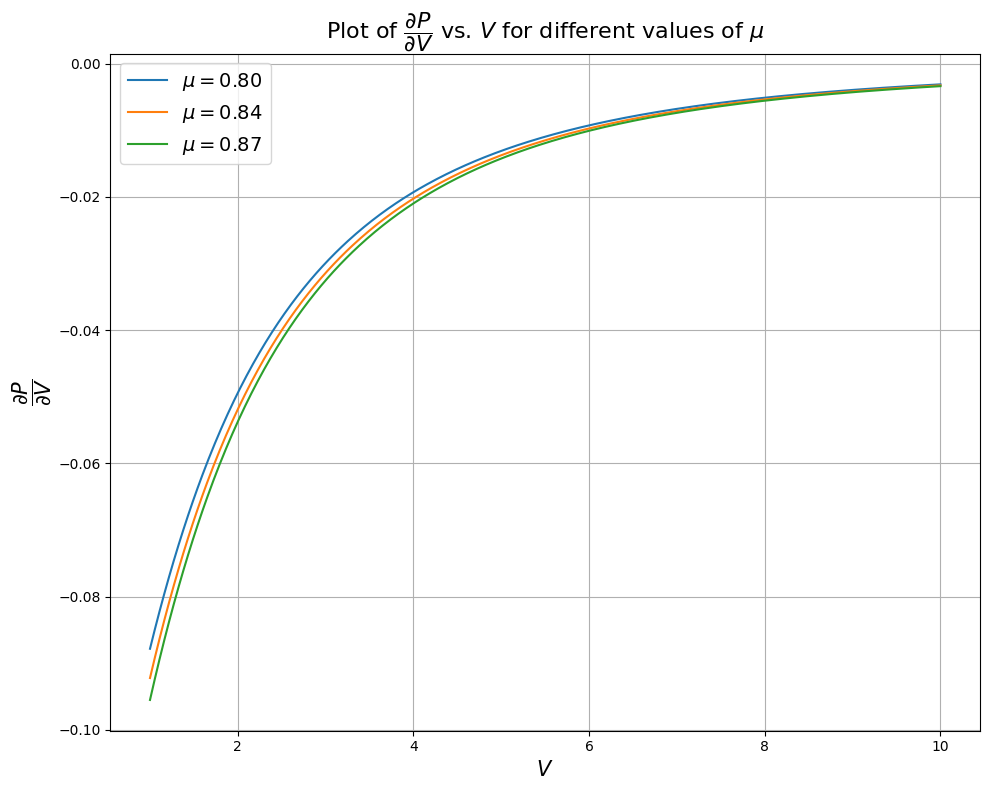}
        \caption{Variation of the $\left(\dfrac{\partial p}{\partial V}\right)_T$ with the volume $V$}
        \label{fig:delPdelV_V}
    \end{subfigure}
    \caption{Here, the first plot shows that the square speed of sound is positive throughout the expansion history of the universe. Thus the proposed model is classically stable against the perturbation. The later plot depicts that $\left(\dfrac{\partial p}{\partial V}\right)_T$ is negative for any volume and thus the model is thermodynamically stable.}
    \label{fig:v2_dPdV_V}
\end{figure}

\begin{figure}
    \centering
    \includegraphics[width=0.75\linewidth]{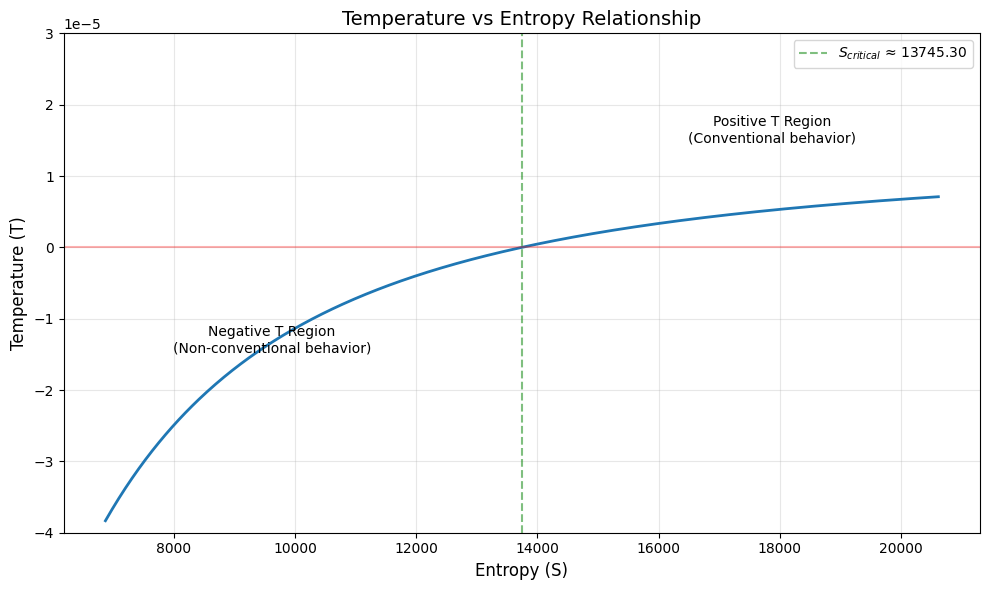}
    \caption{Plot of the variation of Temperature against Entropy}
    \label{fig:T_vs_S}
\end{figure}

Now, from the equation of state of the proposed alternative fluid model~\eqref{Hoava_EoS} and using the thermodynamic relation~\cite{landau1984statistical}, 
\begin{equation}\label{delUdelV_P}
    \left(\dfrac{\partial U}{\partial V}\right)_T=-P,
\end{equation}
we obtain the internal energy $U$ as,
\begin{equation}\label{U_Hova}
    U = \frac{(\mu \pi \rho^0 V)}{2 \cot^{-1}\left[\dfrac{e^{(\mu \pi \rho^0 C_1)}}{V}\right]},
\end{equation}
where, $C_1$ is the integration constant and function of entropy $S$ only. The energy density corresponding to this fluid as a function of volume can be written as,
\begin{equation}\label{rho_V_1}
    \rho = \dfrac{U}{V} = \frac{(\mu \pi \rho^0 )}{2 \cot^{-1}\left[\dfrac{e^{(\mu \pi \rho^0 C_1)}}{V}\right]}.
\end{equation}
Now, defining $V_0 = e^{\mu \pi \rho^0 C_1}$, the above equation can be re-written as,
\begin{equation}\label{rho_V_2}
    \rho = \frac{(\mu \pi \rho^0 )}{2 \cot^{-1}\left[\dfrac{V_0}{V}\right]}.
\end{equation}
Now, using the formula $\frac{1}{a+b}=\frac{1}{a}-\frac{b}{a^2}+\frac{b^2}{a^3}$ and approximating the $\cot^{-1}$ term up to first order, we have arrived at,
\begin{equation}\label{rho_V_Hova}
    \rho = \mu\rho^0 \left[1+\dfrac{1}{\pi}\left(\dfrac{V_0}{V}\right)\right].
\end{equation}
From this equation, it is readily observed that at the very large volume $(V \to \infty)$, the energy density becomes $\rho = \mu\rho^0$ and at the very small volume limit, the energy density becomes very high $(\rho \to \infty)$. 
The corresponding pressure can be written as,
\begin{equation}\label{pressure_Hova}
    P = \dfrac{\mu \rho^0 \left(\pi V + V_0\right) \left(-\pi^2 V + \left(\pi V + V_0\right) \sin\left(\frac{\pi^2 V}{\pi V + V_0}\right)\right)}{\pi^3 V^2}.
\end{equation}
Now, at very high energy density i.e. at very low volume the pressure becomes, $P=0$ and at the very low energy density i.e. at very high volume the pressure becomes $P=-\mu\rho^0$. Thus, the proposed fluid behave as a pressure-less dust-like matter at low volume for earlier epochs and acts as a negative constant mimicking the cosmological constant scenario at later epochs. Thus, this model can successfully unify the dark matter and the dark energy.
\begin{figure}[htbp]
    \centering
    \begin{subfigure}[b]{0.45\textwidth}
        \centering
        \includegraphics[width=\textwidth]{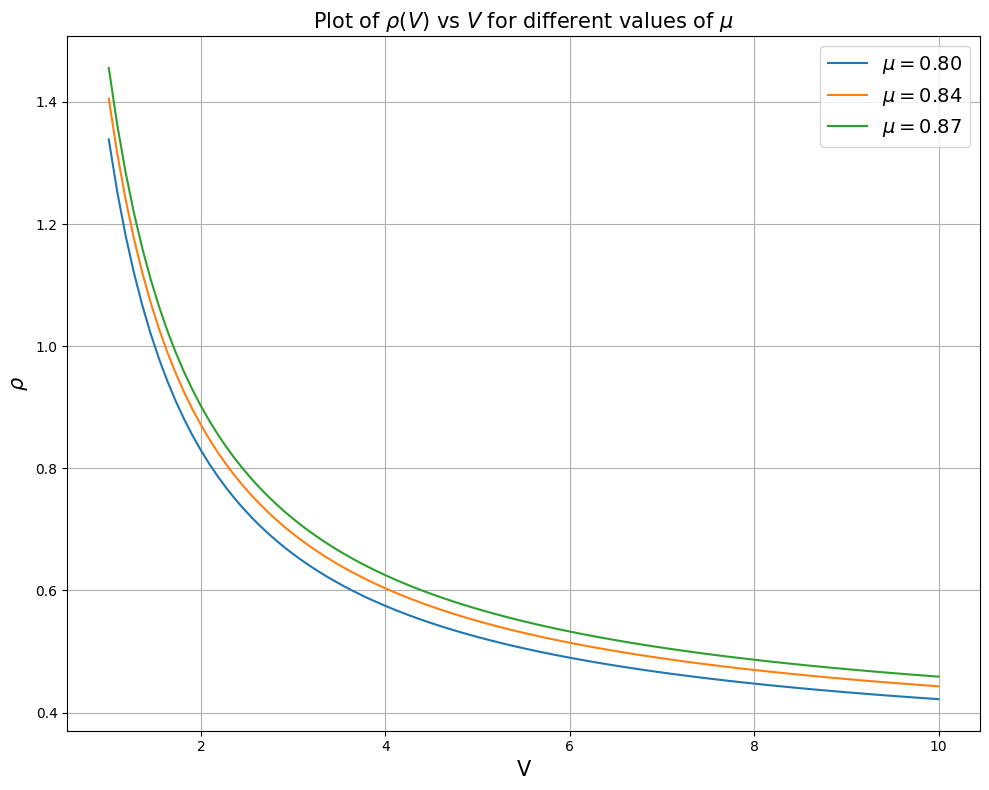}
        \caption{The variation of the energy density $\rho$ with the volume $V$}
        \label{fig:rho_V}
    \end{subfigure}
    \hfill
    \begin{subfigure}[b]{0.45\textwidth}
        \centering
        \includegraphics[width=\textwidth]{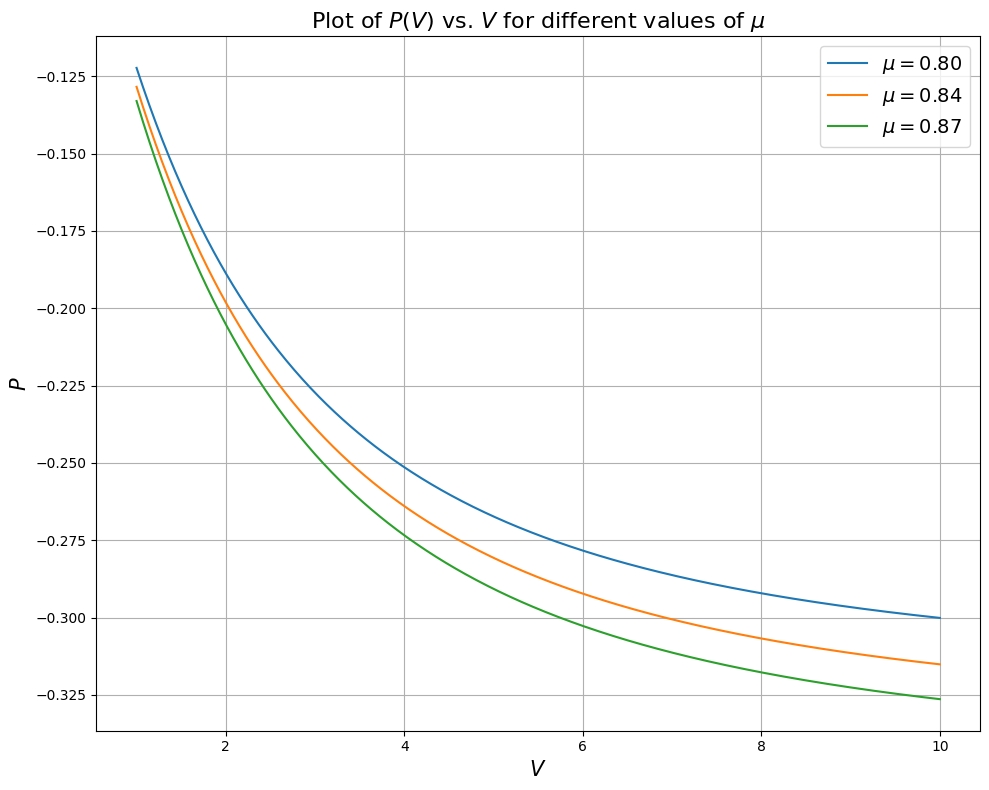}
        \caption{Variation of the pressure $P$ with the volume $V$}
        \label{fig:P_V}
    \end{subfigure}
    \caption{Here, the first plot shows the energy density is positive throughout the evolution and tends to large value at low $V$. At the other hand the pressure shows negative values implying that the fluid can describe the accelerated expansion of the universe. For large volumes, $P$ tends to have some constant negative value. }
    \label{fig:rho_P_V}
\end{figure}
\\The equation of state (EoS) parameter $\omega$ is defined as $\omega=\dfrac{P}{\rho}$. Now, using the corresponding expressions for $P$ and $\rho$ from equation~\eqref{pressure_Hova} and \eqref{rho_V_Hova}, we have obtained
\begin{equation}\label{omega_V}
    \omega(V) = \frac{-\pi^2 V + \left(\pi V + V_0\right) \sin\left(\frac{\pi^2 V}{\pi V + V_0}\right)}{\pi^2 V}.
\end{equation}
Interestingly, at very large volume, the EoS parameter becomes negative constant $(\omega=-1-\frac{1}{\pi})$. This constant EoS parameter resembles closely with the current $\Lambda$CDM model. But for small volume, the EoS parameter becomes volume dependent and acts as a dynamic EoS parameter. Though it is also evident that, at $V \to \infty$, $\omega \to -1$ from the pressure and energy density conditions mentioned above and thus the $\Lambda$CDM scenario is recovered.
\\The deceleration parameter measures the presence of acceleration on the expansion stages of the universe. The negative value of the deceleration parameter $q$ signifies the accelerated expansion of the universe. The deceleration parameter defined as,
\begin{equation}\label{q_definition}
    q = \dfrac{1}{2}+\dfrac{3P}{2\rho}.
\end{equation}
Now, from the equations~\eqref{rho_V_Hova} and \eqref{pressure_Hova} we have obtained,
\begin{equation}\label{q_V}
    q = \frac{-2\pi^2 V + 3\left(\pi V + V_0\right) \sin\left(\frac{\pi^2 V}{\pi V + V_0}\right)}{2\pi^2 V}.
\end{equation}
The deceleration parameter is seen to be positive $(q=\frac{1}{2})$ at very small volume $(V\to 0)$ and is negative, $q=-1$, for large volume limit $(V\to\infty)$. This clearly in agreement with the well known observational fact that the universe is transitioned to an acceleratedly expanding one from the deceleratedly expanding phase. It is worthy to note that the deceleration parameter in eq.~\eqref{q_V} and equation of state parameter $\omega(V)$ in eq.~\eqref{omega_V} do not contain the free parameter $\mu$. So, it can be inferred that, the parameter $\mu$ does not have any special significance in the dynamics of the universe, it just fine tunes the values of pressure and energy density at a specific epoch which exactly seen in the fig.~\ref{fig:rho_P_V}. 
\begin{figure}[htbp]
    \centering
    \begin{subfigure}[b]{0.45\textwidth}
        \centering
        \includegraphics[width=\textwidth]{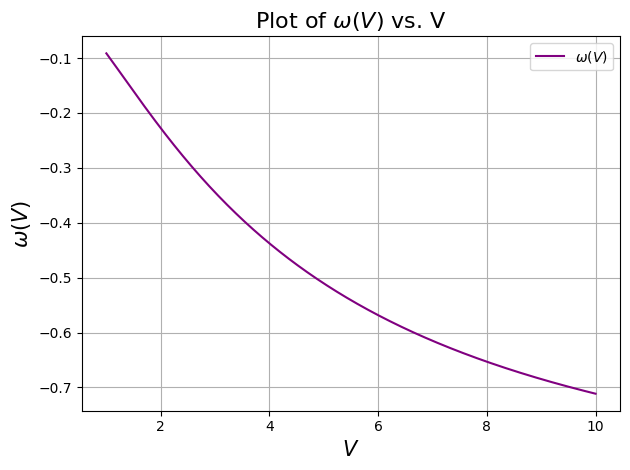}
        \caption{The variation of the EoS parameter $\omega$ with the volume $V$}
        \label{fig:omega_V}
    \end{subfigure}
    \hfill
    \begin{subfigure}[b]{0.45\textwidth}
        \centering
        \includegraphics[width=\textwidth]{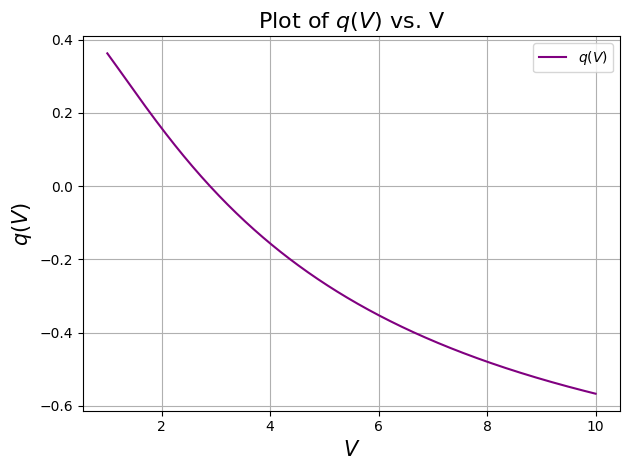}
        \caption{Variation of the deceleration parameter $q$ with the volume $V$}
        \label{fig:q_V}
    \end{subfigure}
    \caption{Here, the first plot shows that the EoS parameter is negative and tends to $-1$ as the universe expands with no evidence of crossing the phantom line $\omega=-1$. The deceleration parameter $q$ is positive at the earlier stages and becomes negative as the universe expands more. }
    \label{fig:omega_q_V}
\end{figure}
The square speed of sound is also employed in this study to test the stability of the model against classical perturbations. The definition is given by, 
\begin{equation}\label{v2_V}
    v_s^2 = \dfrac{dP}{d\rho}= 2\left(1 + \frac{V_0}{\pi V}\right)\sin\left(\frac{\pi}{1 + \frac{V_0}{\pi V}}\right)\frac{1}{\pi} - \cos\left(\frac{\pi}{1 + \frac{V_0}{\pi V}}\right) - 1.
\end{equation}
The figure~\ref{fig:v_square_V} depicts that the square of the speed of sound is positive for the entire expansion stages of the universe and thus this model is classically stable under perturbations. Also, the expression in eq.~\eqref{v2_V} shows that the classical stability of the model is also independent of $\mu$.
Now, to check the thermodynamic stability of the proposed model, we have to analyse that whether these following conditions are being satisfied or not:
\begin{itemize}
    \item During an adiabatic expansion, the pressure should decrease, implying $\left(\dfrac{\partial p}{\partial V}\right)_T < 0$.
    \item The heat capacity at constant volume must be positive, i.e., $C_V > 0$.
\end{itemize}
The $\left(\dfrac{\partial p}{\partial V}\right)_T$ is calculated from the pressure expression in equation~\eqref{pressure_Hova} and given as,
\begin{equation}\label{dPdV_V}
   \left(\dfrac{\partial p}{\partial V}\right)_T= \frac{V_0 \mu \rho_0 \left(-2\pi V \sin\left(\frac{\pi^2 V}{\pi V + V_0}\right) + \pi^2 V \cos\left(\frac{\pi^2 V}{\pi V + V_0}\right) + \pi^2 V - 2V_0 \sin\left(\frac{\pi^2 V}{\pi V + V_0}\right)\right)}{\pi^3 V^3}.
\end{equation}
From the expression it is easily inferable that $V_0 = \mu =\rho^0 \neq 0$ as for this the term $\left(\dfrac{\partial p}{\partial V}\right)_T$ becomes $0$ for any value of the volume $V$. Though the expression in \eqref{dPdV_V} contains the free parameter $\mu$, the figure~\ref{fig:delPdelV_V} indicates that, $\left(\dfrac{\partial p}{\partial V}\right)_T$ will always be negative for any values of $\mu$. For the next part that is to calculate the specific heat at constant volume $C_V$ we have to perform the following analysis. Now, revisiting the equation~\eqref{U_Hova}, we can re-write as,
\begin{equation}\label{U_rewritten}
   U = \frac{(\mu \pi \rho^0 V)}{2 \tan^{-1}\left[e^{(\mu \pi \rho^0 C_1)}V\right]}=\frac{(\mu \pi \rho^0 V)}{2 \tan^{-1}\left[CV\right]} . 
\end{equation}
where $C=e^{(\mu \pi \rho^0 C_1)}$ in this case, and $C$ is just a function of entropy $S$. In the thermodynamic framework, the consideration of the thermodynamics of parameters \(\mu\) and \(C\) as entropy-independent constants creates a fundamental discrepancy that poses a theoretical difficulty. Regardless of the volumetric or pressure conditions of the gas, the system would show a temperature decline to zero under such circumstances. The \(T = 0\) isotherm would then concurrently operate as an isentropic pathway where \(S = \text{const}\). In particular, these behavioural concepts go counter to the third law. Consequently, it is necessary to assume that at least one parameter must show entropy \(S\) dependence in order to conduct a thorough analysis of the thermodynamic stability of the suggested fluid. Using the series approximation we can further write $U$ as,
\begin{equation}\label{U_approximated}
    U \approx \dfrac{\pi\mu\rho^0}{2C}+\dfrac{\pi C\mu\rho^0 V^2}{6}\approx \dfrac{\pi C\mu\rho^0 V^2}{6} .
\end{equation}
Here, the first term is a constant term and does not contribute dimensionally, hence we have ignored it. Now, by performing dimensional analysis of eq.~\eqref{U_approximated} and using the relation $U=TS$ we have obtained,
\begin{equation}\label{C_expression}
    C = \tau S \nu^{-2} .
\end{equation}
Here, $\tau$, $\nu$ are constant parameters with dimension of temperature and volume respectively and $S$ is the entropy. Now, the temperature can be defined as, 
\begin{equation}\label{T_definition}
    T = \dfrac{\partial U}{\partial S} = \Big(\dfrac{\partial U}{\partial C}\Big) \Big(\dfrac{\partial C}{\partial S}\Big).
\end{equation}
The temperature is obtained as, 
\begin{equation}\label{T_value}
    T = \dfrac{\pi \mu \rho^0 (S^2 V^2 \tau^2 - 3\nu^4)}{6S^2 \nu^2 \tau}.
\end{equation}
The plot~\ref{fig:T_vs_S} illustrates the temperature-entropy \((T\)-\(S)\) relationship within our dark energy model, revealing a second-order (continuous) phase transition as the curve smoothly crosses the zero-temperature line at the critical entropy \(S_{\text{critical}}\). Unlike first-order transitions with abrupt thermodynamic changes, this transition involves a gradual shift in the system’s internal state. For \(S < S_{\text{critical}}\), negative temperatures emerge, reflecting non-conventional behaviour where high-energy states are more populated—akin to inverted population systems in statistical mechanics. Beyond \(S_{\text{critical}}\), the temperature becomes positive, signalling a return to conventional thermodynamics consistent with standard cosmic evolution. The absence of latent heat ensures that dark energy’s energy density and pressure evolve continuously, driving cosmic acceleration without disruptions. Near the critical point, second derivatives of thermodynamic potentials diverge, indicating critical phenomena with significant fluctuations in energy density and pressure. This behaviour offers a natural explanation for the universe’s smooth transition from matter-dominated to dark-energy-dominated epochs, reinforcing the thermodynamic consistency of our dark energy model and its role in late-time cosmic acceleration.

\begin{figure}
    \centering
    \includegraphics[width=0.75\linewidth]{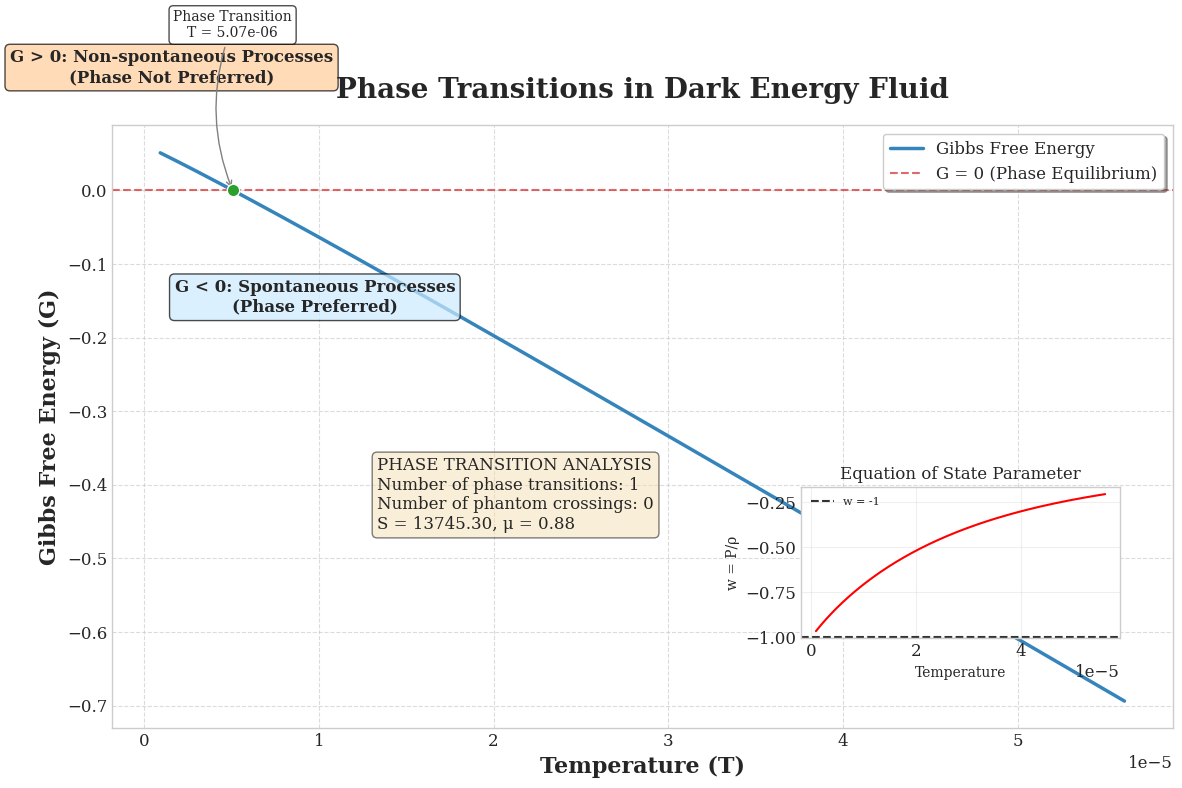}
    \caption{Plot of Gibbs Free Energy with the Temperature $T$}
    \label{fig:G_T}
\end{figure}

We have also calculated Gibbs Free Energy to understand the phase transition more accurately and then plotted it numerically in Fig.~\ref{fig:G_T}. The thermodynamic analysis depicted in Figure~\ref{fig:G_T} reveals a critical phase transition in our proposed dark energy fluid model, distinct from the generalized Chaplygin gas (GCG) model. The phase boundary occurs at $T \approx 5.07 \times 10^{-6}$, where the Gibbs free energy crosses zero. When $G > 0$, the system resides in a metastable state of dark energy, transitioning to a thermodynamically preferred phase as $G$ becomes negative. At $G = 0$, a perfect phase equilibrium is achieved, marking the critical transition temperature. The analysis indicates a single phase transition without phantom crossings ($\omega_{DE} < -1$), ensuring physical consistency throughout cosmic evolution. The entropy parameter $S = 13745.30$ with $\mu = 0.88$ characterizes the microscopic degrees of freedom of the dark energy fluid. The equation of state parameter $\omega_{DE} = P/\rho$ approaches $-1$ asymptotically without crossing the phantom divide, resolving causality issues while providing sufficient negative pressure for cosmic acceleration. As cosmic temperature dropped below the critical value $T_c \approx 5.07 \times 10^{-6}$, dark energy transitioned into its preferred phase, driving the observed late-time accelerated expansion. The smooth Gibbs free energy curve indicates a second-order phase transition, ensuring a continuous evolution that aligns with observational constraints on the universe’s smooth expansion history.
 
Now, the definition of the specific heat at constant volume can be defined as,
\begin{equation}\label{Cv_definition}
    C_V = T\left(\dfrac{\partial S}{\partial T}\right)_V.
\end{equation}
Thus, using the expression of temperature from \eqref{T_value} in the eq.~\eqref{Cv_definition} we have arrived at,
\begin{equation}\label{Cv_value}
    C_V = \dfrac{S^3 V^2 \tau^3}{6\nu^4}-\dfrac{S}{2}.
\end{equation}
Again, from the integrability condition, we know that $T = \dfrac{P+\rho}{S}$ and using this we obtain the temperature as a function of volume, 
\begin{equation}\label{T_V_value}
    T = \dfrac{\mu \rho^0 (\pi V + V_0)^2 \sin\left(\frac{\pi^2 V}{\pi V + V_0}\right)}{\pi^3 S V^2}.
\end{equation}
It is thus clear, as $C_V$ does not contain the parameter $\mu$, it do not affect the thermodynamic stability of the fluid. However, it is still plays a important role to fine tune the parameter values like pressure, energy density, temperature of the fluid at particular epochs.
\begin{figure}[htbp]
    \centering
    \begin{subfigure}[b]{0.45\textwidth}
        \centering
        \includegraphics[width=\textwidth]{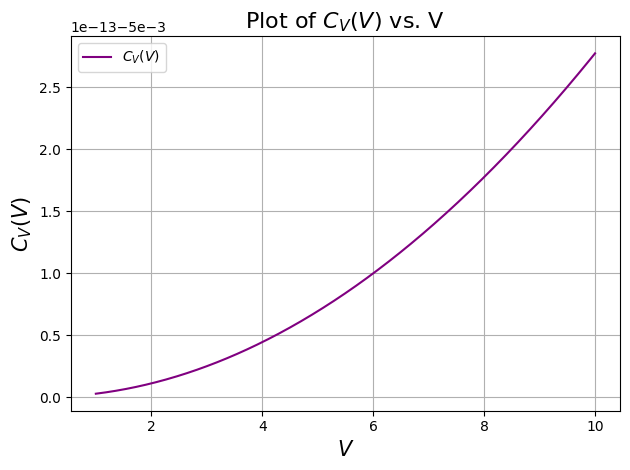}
        \caption{The variation of the specific heat at constant volume $C_V$ with the volume $V$}
        \label{fig:Cv_V}
    \end{subfigure}
    \hfill
    \begin{subfigure}[b]{0.45\textwidth}
        \centering
        \includegraphics[width=\textwidth]{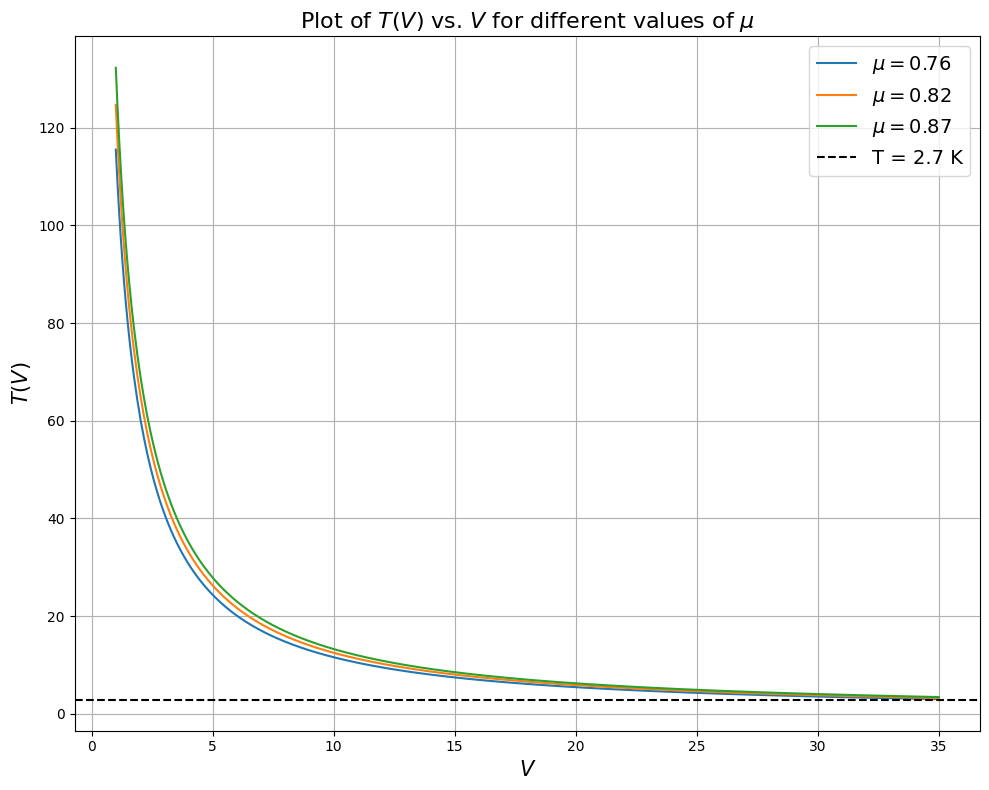}
        \caption{Variation of the temperature $T$ with the volume $V$}
        \label{fig:T_V}
    \end{subfigure}
    \caption{Here, the variation of the first plot shows that the specific heat $C_V$ is positive over any volume. This re-establishes the thermodynamic stability of the model. On the other hand the In the second plot it is shown that the temperature falls down as the universe expands and reaches the value $T=2.7K$ for the later stages of the expansion.}
    \label{fig:Cv_T_V}
\end{figure}
Now, the first law of thermodynamics can be written in the following form~\cite{chakraborty2019evolutionfrwuniversevariable},
\begin{equation}\label{TdS_relation}
    TdS = d\left(\frac{\rho}{m}\right)+Pd \left(\frac{1}{m}\right),
\end{equation}
where, $S$ is the entropy per particle, $m$ is the particle density in the system. Again, one can rewrite this as,
\begin{equation}\label{dS_relation}
    dS = \frac{1}{Tm}d\rho - \frac{P + \rho}{Tm^2}dm.
\end{equation}
This yields two thermodynamic relations,
\begin{gather}
    \left(\frac{\partial S}{\partial \rho}\right)_m = \frac{1}{Tm},\label{dS_dP} \\
    \left(\frac{\partial S}{\partial m}\right)_\rho = -\frac{p + \rho}{Tm^2}.\label{dS_dm}
\end{gather}
Considering the functional dependency $T=T(\rho,m)$, we can readily write another identity,
\begin{equation}\label{dT_relation}
    dT = \left(\frac{\partial T}{\partial m}\right)_\rho dm + \left(\frac{\partial T}{\partial \rho}\right)_m d\rho.
\end{equation}
Apart from that, the integrability condition from the first law can be stated as,
\begin{equation}\label{integrability _condition}
    T\left(\frac{\partial P}{\partial \rho}\right)m = m\left(\frac{\partial T}{\partial m}\right)\rho + (p + \rho)\left(\frac{\partial T}{\partial \rho}\right)_m .
\end{equation}
To solve the above equations~\eqref{dT_relation} and \eqref{integrability _condition}, the condition will be,
\begin{equation}\label{condition_for_solve}
    \dot{m}(\rho+p) - m\dot{\rho}=0
\end{equation}
Now, using this condition along with the previous equations in \eqref{dT_relation} and \eqref{integrability _condition} we can write,
\begin{equation}\label{dT/T_relation}
    \dfrac{dT}{T}=\left(\dfrac{dm}{m}\right)\left(\dfrac{\partial P}{\partial \rho}\right)_m .
\end{equation}
Now, from the physical viewpoint we can assert that the co-moving particle number $(\propto ma^3)$\ is constant in a FLRW universe, we obtain~\cite{BEDRAN2008462,maartens1996causalthermodynamicsrelativity},
\begin{equation}\label{dotm/m_relation}
    \dfrac{\dot{m}}{m}=-3\left(\dfrac{\dot{a}}{a}\right).
\end{equation}
Now, putting this into the above equation~\eqref{dT/T_relation}, the relation is obtained as,
\begin{equation}\label{dotT/T_relation}
    \dfrac{\dot{T}}{T}=-3\left(\dfrac{\dot{a}}{a}\right)\left(\dfrac{\partial P}{\partial \rho}\right)_m .
\end{equation}
This equation provides a temperature profile of the universe filled with the fluid presented by Hoava. Putting the pressure and energy density values from our analysis in eq.~\eqref{dotT/T_relation}, we can obtain the temperature profile in terms of redshift $z$.
\begin{equation}\label{T(z)_Hova}
    T(z) = T_0 \cdot \exp \left( \int_0^z \frac{9 \pi^3}{\mu \rho_0 \left( \frac{V_0 (z'+1)^3}{\pi} + 1 \right)^4} \cdot \frac{dz'}{1+z'} \right).
\end{equation}
Now, this expression is hard to compute. For this sake we will seek the numerical solution and that inspires us to consider some observational dataset of temperature evolution of the universe. The cosmic microwave background radiation (CMBR) emerges as a fundamental remnant of the hot Big Bang, representing the electromagnetic radiation that persisted after matter decoupling during the early evolutionary phases of the Universe. This primordial radiation plays a critical role in astronomical research, particularly in its interactions with interstellar molecules like carbon monoxide (CO), which serve as precise measurement instruments for cosmological investigations.

\begin{figure}
    \centering
    \includegraphics[width=0.75\linewidth]{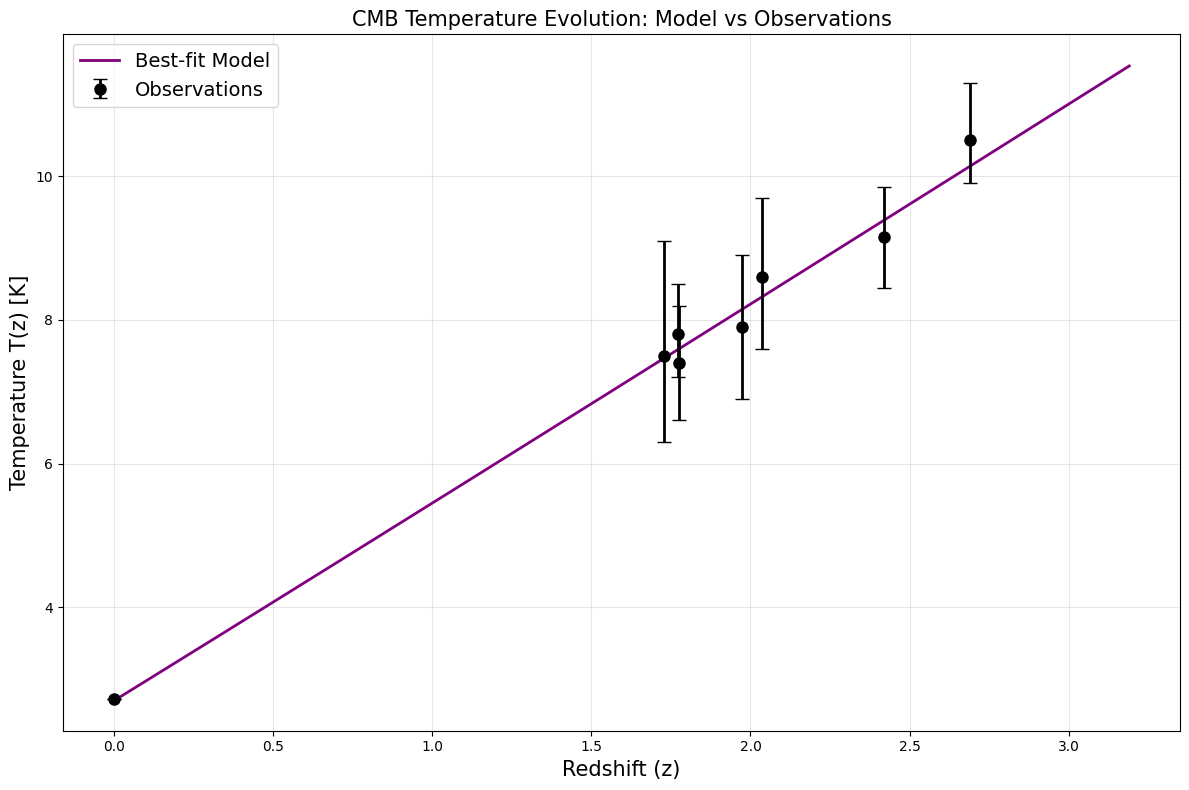}
    \caption{Plot of temperature evolution as seen from the CMBR observations.}
    \label{fig:T_z_CMB}
\end{figure}

Initial groundbreaking research by Songaila et al.~\cite{songaila1994measurement} determined the cosmic microwave background temperature through neutral carbon atom measurements. Specifically, they estimated $T_{CMBR}= 7.4 \pm 0.8K$ at a redshift of $z = 1.776$ in a cloud toward the quasar $Q1331 + 170$. Subsequent refined measurements by Mather et al.~\cite{Mather_1999} established the present CMB black-body temperature at $T_{CMBR} = 2.725 \pm 0.002K$, measured locally at redshift $z = 0$. Lima et al.~\cite{Lima_2000} introduced an innovative perspective challenging standard cosmological models, proposing that the CMB temperature at high redshifts might deviate from predicted standard values. Their research incorporated novel concepts including decaying vacuum energy density and gravitational 'adiabatic' photon creation, complemented by late inflationary models driven by scalar field dynamics. Srianand et al.~\cite{Srianand_2008} conducted a comprehensive investigation using high-resolution spectroscopy, imposing stringent upper limits on $T_{CMB}$ across a large sample of C I fine structure absorption lines. Their most notable finding came from detecting carbon monoxide in a damped Lyman-$\alpha$ system at $z_{abs} = 2.41837$ toward SDSS $J143912.04+111740.5$, from which they determined $T_{CMBR} = 9.15 \pm 0.72K$. Building upon previous research, J. Ge et al.~\cite{Ge_1997} identified absorption lines from ground and excited states of C I in the $z = 1.9731$ damped Lyman-$\alpha$ system of QSO 0013-004. Through careful analysis of population ratios between excited and ground states, they estimated the CMBR temperature as $T = 7.9 \pm 1.0K$ at $0.61$ mm and $z = 1.9731$, which aligned closely with contemporary standard cosmological predictions. Noterdaeme et al.~\cite{Noterdaeme_2010} analysed a sub-damped Lyman-$\alpha$ system with neutral hydrogen column density at $z_{abs} = 2.69$ toward SDSS $J123714.60 + 064759.5$. Their investigation revealed that CO excitation was predominantly influenced by radiative interaction with the CMBR, deriving $T_{ex}(CO) = 10.5K$, corresponding to the expected $T_{CMBR}(z = 2.69) = 10.05K$. A systematic survey utilizing VLT/UVES spectrographs allowed P. Noterdaeme et al.~\cite{Noterdaeme_2011} to constrain the CMBR temperature evolution to $z \approx 3$. By combining their measurements with previous constraints, they formulated a refined temperature model: $T_{CMB} (z) = (2.725 \pm 0.002) \times (1 + z)1-\beta K$, with $\beta = -0.007 \pm 0.027$. These comprehensive investigations have collectively motivated further research into deriving precise expressions for the temperature of the Friedmann-Robertson-Walker (FRW) universe dominated by the proposed fluid matter as a function of redshift, aiming to rigorously validate existing cosmological models through meticulous observational constraints. We have presented our observational datasets in Tab.~\ref{tab:temp_CMB_observations}. The plot of the eq.~\eqref{T(z)_Hova} against the observational temperature values from CMB is presented in fig.~\ref{fig:T_z_CMB}. The plot shows the model is well fitted with the observed temperature values and the best fit model parameters are $T_0=2.725$, $V_0=0.845$, $\mu=0.889$ with the $\chi^2$/dof being $0.7$. The comparison with CMB data serves as a crucial test for the viability of the proposed model. The key insights derived from this comparison include the consistency with the observed thermal history, where the predicted temperature evolution aligns with CMB temperature measurements, validating the model’s compatibility with early universe thermodynamics. This agreement indicates that modified gravity effects do not disrupt the standard thermal history during the recombination era. Additionally, while the primary CMB spectrum remains consistent with standard predictions, subtle deviations in higher-order acoustic peaks suggest modifications to the cosmic fluid’s sound speed and thermal diffusion properties, providing indirect evidence of the fluid’s altered thermodynamic properties. Moreover, the observational comparison imposes constraints on key model parameters, such as the coupling strength of scalar fields and the modification scale of gravitational interactions, ensuring that the proposed phase transitions occur at cosmologically relevant epochs without conflicting with CMB measurements. Methodologically, the study advances the field by integrating thermodynamic analysis with observational constraints from CMB data, bridging the gap between theoretical predictions and empirical observations. This combined approach identifies thermodynamic signatures as observational probes, enabling future CMB experiments and large-scale structure surveys to test the proposed phase transitions and stability conditions. Furthermore, by incorporating thermodynamic constraints, the study refines the parameter space of viable dark energy and modified gravity models, enhancing their predictive power. The approach is also extendable to other cosmological datasets, such as baryon acoustic oscillations (BAO) and supernovae data, providing a robust framework for testing the thermodynamic consistency of alternative cosmological models.
\begin{table}[htbp]
\centering
\begin{tabular}{cccc}
\hline
$z_{\mathrm{obs}}$ & $T_{\mathrm{obs}}$ (K) & $T_{\mathrm{err,up}}$ (K) & $T_{\mathrm{err,down}}$ (K) \\
\hline
0.0000 & 2.725 & 0.002 & 0.002 \\
1.7760 & 7.400 & 0.800 & 0.800 \\
1.7293 & 7.500 & 1.600 & 1.200 \\
1.7738 & 7.800 & 0.700 & 0.600 \\
2.6896 & 10.500 & 0.800 & 0.600 \\
2.4184 & 9.150 & 0.700 & 0.700 \\
2.0377 & 8.600 & 1.100 & 1.000 \\
1.9731 & 7.900 & 1.000 & 1.000 \\
\hline
\end{tabular}
\caption{Observational data showing redshift ($z_{\mathrm{obs}}$), temperature measurements ($T_{\mathrm{obs}}$), and their associated upper ($T_{\mathrm{err,up}}$) and lower ($T_{\mathrm{err,down}}$) uncertainties from the literature~\cite{songaila1994measurement,Mather_1999,Lima_2000,Srianand_2008,Ge_1997,Noterdaeme_2010,Noterdaeme_2011}.}
\label{tab:temp_CMB_observations}
\end{table}
\\Now, if we use another thermodynamic relation,
\begin{equation}\label{dP_dT_definition}
    \dfrac{dP}{dT}=\dfrac{P+\rho}{T},
\end{equation}
and solve this differential equation, we will obtain pressure as a function of temperature given by,
\begin{equation}\label{P_T}
    P(T) = -\left[\mu\pi\rho^0-\left(\dfrac{T}{T^*}\right)^{2}\right]^{\frac{1}{2}}.
\end{equation}
Here, \( T^* \) represents an integration constant, with the condition \( 0 < T < T^* \), indicating that \( T^* \) is the highest temperature that the gas can achieve. Now let us assume the initial values of this system are given by $V = V_i$, $p = p_i$, $\rho = \rho_i$, $T = T_i$.
Using this relation, we can readily obtain the internal energy of the system as function of temperature and volume,
\begin{equation}\label{U_T}
    U = \dfrac{\mu\pi\rho^0 V}{\left[\mu\pi\rho^0-\left(\dfrac{T}{T^*}\right)^{2}\right]^{\frac{1}{2}}}.
\end{equation}
Now, from the equation~\eqref{rho_V_Hova}, we can find the expression of the arbitrary parameter $V_0$ by using the initial values and is given as,
\begin{equation}\label{V0_value}
    V_0 = V_i\pi\left[\dfrac{\rho_i}{\mu\rho^0}-1\right].
\end{equation}
Now, putting it in the pressure and energy density equations, we can obtain $P$ and $\rho$ in terms of initial values given by,
\begin{gather}
    \rho = \rho_i\left[\dfrac{\mu\rho^0}{\rho_i}+\left(1-\dfrac{\mu\rho^0}{\rho_i}\right)\dfrac{V_i}{V}\right],\label{rho_initial} \\
    P = -(\mu\pi\rho^0)\left[\rho_i\left[\dfrac{\mu\rho^0}{\rho_i}+\left(1-\dfrac{\mu\rho^0}{\rho_i}\right)\dfrac{V_i}{V}\right]\right]^{-1}.\label{p_initial}
\end{gather}
In this point, we can define some the reduced parameters, 
\begin{equation}\label{reduced_parameters}
    \epsilon = \dfrac{\rho}{\rho_i}, \quad v = \dfrac{V}{V_i}, \quad p = \dfrac{P}{(\mu\rho^0)^{-3}}, \quad \gamma = \dfrac{\mu\rho^0}{\rho_i}, \quad t = \dfrac{T}{T_i}, \quad t^* = \dfrac{T^*}{T_i}.
\end{equation}
Now, using these definitions, the energy density, pressure can be represented using eqs.~\eqref{rho_initial} and \eqref{p_initial},
\begin{gather}
    \epsilon = \left[\gamma+(1-\gamma)\dfrac{1}{v}\right], \label{rho_reduced}\\
    p = -(\mu\pi\rho^0) \gamma \left[\gamma+(1-\gamma)\dfrac{1}{v}\right]^{-1},\label{p_reduced}\\
     p(T) = -\left[\mu\pi\rho^0-\left(\dfrac{t}{t^*}\right)^{2}\right]^{\frac{1}{2}}.\label{p_T_reduced}
\end{gather}
By definition, when \( p = p_i \), \( V = V_i \), and \( T = T_i \), we have \( t = 1 \) and \( v = 1 \). Therefore, by equating \( p(T) \) form~\eqref{p_T_reduced} and \( p \) from eq.~\eqref{p_reduced}, we obtain,
\begin{equation}\label{p_p(T)}
   \left[\mu\pi\rho^0-\left(\dfrac{1}{t^*}\right)^{2}\right]^{\frac{1}{2}} = (\mu\pi\rho^0)\gamma.
\end{equation}
Solving the equation, we get the value of the $t^*$ as,
\begin{equation}\label{t*_value}
    t^* = \dfrac{1}{(\mu\pi\rho^0)^{\frac{1}{2}}}\left[\dfrac{1}{1-\gamma}\right],
\end{equation}
and the $\gamma$ can be written as,
\begin{equation}\label{gamma_expression}
    \gamma= \left[1-\dfrac{1}{(\mu\pi\rho^0)^2 t^*}\right]^{\frac{1}{2}}.
\end{equation}
In our study, we take into account two important temperature points in the cosmological timeline: the Planck epoch temperature conditions and the theoretical peak temperature of the alternative fluid model to the GCG, represented by the symbol ($T^{*}$)  and  estimated  at
$T^{32}$ 
units. The current cosmic microwave background temperature, which is measured at $T_{i} = 2.7$
units, is contrasted with this. From these boundary conditions, we can proceed with our calculations.
\begin{equation}\label{gamma_value}
    \gamma= \left[1-\dfrac{1}{(\mu\pi\rho^0)^2 10^{32}}\right]^{\frac{1}{2}} \approx 1.
\end{equation}
Now, in this scenario, the energy density of the universe filled with the alternative fluid model at current epoch is $\mu\rho^0$, which is also seen from the equation~\eqref{rho_V_Hova} also, in the large volume limit.

A key aspect of our analysis involves checking the stability of the proposed alternative fluid model under classical perturbations. This is done by computing the squared speed of sound, $v_s^2=\frac{dp}{d\rho}$. In our case, we find that $v_s^2$ remains positive throughout cosmic evolution (see Fig.~\ref{fig:v_square_V}), implying that the model is free from ghost instabilities. Additionally, the specific heat at constant volume, $C_V$, is computed to verify the thermodynamic stability of the model. Our findings indicate that the entropy of the system is always increasing, thus this must satisfy the Generalized Second Law (GSL) at the apparent horizon, though violations can occur at the event horizon at certain epochs. This suggests that the choice of horizon significantly affects the thermodynamic consistency of the model, a feature not commonly explored in previous works on Chaplygin-like fluid models.
\section{Validity of Generalised Second Law On Apparent Horizon and Cosmic Event Horizon}\label{Sect:GSL}
The Generalised Second Law of cosmology asserts that the total entropy of the universe that is the entropy within the universe and the entropy at the horizon must increase. The Friedman equations can be written in the form of first law of thermodynamics as~\cite{Rong_Gen_Cai_2005},
\begin{equation}\label{friedmann_thermodynamic}
    -dE_H = T_HdS_H,
\end{equation}
where, $E_H$ is the energy flux flowing through the horizon, $T_H$ is the horizon temperature and $dS_H$ is the change of entropy at the horizon. Now, assuming the gravitational formulation of first law of thermodynamics holds on the horizons, the mathematical formulation of GSL is as follows,
\begin{equation}\label{GSL_definition}
    \dfrac{dS_T}{dt}=\dfrac{dS_b}{dt}+\dfrac{dS_H}{dt} > 0.
\end{equation}
Here, $S_b$ is the entropy of the bulk fluid within the universe. In terms of horizon length, it can be re-written as~\cite{Chakraborty_2019},
\begin{equation}\label{GSL_expression}
    \dfrac{dS_T}{dt}=\dfrac{4\pi R_H^2(p+\rho)\dot{R}_H}{T_H}.
\end{equation}
Now, if we consider the the cosmological event horizon as future event horizon then\cite{Debnath_2011},
\begin{equation}\label{GSL_event_horizon}
    \dfrac{dS_T}{dt}=16\pi^2 R_H(\dot{H}R_H^2+HR_H-1),
\end{equation}
where, the horizon length $R_H$ and the horizon temperature $T_H$ is given as~\cite{Mohseni_Sadjadi_2010},
\begin{gather}
    R_H=a\int_a^{\infty}\dfrac{da}{a^2 H},\label{R_event_horizon}\\
    T_{EH}=\dfrac{1}{2\pi R_H}.\label{T_event_horizon}
\end{gather}
Using the fact that, $V=a(t)^3$, $3H^2=\rho$ and $a(t)=\dfrac{1}{1+z}$ we have calculated the horizon length $R_H$ in terms of redshift $z$ as,
\begin{equation}\label{R_event_horizon_expression}
    R_H = \frac{\sqrt{3}  {}_2F_1\left(\frac{1}{3}, \frac{1}{2}; \frac{4}{3}; \frac{V_0(z + 1)^3e^{i\pi}}{\pi}\right)}{\sqrt{\mu}\sqrt{\rho_0}},
\end{equation}
where, ${}_2F_1$ represents the hypergeometric function. Then, the rate of change of total entropy is obtained using \eqref{GSL_event_horizon} as,
\begin{align}\label{dSdT_event_horizon}
\frac{dS_T}{dt} = -\frac{8 \sqrt{3} \sqrt{\pi}}{\sqrt{\mu} \sqrt{\rho_0}} 
\Bigg( &
3 \sqrt{\pi} V_0 (z + 1)^3 
\left( {}_2F_1 \left( \frac{1}{3}, \frac{1}{2}; \frac{4}{3}; \frac{V_0 (z + 1)^3 e^{i \pi}}{\pi} \right) \right)^2 \nonumber \\
& - 2 \pi \sqrt{V_0 (z + 1)^3 + \pi} 
{}_2F_1 \left( \frac{1}{3}, \frac{1}{2}; \frac{4}{3}; \frac{V_0 (z + 1)^3 e^{i \pi}}{\pi} \right)\nonumber \\
& + 2 \pi^{3/2} 
\Bigg) 
{}_2F_1 \left( \frac{1}{3}, \frac{1}{2}; \frac{4}{3}; \frac{V_0 (z + 1)^3 e^{i \pi}}{\pi} \right).
\end{align}
Now, we will consider the horizon as the apparent event horizon. The horizon length is given by,
\begin{equation}\label{R_apparent_horizon}
    R_H=\dfrac{1}{H}.
\end{equation}
Next, we considered the apparent horizon temperature to be the Kodama-Hayward temperature and for a flat FLRW universe it is given by~\cite{Binétruy_2015},
\begin{equation}\label{T_apparent_horizon}
    k_BT_{AH}=\left(\dfrac{\hbar G}{c}\right)\left(\dfrac{\rho-3P}{3H}\right).
\end{equation}
The Kodama-Hayward temperature represents a crucial thermodynamic characterization of black hole horizons, offering profound insights into quantum gravitational phenomena near event horizons. This temperature, derived from quantum gravitational corrections, provides a nuanced description of the thermal properties of apparent horizons that transcends classical Hawking radiation models. By incorporating quantum geometric effects, the Kodama-Hayward temperature illuminates the intricate relationship between quantum mechanics and gravitational thermodynamics, suggesting potential resolutions to long-standing paradoxes concerning information preservation and quantum entanglement at black hole boundaries. Its significance lies in bridging fundamental quantum mechanical principles with gravitational spacetime curvature, thereby offering a more comprehensive understanding of black hole thermodynamics beyond traditional semiclassical approximations. It should be noted that we will use the natural unit system where $k_B=G=\hbar=c=1$. Thus the equation~\eqref{GSL_expression} becomes,
\begin{equation}\label{GSL_apparent_horizon}
    \dfrac{dS_T}{dt}=\dfrac{6\pi (P+\rho)^2}{H^3(\rho-3P)}.
\end{equation}
Now, we have deduced the eq.~\eqref{GSL_apparent_horizon} using the eqs.~\eqref{R_apparent_horizon}, \eqref{T_apparent_horizon}, \eqref{rho_V_Hova} and \eqref{pressure_Hova}. As, the expressions of $\dfrac{dS_T}{dt}$ and $T_{AH}$ is very complicated in form, so we do not present those in this paper, rather we have plotted them with respect to redshift $z$ in the following figures~\ref{fig:GSL_apparent_horizon} and \ref{fig:T_apparent_horizon}.
\begin{figure}[htbp]
    \centering
    \begin{subfigure}[b]{0.45\textwidth}
        \centering
        \includegraphics[width=\textwidth]{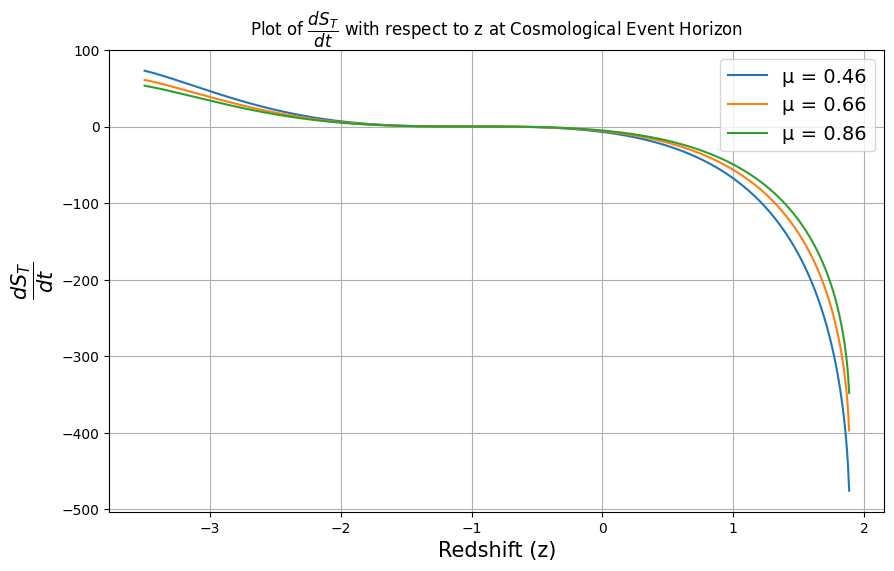}
        \caption{The plot of $\dfrac{dS_T}{dt}$ with redshift $z$ at the cosmological event horizon }
        \label{fig:GSL_event_horizon}
    \end{subfigure}
    \hfill
    \begin{subfigure}[b]{0.45\textwidth}
        \centering
        \includegraphics[width=\textwidth]{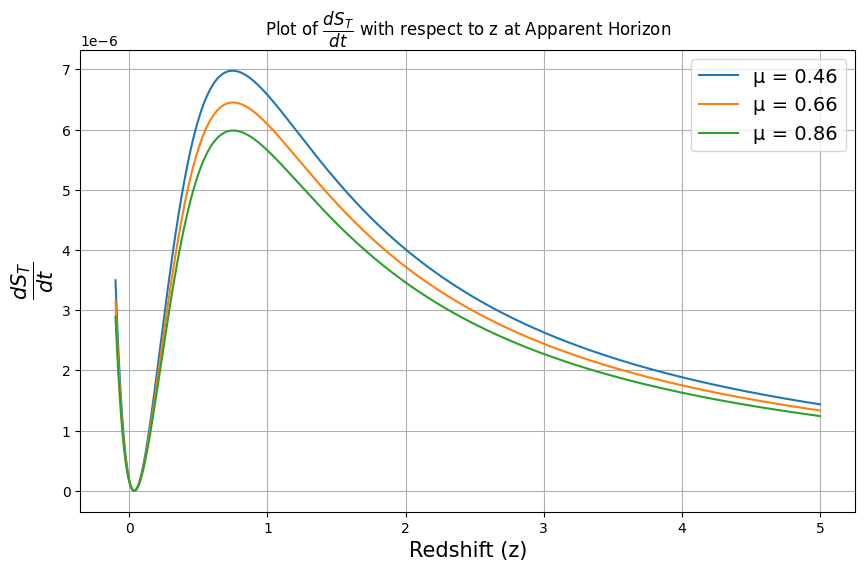}
        \caption{The plot of $\dfrac{dS_T}{dt}$ with redshift $z$ at the apparent event horizon.}
        \label{fig:GSL_apparent_horizon}
    \end{subfigure}
    \hfill
    \begin{subfigure}[b]{0.45\textwidth}
        \centering
        \includegraphics[width=\textwidth]{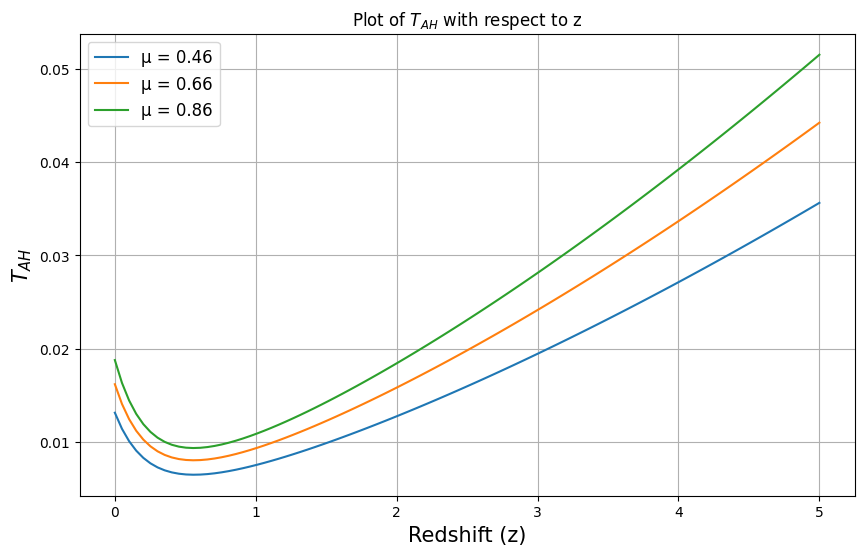}
        \caption{The plot of Kodama-Hayward temperature $T_{AH}$ with redshift $z$ at the apparent event horizon}
        \label{fig:T_apparent_horizon}
    \end{subfigure}
    \caption{Here in this figure. (a) it is seen that the GSL is violated at the cosmological event horizon for the evolutionary epochs of the universe. Though the positivity of $\dfrac{dS_T}{dt}$ at future epochs ensures that GSL will be satisfied in the future epochs of the universe if we consider the future event horizon. (b) it is evident from the positivity of $\dfrac{dS_T}{dt}$ throughout the redshift range that the GSL is always satisfied at the apparent event horizon of the universe. (c) it shows that the horizon temperature is always positive for the whole redshift range ensuring the thermodynamic stability of the model.}
    \label{fig:GSL_T_EH_AH}
\end{figure}
The study is conducted for two horizons: the cosmological event horizon and the apparent horizon. For the event horizon, the entropy rate is determined using Eq.~\eqref{GSL_event_horizon} and plotted in Fig.~\ref{fig:GSL_event_horizon}, indicating GSL violation during certain evolutionary epochs but ensuring future satisfaction. For the apparent horizon, using Eq.~\eqref{GSL_apparent_horizon}, Fig.~\ref{fig:GSL_apparent_horizon} demonstrates that GSL is always satisfied. Additionally, the Kodama-Hayward temperature plotted in Fig.~\ref{fig:T_apparent_horizon} remains positive throughout, ensuring thermodynamic stability.

The sign of $\dfrac{dS_T}{dT}$ plays a crucial role in determining the stability and physical nature of cosmic evolution. At the apparent horizon, where $\dfrac{dS_T}{dT}$ is positive, entropy increases with time, indicating that the system tends to evolve toward a more thermodynamically stable state. This positive gradient suggests that perturbations or deviations from equilibrium are naturally damped, contributing to the long-term stability of the cosmological model. Conversely, at the cosmological event horizon, where $\dfrac{dS_T}{dT}$ is negative, the entropy decreases with time. This implies that the system may evolve away from equilibrium, making it more susceptible to instabilities or non-equilibrium effects. Physically, this reflects the influence of the accelerating expansion of the universe, where the event horizon acts as a boundary that limits causal interactions, potentially affecting the entropy flow and thermodynamic balance. It is important to emphasize that the analysis of GSL not only examines the thermodynamic evolution of the universe but also provides insight into the stability of the underlying cosmological model. The positivity of both $\dfrac{dS_T}{dt}$ and $T_{AH}$ indicates that the model is thermodynamically stable at the apparent horizon, as any deviation would imply instability. In contrast, the negative entropy gradient at the cosmological event horizon highlights the complex interplay between accelerated expansion and thermodynamic evolution, suggesting that additional factors may influence long-term stability. Therefore, explicit considerations of stability are crucial, as thermodynamic stability ensures the robustness and physical viability of the background theory. Hence, the consistency observed in these results strengthens the validity of the proposed cosmological model, aligning with the fundamental principles of thermodynamics and cosmic evolution.

\section{Observational Constraints and Analysis}\label{Sect:Observational_constraints}
In this section, we will constrain our model parameters using Cosmic Chronometers (CC), Baryon Acoustic Oscillation (BAO) dataset, Pantheon+SH0ES compilation and Union 2.1 compilation from Supernova Type Ia. We will also provide a detailed presentation of the used datasets and the methodology. The results of our analysis will then be presented and analysed by comparing the observed quantities using the model with the observed datasets.

\subsection{Methodology}\label{Subsect:Methodology}
In the part that follows, we will provide the observational datasets and methodology to restrict the free parameters of proposed fluid description of dark energy~\cite{Hova_model}. We implemented the MCMC algorithm in Python using the \texttt{emcee} package \cite{foreman2013emcee}. The likelihood function was defined by assuming Gaussian errors in the observational data. Uniform priors were assigned to the model parameters within physically motivated bounds. 

\begin{figure}
    \centering
    \includegraphics[width=0.95\linewidth]{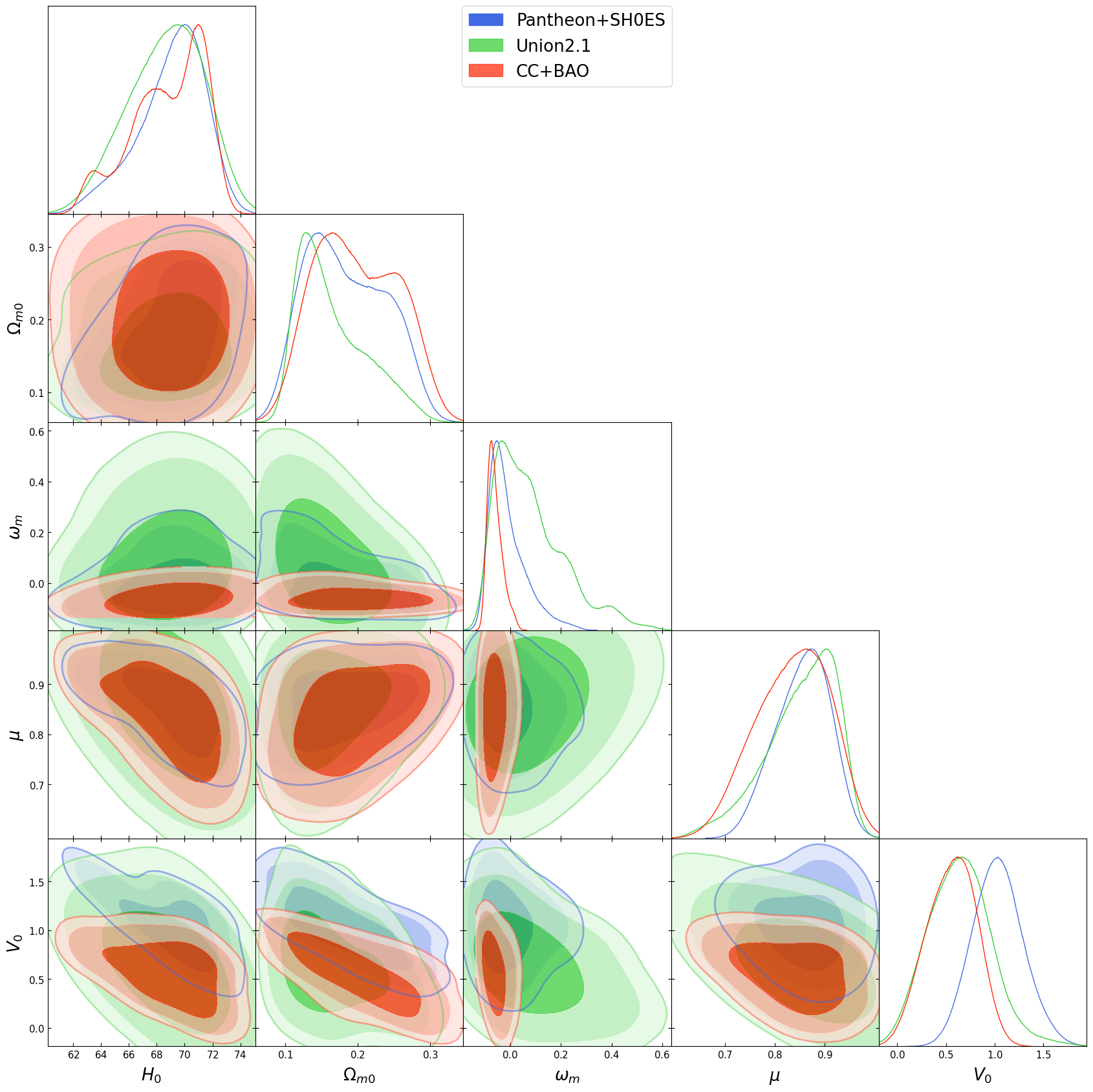}
    \caption{Posterior distribution of the MCMC analysis for the theoretical model at $1\sigma$ and $2\sigma$ confidence level.}
    \label{fig:getdist_model}
\end{figure}

Markov Chain Monte Carlo (MCMC) simulations were conducted with $10000$ iterations, preceded by a $100$-step burn-in period to establish convergence. A swarm of $100$ walkers was utilized to explore the parameter space efficiently. The Gelman-Rubin diagnostic was monitored to verify convergence, ensuring that $R$ remained below $1.1$ for all parameters. The optimal parameter values derived from our analysis are summarized in Table~\ref{tab:best_fit_params_values}. The posterior distributions at 1$\sigma$ and 2$\sigma$ confidence levels are illustrated in Figure~\ref{fig:getdist_model}.

\subsubsection{Cosmic Chronometers (CC) and Baryon Acoustic Oscillation (BAO) Dataset}
Our comprehensive cosmological analysis integrates 26 Baryon Acoustic Oscillation (BAO) measurements with 31 independent Cosmic Chronometer (CC) observations, collectively providing 57 unique cosmological constraints. The CC technique offers a pioneering approach to directly quantify the Hubble parameter $H(z)$ across various cosmic epochs by analyzing differential age variations in passively evolving galactic systems with minimal contemporary star formation. A distinctive merit of CC observations resides in their intrinsic model-agnostic nature, enabling an unbiased exploration of cosmic expansion mechanisms. Our curated CC dataset comprises 31 distinct $H(z)$ measurements spanning redshifts from $z \approx 0.07$ to $z \approx 1.97$, derived through meticulous differential age analyses of mature stellar populations \cite{stern2010cosmic, moresco2012new, moresco2012improved, zhang2014four, moresco2015raising, moresco20166}.

The Baryon Acoustic Oscillation (BAO) represents a primordial acoustic resonance embedded within galactic spatial distributions, functioning as a precise cosmic geometrical standard. Our investigation synthesizes 26 BAO measurements, encompassing isotropic and anisotropic observations from diverse galaxy surveys including SDSS \cite{ross2015clustering, alam2017clustering, gil2020completed, raichoor2021completed, hou2021completed, des2020completed}, DES \cite{abbott2022dark}, DECaLS \cite{sridhar2020clustering}, and 6dFGS \cite{beutler20116df}. These observational data traverse a redshift range of $0.106 < z < 2.36$.

Now, we can write the Hubble parameter expression as a function of redshift $z$ for our model by using \eqref{Hubble_parameter}, \eqref{rho_V_Hova} as,
\begin{equation}\label{Hubble_parameter_model}
    H(z) = H_0 \cdot \sqrt{\Omega_{m0} \cdot (1 + z)^{3(1 + \omega_m)} + (1 - \Omega_{m0}) \cdot \mu \cdot \left(1 + \frac{V_0}{\pi V}\right)}.
\end{equation}
Here, we assume $\Omega_{de0}=1-\Omega_{m0}=\frac{\mu\rho^0}{3H_0^2}$ as the present day dark energy density parameter. The eq.~\eqref{Hubble_parameter_model} is used to compare the theoretical and observational datasets using MCMC analysis.

Given the distinct statistical characteristics of our datasets, we implement a nuanced analytical approach. The 31 Cosmic Chronometer measurements exhibit statistical independence, whereas the 26 BAO measurements demonstrate intrinsic correlations. To accommodate this complexity, we employ a generalized chi-square likelihood formalism that incorporates a full covariance matrix:

\begin{equation}
    \chi^2_{\text{total}} = \sum_i \sum_j \left(H_{\text{obs}}(z_i) - H_{\text{model}}(z_i)\right) \, \mathbf{C}^{-1}_{ij} \, \left(H_{\text{obs}}(z_j) - H_{\text{model}}(z_j)\right).
\end{equation}
Here, $\mathbf{C}^{-1}_{ij}$ represents the inverse of the full covariance matrix, which explicitly accounts for the correlated nature of BAO measurements. For Cosmic Chronometer data, this reduces to a diagonal covariance matrix with independent uncertainties:
\begin{equation}
    \chi^2_{\text{CC}} = \sum_i \left(\frac{H_{\text{obs}}(z_i) - H_{\text{model}}(z_i)}{\sigma_i}\right)^2.
\end{equation}
\\This approach ensures statistically robust integration of heterogeneous cosmological observations, appropriately weighting the correlated BAO measurements while preserving the independent nature of Cosmic Chronometer data.

\subsubsection{Pantheon+SH0ES Compilation}
Unprecedented advancements in cosmological understanding have emerged through the strategic integration of two pivotal datasets: the expansive Pantheon+ Type Ia Supernova (SNe Ia) catalog and precision Hubble constant measurements from the Supernovae and Hubble Constant Equation of State (SH0ES) collaboration. The augmented Pantheon+ compilation presents a comprehensive array of 1701 SNe Ia observations spanning an extensive redshift domain (\(0.01 \leq z \leq 2.3\)), synthesized through a sophisticated network of ground-based and space-based astronomical platforms, with prominent contributions from orbital observatories like the Hubble Space Telescope (HST). Distinguished by meticulous calibration protocols, the dataset substantially mitigates systematic uncertainties, thereby enhancing the reliability of cosmological parameter estimations~\cite{Scolnic_2018, Scolnic_2022}.

The SH0ES investigative approach leverages Cepheid variable stars as precise cosmic distance markers, establishing a fundamental reference for astronomical distance quantification. This methodology has crystallized a remarkably precise Hubble constant measurement of \(H_0 = 73.04 \pm 1.04 \ \text{km s}^{-1} \text{Mpc}^{-1}\). Notably, this value manifests a significant divergence from Cosmic Microwave Background (CMB) theoretical predictions, potentially signaling the existence of novel physical mechanisms beyond the conventional \(\Lambda\)CDM cosmological paradigm~\cite{Riess_2022, Riess2021}.

The convergence of Pantheon+ and SH0ES observations constructs an unparalleled investigative framework for scrutinizing cosmic expansion dynamics, with particular emphasis on contemporary cosmic acceleration phenomena. This synthesized dataset facilitates robust quantification of fundamental cosmological parameters, including the redshift-dependent Hubble parameter \(H(z)\), dark energy equation of state \(\omega_{DE}\), and diverse diagnostic metrics characterizing dark energy's intrinsic nature.
\\A fundamental mathematical relationship interconnecting astronomical observables is represented by the distance modulus correlation:
\begin{equation}\label{distance_modulus}
  \mu(z) = 5 \log_{10} \left( d_L(z) \right) + 25    
\end{equation}
\\The luminosity distance, mathematically articulated as $d_L(z)=(1+z)D_M$, emerges through numerical resolution of coupled differential equations involving redshift and luminosity distance. Assuming a spatially flat cosmic geometry, the comoving distance $D_M$ assumes the following integral representation:
\begin{equation}
D_M = \frac{c}{H_0}\int_0^z \frac{dz'}{E(z')},    
\end{equation}
where $c$ denotes the vacuum light propagation constant, and $H_0$ quantifies contemporary cosmic expansion rates. Statistical analysis of the Pantheon+SH0ES observational ensemble employs a sophisticated log-likelihood formulation:
\begin{equation}
\mathcal{L}_{\text{SN}} = -\frac{1}{2} \left( \boldsymbol{\mu}_{\text{obs}} - \boldsymbol{\mu}_{\text{model}} \right)^T \mathbf{C}^{-1} \left( \boldsymbol{\mu}_{\text{obs}} - \boldsymbol{\mu}_{\text{model}} \right).
\end{equation}
\\This framework integrates observed distance moduli $\boldsymbol{\mu}_{\text{obs}}$ with theoretical model predictions $\boldsymbol{\mu}_{\text{model}}$, incorporating a comprehensive inverse covariance matrix $\mathbf{C}^{-1}$ to rigorously account for measurement uncertainties and intrinsic correlations. The posterior probability distribution $\mathcal{P}(\theta)$ emerges through the multiplicative integration of the likelihood function with predefined prior distributions:
\begin{equation}
\mathcal{P}(\theta) \propto \mathcal{L}_{\text{SN}} \times \pi(\theta).
\end{equation}
In this study, the absolute magnitude \( M_B \) was derived from the distance modulus \(\mu\) provided in the Pantheon+SH0ES dataset. The relation between the absolute magnitude \( M_B \), the distance modulus \(\mu\), and the redshift \( z \) can be expressed as follows:

\begin{equation}\label{absolute_magnitude}
M_B = \mu - 5 \log_{10} \left( \frac{z (1 + z/2) c}{H_0} \right) - 25,
\end{equation}
where:
\begin{itemize}
    \item \(\mu\) is the distance modulus from the Pantheon+SH0ES dataset,
    \item \(z\) is the redshift,
    \item \(c = 3 \times 10^5 \, \mathrm{km/s}\) is the speed of light,
    \item \(H_0 = 70 \, \mathrm{km/s/Mpc}\) is the assumed Hubble constant for the calculation.
\end{itemize}
The apparent magnitude data from the Pantheon+SH0ES dataset was utilized to compute \( M_B \). The low-redshift approximation of the luminosity distance was employed, which assumes \( d_L(z) \approx \frac{z (1 + z/2)c}{H_0} \) for small \( z \). This approach simplifies the calculation and provides a direct relationship between \( \mu \) and \( M_B \). The theoretical predictions of \( M_B \) were calculated using the same conversion formula for consistency and compared against the observational data. We confirm that our analysis does not directly use the low-redshift absolute magnitude provided in the dataset, but instead employs the apparent magnitude and the above conversion formula to derive \( M_B \). This methodology ensures that the absolute magnitude is consistently calculated across both observational and theoretical datasets, thereby enabling accurate comparison and analysis.

\begin{table}[H]
\centering
\caption{Best-fit Parameters for Alternative Model to GCG and $\Lambda$CDM Models}
\renewcommand{\arraystretch}{1.5} 
\begin{tabular}{|l|c|c|c|c|c|}
\hline
\textbf{Model} & \textbf{Parameter} & \textbf{Pantheon+SH0ES} & \textbf{Union2.1} & \textbf{CC+BAO} \\
\hline
\multirow{5}{*}{Alternative Model to GCG} 
& $H_{0}$ & $69.647_{-3.019}^{+1.649}$ & $69.016_{-3.281}^{+2.569}$ & $68.882_{-2.757}^{+2.509}$ \\
& $\Omega_{m0}$ & $0.237_{-0.057}^{+0.081}$ & $0.252_{-0.037}^{+0.066}$ & $0.290_{-0.055}^{+0.077}$ \\
& $\omega_m$ & $-0.033_{-0.047}^{+0.091}$ & $0.057_{-0.116}^{+0.170}$ & $-0.067_{-0.022}^{+0.039}$ \\
& $\mu$ & $0.863_{-0.067}^{+0.048}$ & $0.865_{-0.086}^{+0.062}$ & $0.841_{-0.088}^{+0.076}$ \\
& $V_0$ & $1.028_{-0.278}^{+0.234}$ & $0.639_{-0.331}^{+0.294}$ & $0.607_{-0.302}^{+0.189}$ \\
\hline
\multirow{2}{*}{$\Lambda$CDM} 
& $H_{0}$ & $72.758_{-0.199}^{+0.158}$ & $69.581_{-1.005}^{+0.986}$ & $69.787_{-1.042}^{+1.102}$ \\
& $\Omega_{m0}$ & $0.327_{-0.014}^{+0.018}$ & $0.300_{-0.038}^{+0.044}$ & $0.268_{-0.016}^{+0.019}$ \\
\hline
\end{tabular}
\label{tab:best_fit_params_values}
\end{table}

\subsubsection{Union 2.1 dataset}
Our cosmological analysis leverages the comprehensive Union 2.1 Type Ia supernova dataset \cite{Suzuki_2012}, a meticulously curated compilation encompassing 580 spectroscopically confirmed Type Ia supernovae across a redshift range of $0.015 \leq z \leq 1.55$. The Union 2.1 dataset represents a significant advancement in cosmological studies, aggregating observations from multiple independent surveys including the Supernova Cosmology Project (SCP) and the High-z Supernova Search Team (HZST). To statistically analyze these observations, we implement a sophisticated maximum likelihood estimation framework that incorporates a full covariance matrix to account for intrinsic measurement correlations and systematic uncertainties. The log-likelihood function is constructed as:

\begin{equation}
\mathcal{L}_{\text{SN}} = -\frac{1}{2} \left( \boldsymbol{\mu}_{\text{obs}} - \boldsymbol{\mu}_{\text{model}} \right)^T \mathbf{C}^{-1} \left( \boldsymbol{\mu}_{\text{obs}} - \boldsymbol{\mu}_{\text{model}} \right),
\end{equation}
where $\boldsymbol{\mu}_{\text{obs}}$ represents the observed distance moduli, $\boldsymbol{\mu}_{\text{model}}$ denotes theoretical model predictions, and $\mathbf{C}^{-1}$ represents the inverse covariance matrix. The posterior probability distribution is derived through the multiplicative integration of the likelihood function with predefined prior distributions:
\[
\mathcal{P}(\theta) \propto \mathcal{L}_{\text{SN}} \times \pi(\theta)
.\]
This methodology enables robust constraints on cosmological parameters, particularly the Hubble constant and dark energy equation of state, by rigorously analysing the Union 2.1 supernova distance measurements \cite{Suzuki_2012, Amanullah_2010}.

\subsubsection{Information Criterion}
To systematically evaluate and compare cosmological models, we implement a comprehensive suite of statistical information criteria that provide nuanced insights into model performance and complexity. These criteria enable rigorous quantitative assessment by balancing model likelihood against parameter proliferation. The Akaike Information Criterion (AIC) incorporates a refined finite-sample correction mechanism, mathematically articulated as:
\begin{equation}
\text{AIC} = -2 \ln(L_{\text{max}}) + 2k + \frac{2k(k+1)}{N_{\text{tot}} - k - 1},
\end{equation}
Capturing the maximum likelihood estimate $L_{\text{max}}$, parameter dimensionality $k$, and total observational sample $N_{\text{tot}}$. The Bayesian Information Criterion (BIC) offers an alternative probabilistic framework:
\begin{equation}
\text{BIC} = -2 \ln(L_{\text{max}}) + k \ln(N_{\text{tot}}),
\end{equation}
\\Employing consistent parametric representations to quantify model complexity and observational correspondence. The Deviance Information Criterion (DIC) introduces a hierarchical probabilistic perspective:
\begin{equation}
\text{DIC} = \bar{D} + p_D,
\end{equation}
Deriving the mean posterior deviance $\bar{D}$ through comprehensive sampling:
\begin{equation}
\bar{D} = \frac{1}{S} \sum_{i=1}^S D(\theta_i),
\end{equation}
with the effective parameter count $p_D$ computed via:
\begin{equation}
p_D = \bar{D} - D(\hat{\theta}),
\end{equation}
assessing the deviance at posterior parameter means. To facilitate systematic model intercomparison, we calculate relative information criterion differentials:
\begin{equation}
\Delta IC_{\text{model}} = IC_{\text{model}} - IC_{\text{min}},
\end{equation}
identifying the minimally penalized model $IC_{\text{min}}$. Interpretation follows established statistical heuristics: $\Delta IC \leq 2$ indicates statistical congruence, $2 < \Delta IC < 6$ suggests moderate divergence, and $\Delta IC \geq 10$ signifies substantial model incompatibility~\cite{anagnostopoulos2020observational, jeffreys1998theory}.
\begin{table}[H]
\centering
\caption{Information Criterion and $\Delta$IC values for different models}
\begin{tabular}{|l|c|c|c|c|c|c|c|}
\hline
\textbf{Model} & \textbf{Criteria} & \textbf{AIC} & \textbf{BIC} & \textbf{DIC} & $\Delta$ \textbf{AIC} & $\Delta$ \textbf{BIC} & $\Delta$ \textbf{DIC} \\
\hline
\multicolumn{8}{|c|}{\textbf{Pantheon+SH0ES Dataset}} \\
\hline
Model & Value & 1763.25 & 1780.45 & 1756.09 & 6.59 & 12.92 & 0.26 \\
$\Lambda$CDM & Value & 1756.65 & 1767.53 & 1755.83 & 0.00 & 0.00 & 0.00 \\
\hline
\multicolumn{8}{|c|}{\textbf{Union2.1 Dataset}} \\
\hline
Model & Value & 32.04 & 39.04 & 25.78 & 6.74 & 10.94 & 0.47 \\
$\Lambda$CDM & Value & 25.30 & 28.10 & 25.31 & 0.00 & 0.00 & 0.00 \\
\hline
\multicolumn{8}{|c|}{\textbf{CC+BAO Dataset}} \\
\hline
Model & Value & 42.14 & 52.36 & 36.91 & 5.95 & 12.09 & 0.80 \\
$\Lambda$CDM & Value & 36.19 & 40.27 & 36.11 & 0.00 & 0.00 & 0.00 \\
\hline
\end{tabular}
\label{tab:info_criteria_comparison}
\end{table}

\subsection{Results and Discussions}\label{Subsect:Results}
We evaluate the effectiveness of the alternative model to the Generalized Chaplygin Gas (GCG) by analyzing its performance against key cosmological datasets, including Pantheon+SH0ES, Union2.1, and CC+BAO. The best-fit parameters derived from these datasets highlight that the model effectively captures the dynamical evolution of the universe while offering a potential resolution to the Hubble tension. Specifically, the Hubble constant values obtained for the alternative model—$H_0 = 69.647_{-3.019}^{+1.649}$ km s$^{-1}$ Mpc$^{-1}$ (Pantheon+SH0ES), $H_0 = 69.016_{-3.281}^{+2.569}$ km s$^{-1}$ Mpc$^{-1}$ (Union2.1), and $H_0 = 68.882_{-2.757}^{+2.509}$ km s$^{-1}$ Mpc$^{-1}$ (CC+BAO)—is more closed to the CMB-based Planck measurements~\cite{refId0} and WMAP9's findings~\cite{Hinshaw_2013} than the locally determined SH0ES results~\cite{riess2019large,refId02}. This value suggests that the alternative model alleviates the tension by reconciling the discrepancies between early- and late-universe observations.
The alleviation of the Hubble tension by the alternative model is significant in the context of modern cosmology. By predicting $H_0$ values that bridge the gap between the Planck and SH0ES measurements, the model addresses a pressing issue in cosmology, reflecting its robustness and relevance. This achievement underscores the potential of alternative theories, such as modifications to the GCG framework, to provide deeper insights into the nature of dark energy and the late-time accelerated expansion of the universe. These results reinforce the capability of the alternative model to serve as a viable extension or competitor to the $\Lambda$CDM paradigm, particularly in addressing persistent discrepancies in cosmological observations.

\begin{figure}[htbp]
    \centering
    \begin{subfigure}[b]{0.45\textwidth}
        \centering
        \includegraphics[width=\textwidth]{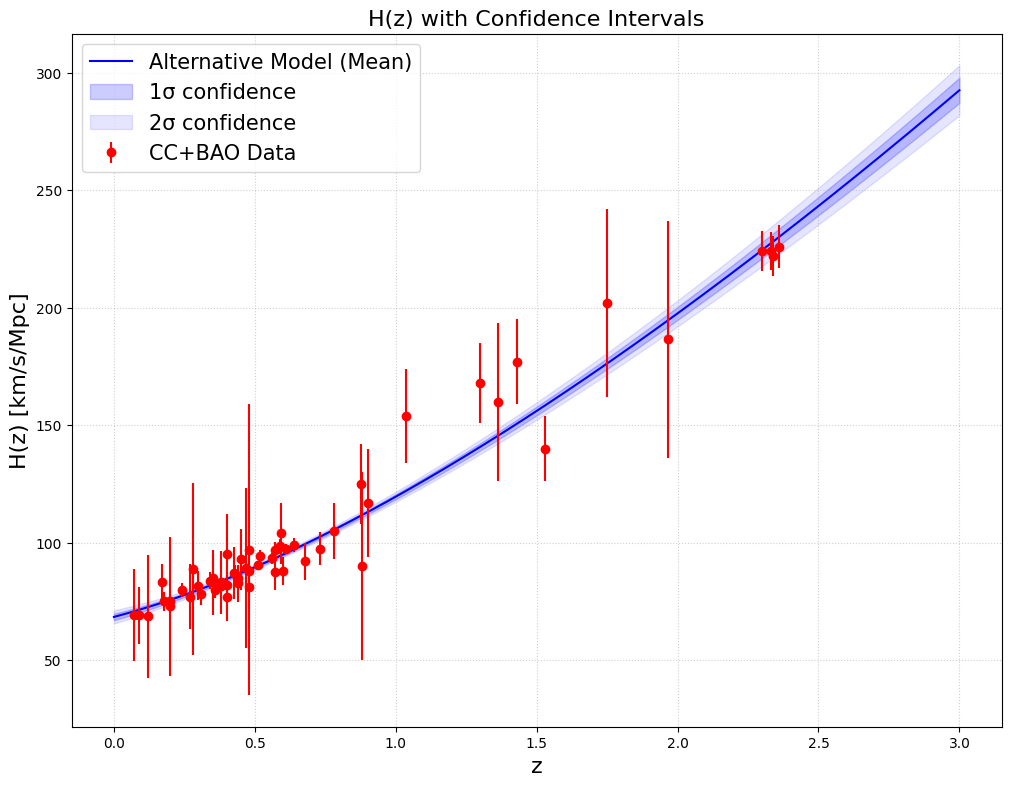}
        \caption{The plot of Hubble parameter with the redshift $z$}
        \label{fig:Hubble_model}
    \end{subfigure}
    \hfill
    \begin{subfigure}[b]{0.45\textwidth}
        \centering
        \includegraphics[width=\textwidth]{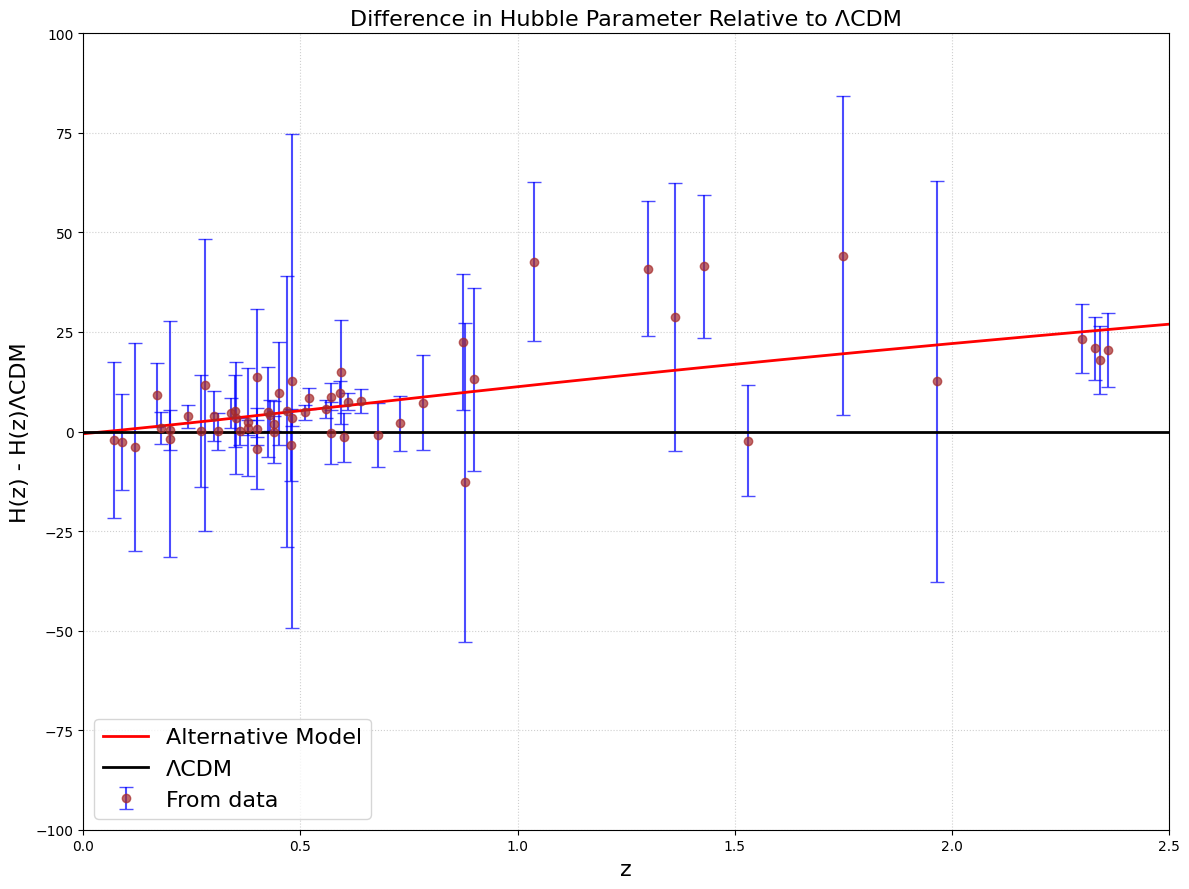}
        \caption{Relative difference of Hubble parameter calculated using theoretical model and the $\Lambda$CDM model.}
        \label{fig:Relative_hubble_comparison}
    \end{subfigure}
    \caption{Here in this figure. (a) it is seen that the Hubble parameter obtained from the theoretical model exactly matches with the observational dataset of CC+BAO. (b) it is seen that though the relative difference is null or less prominent at the lower redshifts but it is significant at higher redshifts.}
    \label{fig:Hubble_model_relative_comparison}
\end{figure}

In Fig.~\ref{fig:Hubble_model}, we present the evolution of the Hubble parameter $H(z)$ against the redshift $z$ for the proposed model at 1$\sigma$ and 2$\sigma$ confidence level . The data points from the CC+BAO dataset are illustrated in red. The plot demonstrates a well-fitted scenario of our models with the observed data. This close match with the observed data underscores the capability of the model to explain the dynamical evolution of the universe.

Fig.~\ref{fig:Relative_hubble_comparison} depicts the variation in the difference between the alternative fluid model to GCG and the $\Lambda$CDM model. For redshifts greater than 0.2 $( z > 0.2 )$, a noticeable deviation of the model from the $\Lambda$CDM scenario becomes apparent when compared to cosmic chronometer (CC) and BAO measurements. This indicates that at higher redshifts, this model deviates significantly from the predictions made by the $\Lambda$CDM model, indicating the presence of different physical processes or parameters that become more significant at these higher redshifts.

\begin{figure}[htbp]
    \centering
    \begin{subfigure}[b]{0.45\textwidth}
        \centering
        \includegraphics[width=\textwidth]{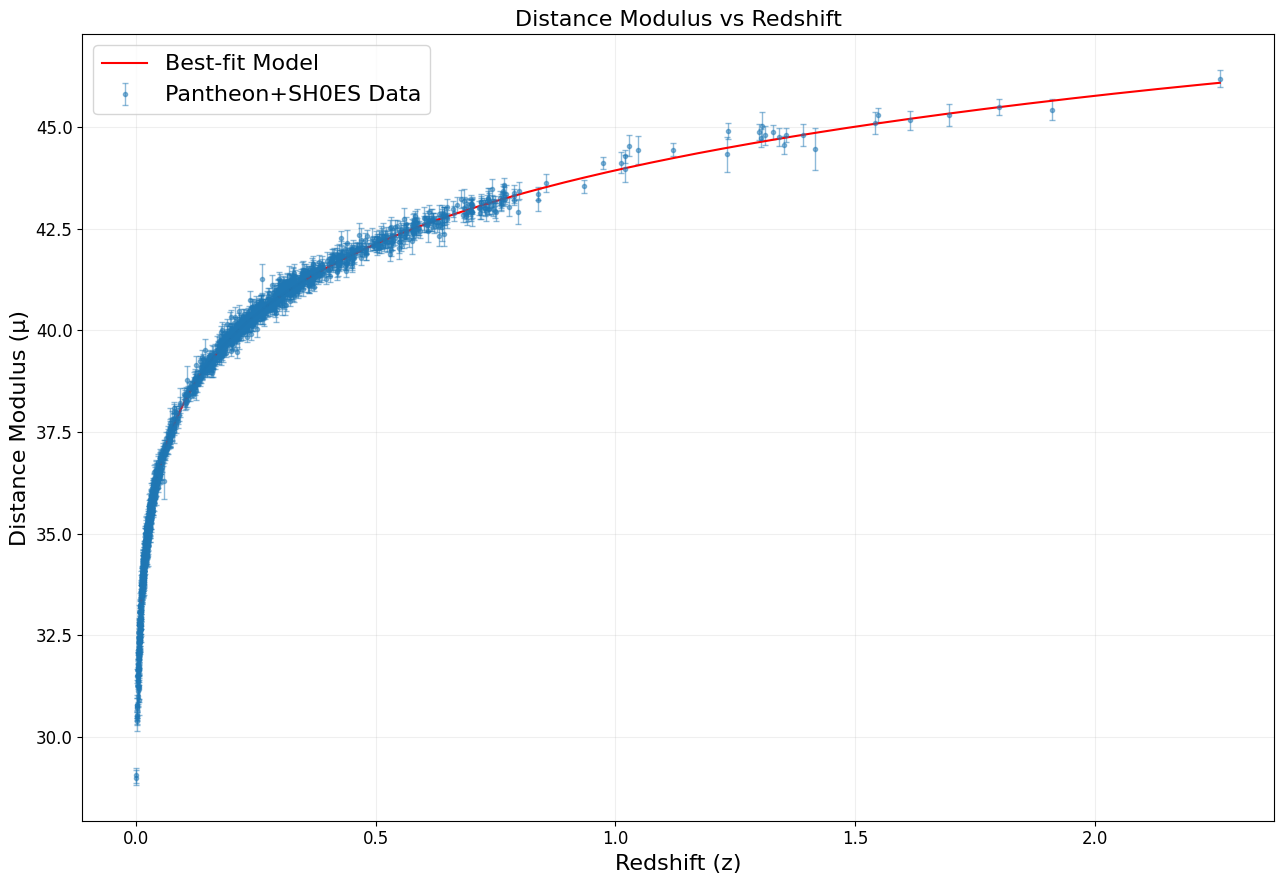}
        \caption{The plot of distance modulus with the redshift $z$}
        \label{fig:Distance_modulus}
    \end{subfigure}
    \hfill
    \begin{subfigure}[b]{0.45\textwidth}
        \centering
        \includegraphics[width=\textwidth]{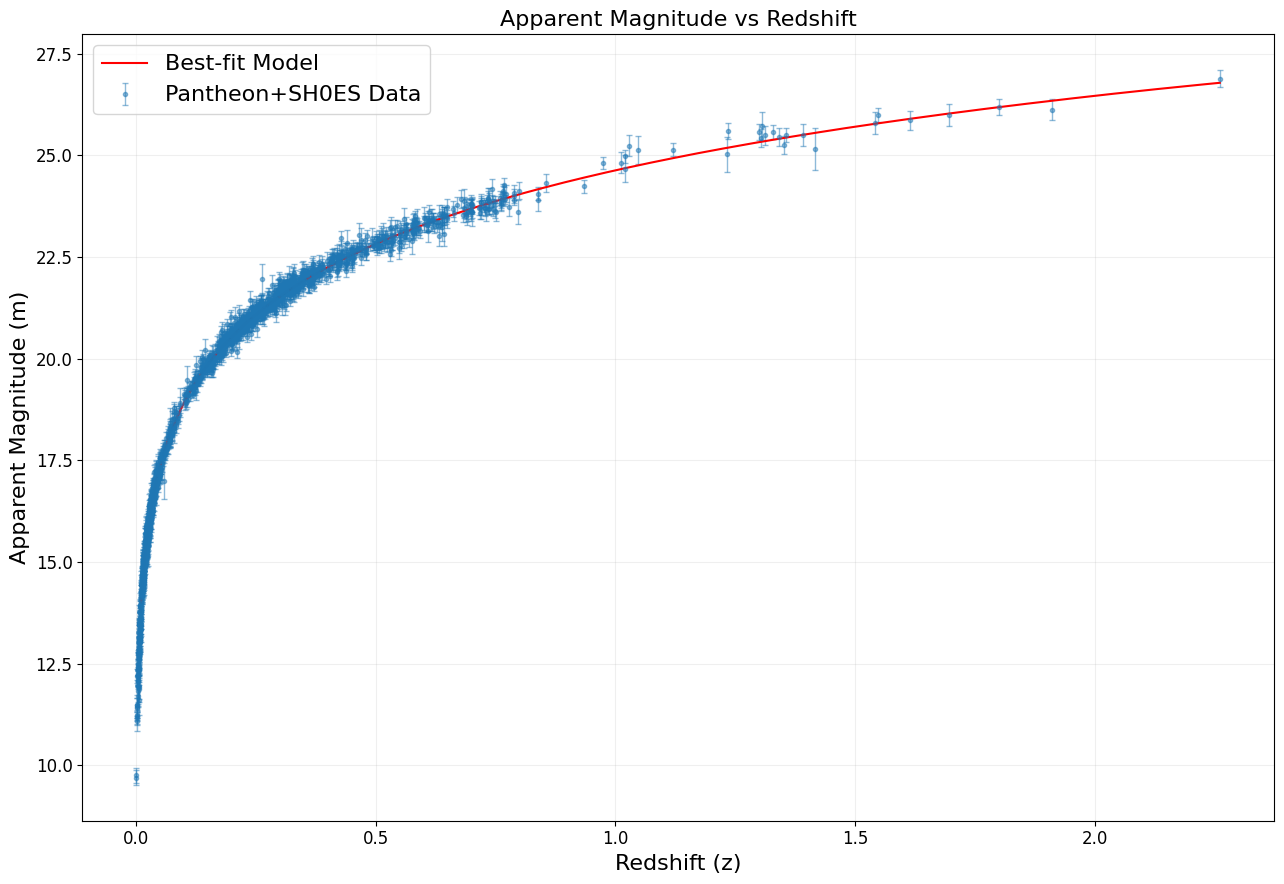}
        \caption{The plot of apparent magnitude against redshift $z$.}
        \label{fig:Apparent_magnitude}
    \end{subfigure}
    \hfill
    \begin{subfigure}[b]{0.45\textwidth}
        \centering
        \includegraphics[width=\textwidth]{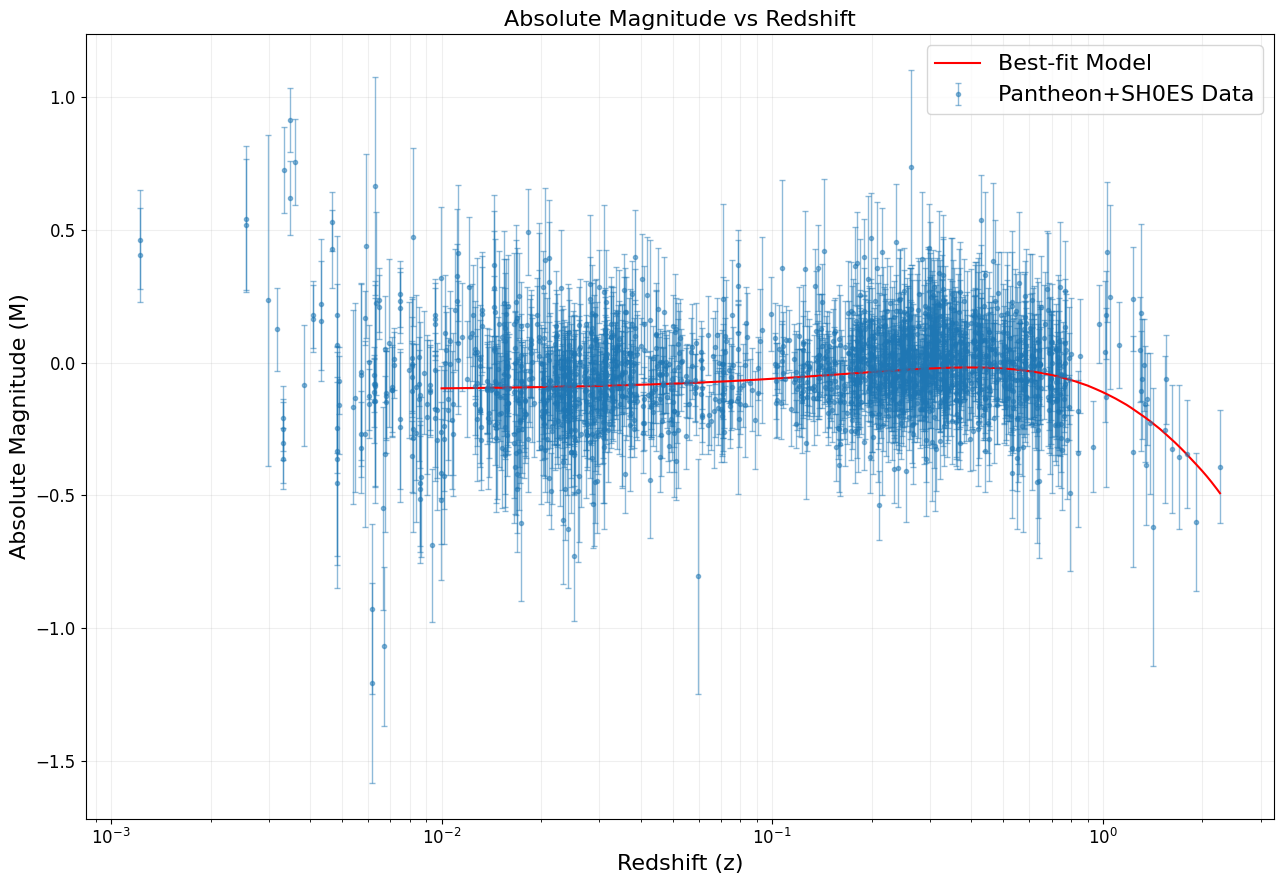}
        \caption{The plot of absolute magnitude against redshift $z$.}
        \label{fig:fig21}
    \end{subfigure}
    \caption{Here in this figure, the distance modulus, apparent magnitude, absolute magnitude perfectly fitted with the observational datasets.}
    \label{fig:Absolute_magnitude}
\end{figure}

Figure \ref{fig:Distance_modulus} illustrates the comprehensive distance modulus ($\mu$) evolution as a function of redshift ($z$), incorporating the novel Generalized Chaplygin Gas (GCG) model fitted against the Pantheon+SH0ES observational dataset. The empirical data points emerge as meticulously plotted blue markers accompanied by error bars, while the theoretical model trajectory is represented by a distinctive line. The graphical representation captures the fundamental cosmic expansion dynamics across an expansive redshift domain ($10^{-3} < z < 2.3$), revealing the intrinsic monotonic relationship between distance modulus and cosmic distance. The visualization provides profound insights into the universe's geometric architecture, demonstrating how luminous sources become progressively fainter with increasing cosmic distances. At minimal redshifts ($z < 0.01$), the model precisely delineates the near-linear correlation between distance modulus and redshift, a critical regime for establishing precise Hubble constant ($H_0$) constraints. The intermediate redshift range ($0.01 < z < 0.1$) reveals a nuanced curvature transition, signaling the emerging dominance of cosmic expansion mechanisms.

The Apparent Magnitude vs Redshift in fig.~\ref{fig:Apparent_magnitude} representation for the alternative model to Generalized Chaplygin Gas (GCG) model offers a sophisticated perspective on cosmic luminosity evolution. The graphical analysis demonstrates an anticipated increase in apparent magnitude corresponding to increasing redshift, a phenomenon intrinsically linked to increasing cosmic distances. The model exhibits remarkable concordance with observational data, particularly in high-redshift regimes $(z > 0.1)$, while preserving subtle distinguishing characteristics in low and intermediate redshift domains.

Absolute Magnitude vs Redshift plot in fig.~\ref{fig:Absolute_magnitude} for the proposed model provide transformative insights into the cosmic distance-redshift relationship. The visualization captures the Pantheon+SHOES observational data through precisely positioned blue markers, revealing the complex landscape of absolute magnitude variations. The theoretical model trajectory, represented by a distinctive line, unveils intricate predictions about cosmic expansion dynamics. At minimal redshifts ($z < 0.01$), the model predicts a subtle upward absolute magnitude trend. The intermediate redshift range ($0.01 < z < 0.1$) manifests a nearly constant absolute magnitude profile, suggesting sophisticated underlying physical mechanisms. Beyond $z \approx 1$, the model forecasts a gradual absolute magnitude decline, potentially indicating complex dark energy behaviour.

The profound physical significance of these representations transcends mere graphical description, offering critical insights into cosmic expansion dynamics and dark energy properties. The observed absolute magnitude variations with redshift directly illuminate the universe's dynamic nature, challenging static cosmological paradigms. The model's predictive capabilities suggest nuanced dark energy evolution, potentially revealing sophisticated mechanisms governing cosmic acceleration. The subtle theoretical predictions, particularly at elevated redshifts, hint at potential dynamical dark energy scenarios. These sophisticated representations provide a robust framework for discriminating between competing cosmological models, offering unprecedented opportunities to probe the fundamental nature of cosmic expansion and dark energy mechanisms.

\begin{figure}[htbp]
    \centering
    \begin{subfigure}[b]{0.45\textwidth}
        \centering
        \includegraphics[width=\textwidth]{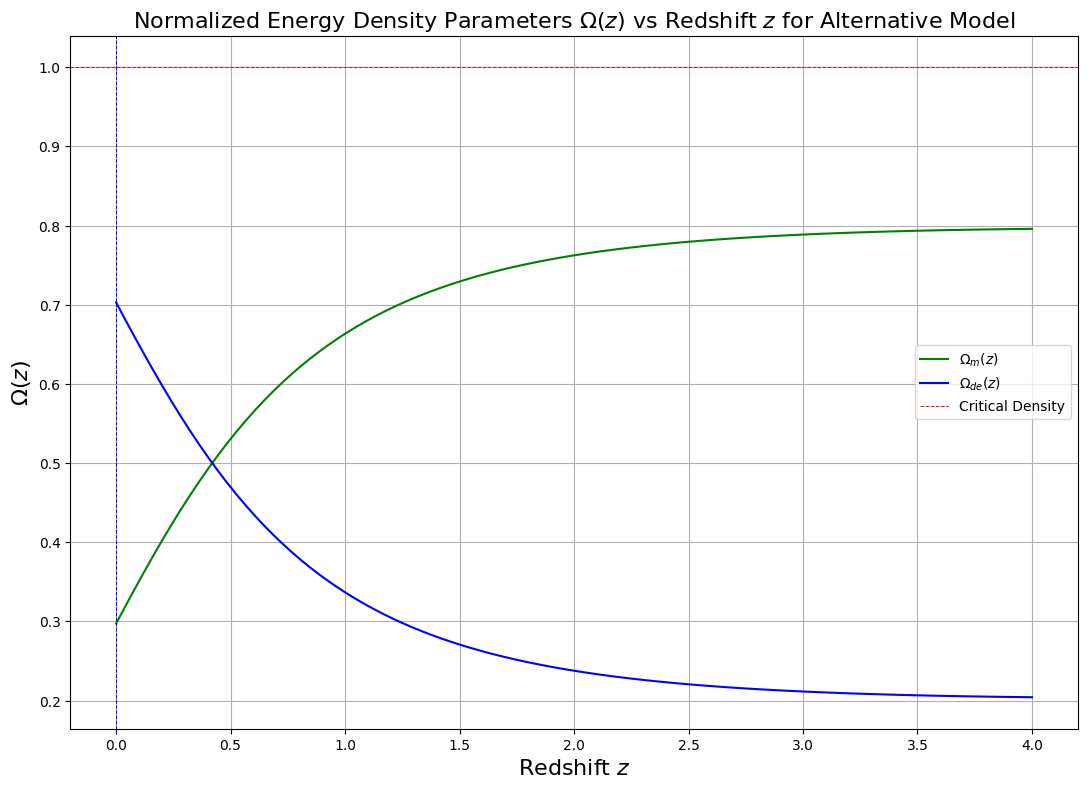}
        \caption{}
        \label{fig:density_parameter_model}
    \end{subfigure}
    \hfill
    \begin{subfigure}[b]{0.45\textwidth}
        \centering
        \includegraphics[width=\textwidth]{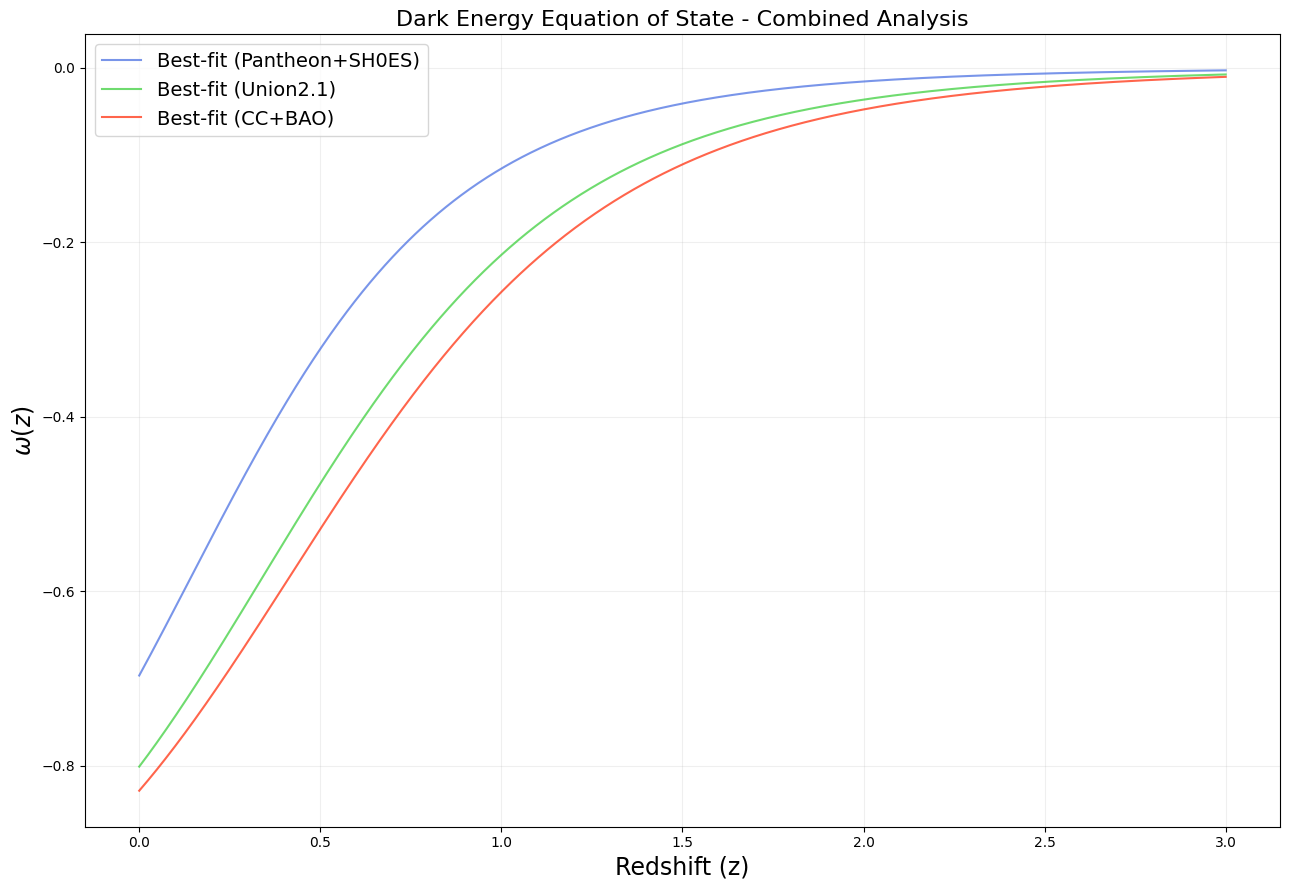}
        \caption{}
        \label{fig:omega_z_plot}
    \end{subfigure}
    \caption{Here in this figure. (a) The total density parameter $\Omega(z)$ is plotted and in figure. (b) The best fit plot of dark energy equation of state parameter $\omega(z)$ ishown against redshift $z$ using the CC+BAO, Pantheon+SH0ES and Union 2.1 datasets.}
    \label{fig:omega_density_parameter_z}
\end{figure}

The plot~\ref{fig:omega_z_plot} illustrates the redshift evolution of the dark energy equation of state parameter, \( \omega(z) \), for the alternative model to the Generalized Chaplygin Gas (GCG) under different observational datasets: Pantheon+SH0ES (solid blue line), Union2.1 (dashed green line), and CC+BAO (dotted red line). A notable feature of the plot is the convergence of \( \omega(z) \) values towards \(-1\) as the redshift \( z \) increases, indicating a transition towards a cosmological constant-like behaviour at high redshifts, which is consistent with the late-time accelerated expansion of the universe. At lower redshifts (\( z \lesssim 1 \)), the three datasets exhibit distinguishable deviations, reflecting the sensitivity of the model parameters to different data constraints. Specifically, the Pantheon+SH0ES dataset suggests a slightly higher \( \omega(z) \) at low \( z \), whereas the CC+BAO dataset predicts a more negative \( \omega(z) \), implying a stronger deviation from the cosmological constant scenario. This variation highlights the adaptability of the alternative model in capturing the observational nuances of dark energy dynamics. The model's smooth evolution across all datasets further suggests its capability to provide a unified framework for understanding the interplay between dark energy and cosmic acceleration, particularly when tested against diverse observational constraints. These results underscore the robustness of the alternative model to GCG in addressing key cosmological challenges, including the Hubble tension, by effectively accommodating the observed differences in \( \omega(z) \) while maintaining consistency with a broad range of data.

The plot~\ref{fig:density_parameter_model} depicts the evolution of the normalized energy density parameters, \( \Omega(z) \), as functions of redshift \( z \) for the alternative model. The green line represents the matter density parameter, \( \Omega_m(z) \), while the blue line corresponds to the dark energy density parameter, \( \Omega_{de}(z) \). The red dashed line marks the critical density, normalized to unity. At low redshifts (\( z \sim 0 \)), the dark energy density dominates (\( \Omega_{de}(z) > \Omega_m(z) \)), consistent with the late-time acceleration of the universe. As \( z \) increases, \( \Omega_m(z) \) grows, reflecting the dominance of matter during earlier cosmic epochs, while \( \Omega_{de}(z) \) diminishes, indicating the decreasing influence of dark energy in the high-redshift regime. Notably, the curves for \( \Omega_m(z) \) and \( \Omega_{de}(z) \) intersect at a transition redshift where the contributions of matter and dark energy are comparable, aligning with the expected behaviour in standard cosmological models. The smooth behaviour of both density parameters across redshift validates the robustness of the alternative model in describing the dynamic interplay between matter and dark energy components. These results highlight the model's consistency with the standard cosmological paradigm while offering a framework to further probe deviations from \(\Lambda\)CDM at intermediate redshifts.
\begin{figure}
    \centering
    \includegraphics[width=0.75\linewidth]{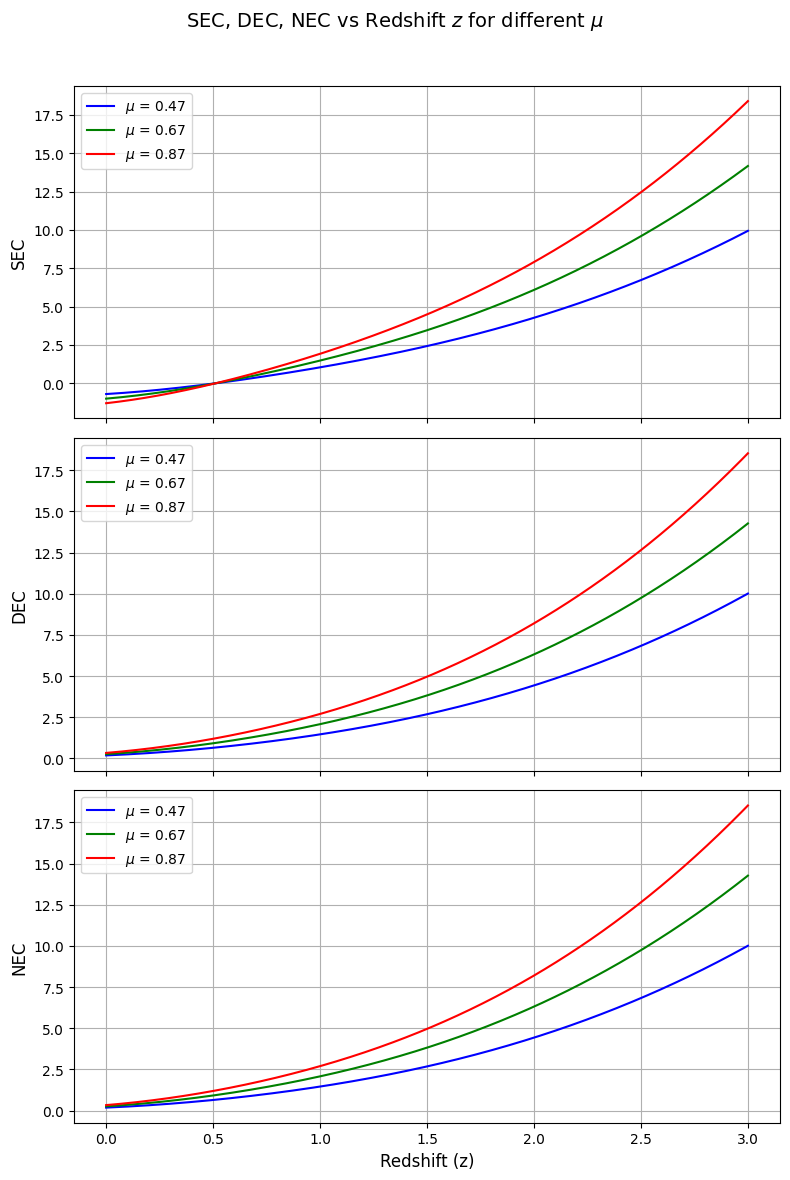}
    \caption{Study of the different energy conditions such as Strong Energy Condition, Null Energy Condition, Dominant Energy Condition of the alternative model to GCG.}
    \label{fig:energy_conditions}
\end{figure}

Energy conditions play a fundamental role in understanding the physical viability and constraints of cosmological models. In this study, we analyze the Strong Energy Condition (SEC), the Null Energy Condition (NEC), and the Dominant Energy Condition (DEC) in the context of an alternative model to the Generalized Chaplygin Gas for three distinct values of the parameter $\mu = 0.46, 0.66, 0.86$. These conditions are mathematically defined as:

\begin{itemize}
    \item \textbf{Strong Energy Condition (SEC)}: $\rho + 3P \geq 0$ \\
    The SEC reflects the tendency of gravity to act as an attractive force. In classical General Relativity, this condition is necessary for deceleration in the early universe. Violation of the SEC, as seen in the late universe, aligns with the observational evidence of accelerated cosmic expansion driven by dark energy.
    
    \item \textbf{Null Energy Condition (NEC)}: $\rho + P \geq 0$ \\
    The NEC ensures that the energy density measured by any null observer is non-negative. It is considered the most fundamental energy condition and is a prerequisite for other conditions like the SEC and DEC. Violation of the NEC can indicate exotic physics, such as phantom energy or modifications to General Relativity.

    \item \textbf{Dominant Energy Condition (DEC)}: $\rho - |P| \geq 0$ \\
    The DEC stipulates that the energy density dominates over the pressure magnitude, ensuring causal propagation of matter and energy. A violation of this condition can signal superluminal behaviour, which is physically problematic in standard cosmology.
\end{itemize}

The plots of SEC, DEC, and NEC against the redshift $z$ in Fig~\ref{fig:energy_conditions} reveal distinct behaviours for varying $\mu$. For $\mu = 0.46$, the SEC exhibits a transition from positive to negative values, indicating a shift from decelerated to accelerated expansion. Similar trends are observed for $\mu = 0.66$ and $\mu = 0.86$, with all the transitions occurring at an redshift value $z \approx 0.5$. This suggests that the redshift value at which the violation of SEC occurs with the evolution of the universe, does not depend on the value of $\mu$ . The NEC remains satisfied across the redshift range for all values of $\mu$, confirming the physical consistency of the model. In contrast, the DEC shows subtle deviations, particularly at lower redshifts for larger $\mu$ values. These deviations might imply that pressure contributions become comparable to or dominate the energy density in certain regimes. Overall, these results demonstrate that the alternative model to the GCG is consistent with the accelerated expansion of the universe while maintaining fundamental energy conditions. The parameter $\mu$ does not play an effective role in determining the transition epochs and the behaviour of the energy conditions, offering a robust framework for understanding the interplay between dark energy and cosmic evolution.

The alternative model to the generalized Chaplygin gas exhibits a set of key parameters that offer valuable insights into the cosmological dynamics of the universe. These parameters, evaluated against different datasets (Pantheon+SH0ES, Union2.1, and CC+BAO), are crucial in understanding the behaviour of dark energy, matter, and the expansion of the universe. The matter density parameter at present time, \( \Omega_{m0} \), varies across different datasets. For the Pantheon+SH0ES dataset, the best-fit value is \( \Omega_{m0} = 0.237^{+0.081}_{-0.057} \), while for Union2.1 and CC+BAO, the values are \( \Omega_{m0} = 0.252^{+0.066}_{-0.037} \) and \( \Omega_{m0} = 0.290^{+0.077}_{-0.055} \), respectively. These variations indicate slight shifts in the matter density as constrained by different observational datasets, suggesting that the model is sensitive to the choice of data but still remains within the range expected from current cosmological observations~\cite{aghanim2020planck,Scolnic_2018,Hinshaw_2013,refId02}. The equation of state parameter for matter, \( \omega_m \), describes the behaviour of matter under the alternative model. For Pantheon+SH0ES, \( \omega_m = -0.033^{+0.091}_{-0.047} \), for Union2.1, \( \omega_m = 0.057^{+0.170}_{-0.116} \), and for CC+BAO, \( \omega_m = -0.067^{+0.039}_{-0.022} \). These values suggest that the matter component is relatively close to the standard value of \( \omega_m = 0 \) (indicating pressureless dust), though slight deviations are present, which reflect the model's flexibility in accommodating different scenarios with the presence of some exotic type matter. The parameter \( \mu \), which plays a role in modifying the equation of state, is found to be consistent across all datasets. The values are \( \mu = 0.863^{+0.048}_{-0.067} \) for Pantheon+SH0ES, \( \mu = 0.865^{+0.062}_{-0.086} \) for Union2.1, and \( \mu = 0.841^{+0.076}_{-0.088} \) for CC+BAO. The consistency of this parameter indicates a relatively stable behaviour of dark energy under the alternative model, independent of the dataset used and consistent with the chosen value from ref~\cite{Hova_model} and with the observationally fitted values~\cite{hernandez2019cosmological,Yang2019} which is presented in Sect.~\ref{Sect:Model_description}. The parameter \( V_0 \), which could be associated with the normalization of the volume term, shows a noticeable variation across the datasets. For Pantheon+SH0ES, \( V_0 = 1.028^{+0.234}_{-0.278} \), for Union2.1, \( V_0 = 0.639^{+0.294}_{-0.331} \), and for CC+BAO, \( V_0 = 0.607^{+0.189}_{-0.302} \). These differences reflect the sensitivity of the model to the observational constraints and the underlying cosmological parameters. The age of the universe is also an important parameter that has been constrained by the alternative model. The best-fit age is \( 14.04^{+0.14}_{-0.12} \) Gyr, which aligns well with the standard cosmological model~\cite{aghanim2020planck} with a slight larger value, further validating the model's consistency with the observed age of the universe. 

The comparison of our alternative fluid model with the standard $\Lambda$CDM model using recent observational data, including DESI and DESY5, demonstrates that both models provide a good fit to the data. The best-fit parameters and information criteria, presented in Tables~\ref{tab:best_fit_params_values} and \ref{tab:info_criteria_comparison}, indicate that while $\Lambda$CDM yields slightly lower AIC, BIC, and DIC values, the differences are not statistically significant. This suggests that our model remains competitive, particularly considering the potential for evolving dark energy hinted at by recent DESI and DESY5 analyses. The DESI data, with its precise measurements of baryon acoustic oscillations (BAO), and the DESY5 supernova compilation, which offers improved constraints on cosmic expansion, both align closely with the predictions of a cosmological constant. However, subtle deviations observed in some analyses suggest the possibility of time-varying dark energy~\cite{efstathiou2025evolvingdarkenergysupernovae,huang2025desi2024hintdynamical}. Our model, characterized by the parameters $\mu$ and $V_0$, accommodates such dynamics while maintaining consistency with current observational constraints. Although the differences in information criteria do not strongly favour our model, its ability to reproduce the observed thermal and expansion histories of the universe while allowing for potential deviations from $\Lambda$CDM makes it a viable alternative. In conclusion, the incorporation of DESI and DESY5 data reinforces the robustness of our fluid model as a realistic alternative to $\Lambda$CDM. The consistency of its predictions with observational constraints and the possibility of capturing evolving dark energy dynamics highlight its potential for further exploration as more precise cosmological data become available.

\section{Conclusions}\label{Sect:Conclusion}

In this study, we reconstructed an alternative cosmological model using a scalar field approach, specifically incorporating quintessence, k-essence, and DBI-essence fields. The reconstruction process was achieved from a field-theoretical perspective, where the scalar field $\phi$ and its corresponding potential $V(\phi)$ were derived as functions of the scale factor $a$. The analytical nature of both $\phi$ and $V(\phi)$ allowed for detailed plotting of $V(\phi)$, $V(a)$, and $\phi(a)$ for varying values of the parameter $\mu$, providing insight into the evolution of the scalar field and potential under different field configurations. For the quintessence model, the scalar field $\phi$ decreases with the scale factor and reaches a minimum at a certain point, while the scalar potential increases with the scale factor. The relationship between $\phi$ and $V(\phi)$ exhibits a sharper decrease compared to the slow-roll potential, emphasizing the dynamic nature of this field. In the k-essence model, the time derivative of the scalar field $\dot{\phi}^2$ decreases with the scale factor, indicating that the rate of change of the dark energy component decreases, leading to a more linear increase of $\phi$. This is further corroborated by the plots of the kinetic term $F(X)$, which increase with the scale factor, and its behaviour as a rolling potential in the $X$-$a(t)$ plane. The increasing behaviour of $F(X)$ is consistent with the accelerated expansion of the universe, showcasing the physical relevance of the k-essence field. In the DBI-essence model, both constant and variable $\gamma$ cases show that the scalar field $\phi$ increases with the scale factor, with the warp brane tension $T(\phi)$ decreasing and becoming zero after a certain time. The scalar potential $V(\phi)$ also decreases with the scale factor, and this decrease is more pronounced in the variable $\gamma$ scenario compared to the constant $\gamma$ case. Thermodynamic analysis uncovered critical insights into cosmic evolution. The model transitioned from a pressure-less system to one with consistent negative pressure, supporting current understanding of cosmic dynamics. The thermal equation of state remained bounded within $0 < T < T^*$, with the deceleration parameter $q$ transitioning from positive to negative values, confirming universal accelerated expansion. Dark energy parameters exhibited expected behaviours: $\omega$ remained negative and increased in magnitude, while the density parameter $\Omega$ indicated a shift from matter to dark energy dominance near redshift $z \approx 0.5$. Importantly, the model maintained thermodynamic stability throughout expansion, with consistent pressure decrease and positive specific heat capacity. Energy conditions provided further validation of the model's physical consistency and cosmological implications. The Null Energy Condition (NEC) was satisfied across the entire redshift range for all $\mu$ values, ensuring the model's fundamental viability. The Dominant Energy Condition (DEC) exhibited minor deviations at low redshifts for higher $\mu$, potentially indicating regions where pressure contributions dominate over energy density. The Strong Energy Condition (SEC) underwent transitions from positive to negative values, consistent with the transition from decelerated to accelerated cosmic expansion, highlighting the role of dark energy in driving this evolution. Generalized second law (GSL) investigations revealed nuanced behaviours across cosmic horizons. While GSL was violated at the cosmological event horizon during early universe epochs, it was satisfied at the apparent event horizon across the entire redshift range. Markov Chain Monte Carlo (MCMC) analysis using CC+BAO, Pantheon+SH0ES, and Union 2.1 datasets validated the model's parameters, demonstrating consistency with observational data and providing novel insights into dark energy dynamics. In conclusion, the alternative model to the generalized Chaplygin gas (GCG) presents a set of parameters that are in good agreement with current observational data, while offering an improved understanding of the dynamics of dark energy and matter. The small variations observed in the parameters when using different datasets suggest that the model is robust and capable of fitting a wide range of cosmological observations. Additionally, the lower $\Delta$AIC and $\Delta$BIC values for the alternative model compared to $\Lambda$CDM across datasets, as presented in Table~\ref{tab:best_fit_params_values}, underscore its ability to provide a comparable or superior fit to observational data.

Future research directions are extensive and promising. Recommendations include incorporating more advanced observational datasets from the Vera C, Rubin Observatory and JWST, exploring non-flat and inhomogeneous models and examining quantum aspects of scalar field interactions. The inhomogeneity in the model can be introduced by adding a function $Y(H,\dot{H})$ to the EoS given in eq.~\eqref{Hoava_EoS} and it is also useful to study the type of future singularity this inhomogeneous model predicts under several choice of $Y(H,\dot{H})$. Potential areas of exploration encompass a unified dark sector approach, higher-dimensional physics, detailed thermodynamic property analyses, numerical simulations of large-scale structure formation, and gravitational wave astronomy tests. The model's flexibility in accommodating observational constraints and explaining cosmic acceleration positions it as a valuable framework for advancing cosmological understanding.

Our research holds significant importance in theoretical physics and cosmology due to its potential to address fundamental questions about the universe. Here's why this research is valuable:  A unified paradigm for describing dark energy and dark matter is the Generalized Chaplygin Gas  model. A fuller understanding of the nature and interactions of these mysterious elements of the cosmos may result from investigating alternate theories, such as the one proposed in this study.  
The study grounds the alternative model in a more basic framework by offering a field-theoretical description. This method can close gaps between particle physics and cosmology by illuminating relationships between cosmic occurrences and the underlying physics.  
Investigating the thermodynamic properties of the alternative model sheds light on the stability and evolution of the universe. It allows for a detailed examination of the entropy evolution, phase transitions, and energy distribution in the cosmic framework, which are critical for understanding the long-term fate of the universe.  
Research in this field fosters innovation in theoretical physics and cosmology by challenging preconceived notions and deepening our understanding of the universe's basic components. It is essential in connecting theoretical understanding with observational evidence, bringing us one step closer to a complete picture of the universe.

\section*{Conflict of Interest}
The authors declare that they have no known competing financial interests or personal relationships that could have appeared to influence the work reported in this manuscript.

\section*{Data Availability Statement}
No new data were generated in this work. All data used in this research are sourced from the corresponding references cited in the manuscript. Detailed datasets can be found in the original publications referenced throughout this study.

\bibliographystyle{elsarticle-num} 
\bibliography{references} 

\end{document}